\documentclass[12pt,letterpaper]{article}
\pdfoutput=1

\usepackage{amsmath,amssymb,calc, amsthm,bbm, epsfig,psfrag, mathtools,comment}
\usepackage{mathrsfs}
\usepackage{graphicx, enumerate}
\usepackage{float}

\usepackage[numbers,sort&compress]{natbib}

\usepackage[dvipsnames]{xcolor}
\usepackage{tikz}
\usepackage{tikz-cd}
\usepackage{physics}
\usepackage{tensor}

\usepackage[utf8]{inputenc} 
\definecolor{azure}{rgb}{0.0, 0.5, 1.0}
\definecolor{darkblue}{rgb}{0.15,0.35,0.7}
\definecolor{reddish}{rgb}{0.65, 0.2, 0.2}
\definecolor{brandeisblue}{rgb}{0.0, 0.44, 1.0}
\definecolor{ceruleanblue}{rgb}{0.16, 0.32, 0.75}
\definecolor{indigo(dye)}{rgb}{0.0, 0.25, 0.42}
\usepackage[linktocpage=true]{hyperref}
\hypersetup{
colorlinks=true,
citecolor=ceruleanblue,
linkcolor=ceruleanblue,
urlcolor=ceruleanblue,
pdfauthor={},
pdftitle={},
pdfsubject={}
}

\definecolor{indigo(dye)}{rgb}{0.0, 0.25, 0.42}

\newcommand{\zm}[1]{\mkern 1.5mu\ov{\mkern-1.5mu#1\mkern-1.5mu}\mkern 1.5mu}

\newcommand\scri{\mathscr{I}} 

\newcommand\scrip{\mathscr{I}^+} 

\newcommand\wcE{\widetilde{\mathcal{E}}}

\newcommand\hR{{\hat{R}}}

\newcommand\M{M}

\newcommand{\I}{\mathcal{M}_{\textrm{int}}}

\newcommand{\B}{\mathbb{R}^{4}}

\newcommand{\Sp}{S}

\newcommand{\G}{\hat{g}}

\newcommand{\Imet}{\hat{g}}

\newcommand{\Smet}{q}

\newcommand{\Dint}{\bm{D}}

\newcommand{\Ds}{\mathscr{D}}

\newcommand{\IRiem}{\mathcal{R}}

\newcommand{\dm}{{d_{\textrm{L}}}}

\newcommand{\dims}{{\hat{d}_{\textrm{L}}}}

\usepackage{amsthm}
\newtheorem{lem}{Lemma}
\newtheorem{thm}{Theorem}

\newtheorem{rem}{Remark}

\newtheorem{prop}{Proposition}
\usepackage{cleveref}
\usepackage{bm}
\crefname{lem}{lemma}{lemmas}
\crefname{thm}{theorem}{theorems}
\crefname{cor}{corollary}{corollaries}
\crefname{rem}{remark}{remarks}
\crefname{prop}{proposition}{propositions}

\makeatletter
\renewcommand\section{\@startsection {section}{1}{\z@}%
                               {-3.5ex \@plus -1ex \@minus -.2ex}
                               {2.3ex \@plus.2ex}%
                               {\normalfont\large\bfseries}}
\renewcommand\subsection{\@startsection{subsection}{2}{\z@}%
                                 {-3.25ex\@plus -1ex \@minus -.2ex}%
                                 {1.5ex \@plus .2ex}%
                                 {\normalfont\bfseries}}
\makeatother

\let\non\nonumber

\newcommand{\m}{\mu}

\newcommand{\news}{{\mathcal{N}}}

\def\bea#1\eea{\begin{align}#1\end{align}}

\def\bes #1\ees{\begin{split}#1\end{split}}

\newcommand{\be}{\begin{equation}}
\newcommand{\ee}{\end{equation}}

\newfont{\goth}{ygoth.tfm scaled 1200}                   

\numberwithin{equation}{section}

\newcommand{\C}[1]{$(\ref{#1})$}

\newcommand{\ov}{\overline}
\newcommand{\ul}{\underline}

\begin{document}
\begin{titlepage}

\begin{center}

{September 23, 2021}
\hfill         \phantom{xxx}  EFI--20-5

\vskip 2 cm {\Large \bf Gravitational Memory and Compact Extra Dimensions} 

\vskip 1.25 cm {\bf Christian Ferko$^{1,2}$, Gautam Satishchandran$^{1}$ and Savdeep Sethi$^{1}$}\non\\

\vskip 0.2 cm
{\it $^1$ Enrico Fermi Institute \& Kadanoff Center for Theoretical Physics \\ University of Chicago, Chicago, IL 60637, USA}

\vskip 0.2 cm
 {\it $^2$ Center for Quantum Mathematics and Physics (QMAP) \\ Department of Physics \& Astronomy, University of California, Davis, CA 95616, USA}

\vskip 0.2 cm

\end{center}
\vskip 1.5 cm

\begin{abstract}

\baselineskip=18pt

We develop a general formalism for treating radiative degrees of freedom near $\scrip$ in theories with an arbitrary Ricci-flat internal space. These radiative modes are encoded in a generalized news tensor which decomposes into gravitational, electromagnetic, and scalar components.
We find a preferred gauge which simplifies the asymptotic analysis of the full nonlinear Einstein equations and makes the asymptotic symmetry group transparent. This asymptotic symmetry group extends the BMS group to include angle-dependent isometries of the internal space. We apply this formalism to study memory effects, which are expected to be observed in future experiments, that arise from bursts of higher-dimensional gravitational radiation. We outline how measurements made by gravitational wave observatories might probe properties of the compact extra dimensions.

\end{abstract}

\end{titlepage}

\tableofcontents
\section{Introduction} \label{intro}

Perhaps the most robust prediction of string theory is the existence of extra spatial dimensions. Perturbative string theory requires ten spacetime dimensions while non-perturbative string theory predicts an eleventh dimension.  In this era of gravitational wave astronomy, it is exciting to explore ways of probing the extra dimensions found in either string theory, or other theories of higher-dimensional gravity. Gravitational wave observatories, like LIGO, measure features of the gravitational radiation produced by mergers of compact objects like black holes, neutron stars or even more exotic possibilities. The goal of this work is to begin to explore which features of the internal compactification space might be accessible through gravitational signatures. Probing the structure of compactified dimensions usually requires high energies. Unlike our usual intuition from particle physics correlating high energy with small wavelengths, gravity offers potential probes of short distance physics via black holes, where higher energy means larger objects.

The goal of this work is two-fold: first we will describe how LIGO and future gravitational wave observatories can see universal signatures of new physics at very low frequencies. By new physics we mean sources of stress-energy which can be treated as effectively null; for example, highly energetic low mass particles. At zero frequency, there is an observable called gravitational memory which is sensitive to new sources of stress-energy. Future experiments have a reasonable likelihood of measuring the memory effect~\cite{Lasky:2016knh, Wang:2014zls, vanHaasteren:2009fy}. This is certainly not the only potential observable of interest! The gravitational waveform itself encodes more data about new physics, including any potential extra dimensions. However, analyzing the full waveform typically requires more model-dependent inputs and a numerical study. 

The second goal is defining gravitational radiation in a reasonably precise way in compactified spacetimes. Defining gravitational radiation is a non-trivial exercise which was solved in four-dimensional asymptotically flat spacetime in classic work of Bondi, Metzner and Sachs~\cite{Bondi:1960jsa, Bondi:1962px, Sachs:1962wk}. One of the outcomes of that work was the enlargement of the asymptotic Poincar\'e group to the infinite-dimensional BMS group that includes supertranslations, which we will review shortly.\footnote{These supertranslations have no connection to supersymmetry. This is just an unfortunate clash of nomenclature.} A complete analysis of gravitational radiation in all non-compact spacetime dimensions appears in~\cite{Satishchandran:2019pyc}, building on the earlier work of~\cite{Hollands:2003ie, Hollands:2004ac}. Somewhat surprisingly, gravitational radiation for spacetimes with compact dimensions has not yet been studied beyond linearized gravity, or in the special case of a circle compactification \cite{Andriot:2017oaz, Lu:2019jus, Tan:2020hvy}. As in the non-compact case, a full nonlinear analysis is needed to define a notion of radiated power per unit angle, which gives energy-momentum loss as well as the null memory contribution to the total memory effect~\cite{Christodoulou:1991cr}. 

The simplest compactified space we might imagine is a circle or a torus. From that example studied in section~\ref{sec:circle} we will unify scalar~\cite{Tolish:2014bka}, electromagnetic~\cite{Bieri:2013hqa,Pasterski:2015zua,Bieri:2011zb} and gravitational~\cite{Zeldovich:1974gvh,Christodoulou:1991cr} notions of memory in the spirit of Kaluza and Klein. In section~\ref{colormemory} we sketch how this approach can be used to derive memory for non-abelian gauge theories, discussed for example in~\cite{Pate:2017vwa}, from a higher-dimensional gravity theory compactified on a space with a non-abelian isometry group.
String theory suggests a richer class of compactification spaces, described below in section~\ref{cmpctst}, with a first generalization from tori to Ricci-flat spaces. In their full glory, however, the vacuum solutions are quite intricate warped spacetimes.  In this analysis we largely focus on the case of unwarped Ricci-flat spacetimes where the analysis is more tractable. Well-known examples of this type include manifolds of special holonomy like $G_2$ manifolds used in M-theory compactifications and Calabi-Yau $3$-folds used in string compactifications. However we are not restricting to supersymmetric vacuum configurations in this analysis. We consider general Ricci-flat compactifications, which do not necessarily have special holonomy. For a recent discussion about Ricci-flat spaces which do not have special holonomy, see~\cite{Acharya:2019mcu}.\footnote{While less familiar than the special holonomy Ricci-flat spaces which preserve supersymmetry, it is not hard to construct non-supersymmetric examples along the following lines: take a $K3$ surface that admits an involution which does not preserve the holomorphic $2$-form and may have fixed points. Consider the space $(K3 \times \mathbb{T}^k)/G$ where the quotient group $G$ acts on the $K3$ surface as just described, and simultaneously on the torus by translations so that $G$ is freely acting. Similar examples can be constructed without tori, sometimes at the expense of the spin structure, by taking special holonomy spaces that admit fixed-point free involutions and considering the resulting quotient space; the Enriques surface, constructed as a ${\mathbb Z}_2$ quotient of a $K3$ surface, is of that type.} For warped compactifications where four-dimensional effective field theory still makes sense, we expect a qualitatively similar picture to the Ricci-flat case with a suitable change in the effective null stress-energy generated from the compact dimensions. 

\begin{figure}[H]
    \centering
    \includegraphics[width=.6\textwidth]{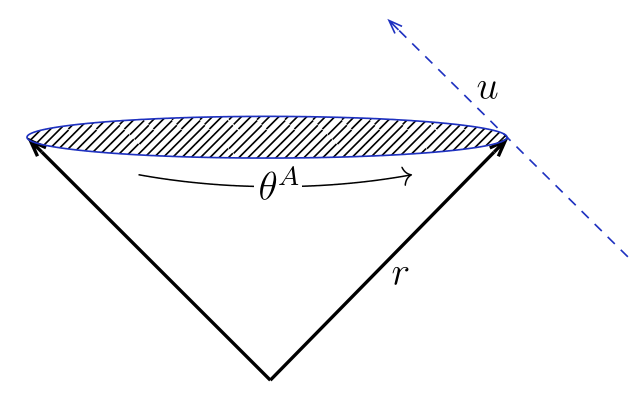}
    \caption{A depiction of Bondi coordinates.}
    \label{bondicoordinates}
\end{figure}

To introduce the memory observable, consider $3+1$ spacetime dimensions and pure Einstein-Hilbert gravity with no additional sources of stress-energy: 
\begin{equation}
    S = {1\over 16 \pi G}\int d^4x \sqrt{-g}R. 
\end{equation}
An asymptotically flat metric is conveniently written in terms of Bondi coordinates $(u,r,\theta)$ adapted to  outgoing null directions. This coordinate system is depicted in figure~\ref{bondicoordinates}. The $\theta^{A}$ are coordinates for the two-sphere at null infinity with unit round metric $q_{AB}$. In Bondi gauge, $g_{rr}=g_{rA}=0$ and $\partial_{r}\{ \det(g_{AB}) \}=0$. The metric with signature $(-,+,+,+)$ then takes the form 
\begin{align}
\label{etabondi}
    ds^2 =& \left\{ \eta_{\mu\nu} + \sum_n {h_{\mu\nu}^{(n)} \over r^n}\right\} dx^\mu dx^\nu, \cr 
    = & -du^{2}-2dudr + q_{AB} e^{A} e^{B} + {2m_{\text B}(u,\theta) \over r} du^2 + {h_{AB}^{(1)}(u,\theta) \over r} e^A e^B + O\bigg({1\over r^2}\bigg),  
\end{align}
where $e^A = r d\theta^{A}$ and $m_{\text B}$ is the Bondi mass aspect. The radiative degrees are encapsulated by the ``news'' tensor which is given by
\begin{equation}
    N_{AB}(u,\theta) = \bigg( q_{A}{}^{C}q_{B}{}^{D}-\frac{1}{2}q_{AB}q^{CD}\bigg)\partial_{u}h_{CD}^{(1)}(u,\theta).
\end{equation}
Memory can be viewed as the displacement of an array of freely floating test masses 
located near null infinity created by the passage of a gravitational wave. The full memory effect is given in terms of the news tensor:
\begin{align}\label{definememory}
    \Delta_{AB}(\theta)\equiv \frac{1}{2} \int_{-\infty}^{\infty}du^{\prime }N_{AB}(u^{\prime },\theta).
\end{align}
Memory can be decomposed into two contributions~\cite{Tolish:2014oda}: the first is an ``ordinary'' contribution produced by the change in the mass multipole moments of the radiation source; for example, a black hole binary merger. This contribution can be seen in a weak field linearized gravity approximation~\cite{Zeldovich:1974gvh}. There is also a more subtle ``null'' memory effect produced by the energy flux that reaches null infinity~\cite{Christodoulou:1991cr}. 

\subsection{Four-dimensional effective field theory}\label{fourdeffectivefieldtheory}

The first question we might ask is how a gravitational wave detector might see a sign of new physics. Let us suppose that far away from sources and near the detector, the vacuum Einstein equations are applicable. On the one hand, the memory effect is given by the news tensor via~\C{definememory}. Let us model the detector as a collection of test particles near null infinity. At leading order in $1\over r$, the displacement of the test particles in the angular directions is given by
\begin{align}\label{displacement-exp}
 \xi_{A} = \xi^{(0)}_{A}(\theta) +  {\xi_{A}^{(1)}(u,\theta)\over r}+O\bigg({1\over r^{2}}\bigg),  
\end{align}
where the initial positions are given by $\xi^{(0)}_{A}$. Near null infinity, $\xi_{A}^{(1)}(u,\theta)$ is determined by the geodesic deviation equation which implies that the relative accelerations of the test particles with respect to retarded time is given by:
\begin{equation}\label{geodesiceinstein}
   \frac{\partial^{2} \xi^{(1)}_{A}}{\partial u^{2}}  = - R^{(1)}_{uAuB} \xi_{(0)}^B. 
\end{equation}
This component of the Riemann tensor at leading order in $1 \over r$ can be expressed in terms of the Bondi news giving the relation, 
\begin{equation}
\frac{\partial^{2} \xi^{(1)}_{A}}{\partial u^{2}} =\frac{1}{2}\partial_{u}N_{AB}(u,\theta)\xi^{B}_{(0)}(\theta). 
\end{equation}  
An elementary derivation of this formula can be found in section~\ref{sec:memory_effect_compactified}. The displacement of the ``arms'' of the detector as a function of retarded time is 
\begin{equation}\label{armmotion}
\Delta \xi_{A}^{(1)}(u,\theta) = \frac{1}{2}\int_{-\infty}^{u}du^{\prime} \, N_{AB}(u^{\prime },\theta)\xi_{(0)}^{B}(\theta). 
\end{equation}
For convergence of this integral for all retarded time, we assume the news tensor decays in the far past/future as $N_{AB}\sim O\left({1\over |u|^{1+\epsilon}} \right)$ for $\epsilon>0$. The memory effect is given by,  
\begin{equation}
\lim_{u\rightarrow \infty}\Delta\xi^{(1)}_A(u,\theta) = \Delta_{AB}(\theta)\xi^{B}_{(0)}(\theta). 
\end{equation}
On the other hand, assuming the vacuum Einstein equations one finds that
\begin{align} \label{derivmemory}
    \Ds^{A}\Ds^{B}\Delta_{AB} = 2\Delta m_{\text{B}}(\theta) + \frac{1}{4}\int_{-\infty}^{\infty}du \,  N_{AB}(u,\theta)N^{AB}(u,\theta) ,
\end{align}
 where $\Ds_{A}$ is the covariant derivative on the unit $2$-sphere. In principle this formula can be inverted to get the memory tensor $\Delta_{AB}$.  The first term on the right hand side of~\C{derivmemory}\ is the change in the Bondi mass aspect, which captures the ordinary memory contribution. In principle, the ordinary memory can be determined from data by comparison with simulated wave-forms. The second term is the null memory contribution. This is proportional to the power radiated per unit angle. For a binary black hole merger the contribution of the null memory is roughly $\sim 10^{3}$ times larger than the ordinary memory~\cite{Favata_2009}. Therefore, the dominant contribution to \cref{derivmemory} is the null memory term. 

The upshot is that the news can be extracted from the arm motion via~\C{armmotion}\ and then used for a second evaluation of the expected memory using~\C{derivmemory}, which assumes the vacuum Einstein equations. If this computation of the memory disagrees with observation, there must be some other physics affecting the detector. 

\subsubsection*{\ul{\it Minimally-coupled stress-energy}}

First imagine a situation with a single distinguished metric, namely the Einstein-frame metric $g$, and some matter stress-energy $T_{\mu\nu}$ which might, for example, be governed by an action $S_M$ coupled to this metric: 
\begin{equation}
    S = {1\over 16 \pi G}\int d^4x \sqrt{-g}R + S_M(g). 
\end{equation}
As usual, the Hilbert stress tensor is given by $T_{\mu\nu}=-\left({2\over \sqrt{-g}}\right){\delta S_{M}\over \delta g^{\mu\nu}}.$ 
In this situation, \C{derivmemory} is augmented by a contribution from null stress-energy given below, 
\begin{align} \label{derivmemory-null}
    \Ds^{A}\Ds^{B}\Delta_{AB}(\theta) = 2\Delta m_{B}(\theta) + 8\pi \int_{-\infty}^{\infty}du \, \bigg(T^{(2)}_{uu}+\frac{1}{32\pi}N^{AB}N_{AB}\bigg) ,
\end{align}
where $T_{uu}^{(2)}(u,\theta)\equiv \displaystyle{\lim_{r \to \infty}} r^{2}  \, T_{uu}(u,r,\theta)$. In addition to~\C{etabondi}, the derivation of~\C{derivmemory-null} assumes that the stress-tensor decays like $O\left({1\over r^2}\right)$ and obeys the dominant energy condition: namely, that $T_{\mu\nu} v^\nu$ is time-like or null for any time-like or null vector $v^\mu$.  
This modified relation has been proposed as a way of detecting the contribution of neutrino radiation to the memory effect~\cite{Bieri:2013gwa}. 

\subsubsection*{\ul{\it Jordan-frame stress-energy}}

The other case of interest to us is the situation where there are scalar fields, collectively denoted $\phi$, and the matter sector couples to a Jordan-frame metric $g^{(J)}$ distinct from the Einstein metric. We can model this situation by the action,
\begin{equation}
    S = {1\over 16 \pi G}\int d^4x \sqrt{-g}R + \int d^4x \left( -{1\over 2} \partial^\mu \phi \partial_\mu \phi  - V(\phi)\right) +{S}_M(g^{(J)}), 
\end{equation}
where $g_{\mu \nu}^{(J)} = e^{\omega(\phi)}g_{\mu \nu} $ and $\omega(\phi)$ is a scale factor that depends on the scalar fields $\phi$. For example, Brans-Dicke theory is of this type with a single scalar field $\phi$, and a function $\omega$ proportional to $\phi$; a nice discussion of memory and asymptotically-flat solutions for Brans-Dicke theories can be found in~\cite{tahura2020bransdicke}. The choice of Jordan frame metric is ambiguous up to a shift of the scale factor $\omega$ by a constant. For convenience we will choose this constant so that $\omega(\phi)$ vanishes as $r \rightarrow\infty$. 

It is worth commenting on masses at this point. Any real detector is obviously not located at $\scrip$ so a sufficiently energetic flux of low mass particles will effectively behave like null stress-energy. With this caveat in mind, our analysis will usually assume an idealized situation where the detector lives near $\scrip$ and we can treat particles near $\scrip$ as massless. To derive an expression for memory, we again assume that the stress tensor obeys the dominant energy condition with $O({1\over r^2})$ decay for large $r$. Similarly any scalar field $\phi$ has the following expansion near $\scrip$, 
\begin{equation}\label{scalarexpansion}
    \phi \sim \phi^{(0)} + {\phi^{(1)}(u,\theta)\over r} + O\left({1\over r^2}\right) ,
\end{equation}
where $\phi^{(0)}$ is a constant. Our detector is constructed from the matter sector governed by ${S}_M(g^{(J)})$. Geodesic deviation determines how the detector reacts to a burst of gravitational radiation. For stationary test particles situated near $\scrip$, the geodesic deviation is again described by
\begin{equation}\label{geodesicjordan}
   \frac{\partial^{2} \xi^{(1;J)}_{A}}{\partial u^{2}}  = - R^{(1;J)}_{uAuB} \, \xi_{(0; J)}^B. 
\end{equation}
Here the two superscripts denote the power in the $1/r$ expansion and Jordan-frame.
Although the Jordan-frame metric is not in Bondi gauge described in~\cref{etabondi}, it is still true that $h_{rr}^{(1;J)}$ and $h_{rA}^{(1;J)}$ vanish.  For metrics of this form, 
the relevant component of the Riemann tensor takes the form
\begin{align}
    R^{(1;J)}_{uAuB} &= -\frac{1}{2}\partial_{u}^{2}h_{AB}^{(1;J)} \cr
     & = -\frac{1}{2}\partial_{u}^2 \left( h_{AB}^{(1)} + \omega^{(1)} q_{AB} \right) \cr
    & =-\frac{1}{2}\partial_{u} \left( N_{AB}+\partial_{u}\omega^{(1)} q_{AB} \right) ,
\end{align}
where in the last line we used the fact that $q^{AB}h_{AB}^{(1)}=0$ in Bondi gauge. The arm displacement is now given by
\begin{equation}\label{armmotionjordan}
\Delta \xi_{A}^{(1;J)}(u,\theta) = \frac{1}{2}\int_{-\infty}^{u}du^{\prime}\left( N_{AB}(u^{\prime },\theta) + \partial_u \omega^{(1)} q_{AB} \right)\xi_{(0;J)}^{B}(\theta). 
\end{equation}
\Cref{armmotionjordan} gives the motion of the arms of the detector moving on a geodesic of the Jordan frame metric. This motion has a transverse piece due to the contribution of $N_{AB}$ and a longitudinal piece due to the contribution of the conformal mode $\partial_{u}\omega^{(1)}$. This extra piece is also known as the breathing mode of the gravitational radiation. 

If the scalar charge, defined by $\omega^{(1)}(u,\theta)$ in analogy with~\C{scalarexpansion}, does not change then the second term in~\cref{armmotionjordan} vanishes. In Jordan frame, the memory effect is again given by:  
\begin{equation}
\lim_{u\rightarrow \infty}\Delta\xi^{(1;J)}_A(u,\theta) = \Delta^{(J)}_{AB}(\theta)\xi^{B}_{(0;J)}(\theta). 
\end{equation}
The news tensor appearing in~\C{armmotionjordan} can again be related to the square of the news tensor via Einstein's equations, 
\begin{align} \label{derivmemory-jordan-null}
    \Ds^{A}\Ds^{B}\Delta^{(J)}_{AB} =2 \Delta m^{(J)}(\theta) + 8\pi \int_{-\infty}^{\infty}du \, \bigg(T^{(2)}_{uu}(u,\theta)+\frac{1}{32\pi}N^{2}(u,\theta)\bigg) ,
\end{align}
where $T_{uu}^{(2)}(u,\theta)$ is again defined by $\displaystyle{\lim_{r \to \infty}} r^{2}  \, T_{uu}(u,r,\theta)$ and $m^{(J)}(\theta)=m_B(\theta)+\frac{1}{2}\Ds^{2}\omega^{(1)}$. The frame-dependence can therefore contribute to the memory in competition with null stress-energy as long as the associated scalar fields can be treated as massless.
 
\subsubsection*{\ul{\it Higher-derivative interactions}}

Any effective description for a theory of quantum gravity will have higher derivative interactions. These interactions are crucial for constructing vacuum solutions with flux in string theory, which we will discuss in section~\ref{cmpctst}. In this work, we will not take into account higher derivative interactions in the full higher-dimensional theory. That is a very difficult problem to address. Rather we will consider higher derivative interactions in the four-dimensional effective theory. As long as we can reduce to an effective four-dimensional description, this should cover any possible observable effects from these couplings. 

Let us consider purely gravitational corrections to the Einstein-Hilbert action, which take the schematic form:
\begin{equation}
     S = {1\over 16 \pi G}\int d^4x \left( \sqrt{-g}R + O(R^2) + O(R^3) + \ldots\right). 
\end{equation}
The higher derivative corrections are suppressed by some scale. We want to answer the question: which combinations of curvatures could possibly affect memory? Memory is determined by terms that decay at $O({1\over r^2})$ near $\scrip$. The Riemann tensor for the metric~\C{etabondi} decays like ${1\over r}$. Any contractions of Riemann with metrics will also decay at $O({1\over r})$ or faster. This means that terms of $O(R^3)$ are already decaying too fast to affect memory. On the other hand, terms of  $O(R^2)$ deserve further investigation. 

At the four derivative order there are two topological couplings, the Pontryagin density and the Euler density,  proportional to 
\begin{equation}
    \int \Tr \left( \hR\wedge \hR\right), \quad \int \Tr \left( \hR\wedge \ast \hR\right), 
\end{equation}
where $\hR$ is the curvature $2$-form. These terms do not affect either the equations of motion, or memory. One might imagine adding an axion coupling of the sort $\int \hat\phi \Tr \left( \hR\wedge \hR\right)$ for an axion ${\hat\phi}$, but such a coupling decays at $O({1\over r^3})$ because the non-constant behavior of the axion is $O({1\over r})$. That leaves the combinations
\begin{equation}
    \int \sqrt{-g} R^2, \quad \int \sqrt{-g} R_{\mu\nu}R^{\mu\nu}, \quad \int \sqrt{-g} R_{\mu\nu\lambda\rho}R^{\mu\nu\lambda\rho}. 
\end{equation}
However the first two terms can be field redefined away. The third term is related to the Euler density, which is proportional to $R^2 - 4R_{\mu\nu}R^{\mu\nu} +R_{\mu\nu\lambda\rho}R^{\mu\nu\lambda\rho}$, and therefore the third term can also be ignored. 
Based on this discussion, it appears that memory is insensitive to higher derivative corrections.

\subsection{Compactified spacetimes }
\label{cmpctst}
There are really three separate facets to the question of exploring compactified dimensions using gravitational radiation. The first question one might ask is what class of spacetimes should we consider? The simplest Kaluza-Klein spacetime is higher-dimensional Minkowski space compactified on a torus; for example, five-dimensional Minkowski space compactified on a circle of radius $R$. This is a very useful example for exploring basic phenomena encountered in higher dimensions. String theory, however, suggests a richer class of spacetimes used in the construction of the string landscape. While there is much debate about the string landscape, we will stick with elements of the underlying string constructions that are most likely to survive in the future. 

The main surprise that string theory offers to a general relativist interested in radiation is the need to consider warped compactifications to four dimensions with vacuum configurations of the form, 
\begin{align}\label{warpedmetric}
ds^2 = e^{-\varphi(y)} \eta + e^{\varphi(y)} ds_{\I}^2(y),
\end{align}
where $\eta$ is the $D=4$ Minkowski metric, $ds_{\I}^2$ is the metric for a Ricci-flat internal space $\I$ with coordinates $y$, and $\varphi(y)$ is the warp factor~\cite{Dasgupta:1999ss}. There are also higher form flux fields that thread both the internal space and spacetime, which can be viewed as conventional sources of stress-energy. Gravitational waves in warped backgrounds of this type have been studied in~\cite{Andriot:2019hay, Andriot:2021gwv}. For a compact $\I$, this metric does not solve the spacetime Einstein equations without the inclusion of exotic ingredients like orientifold planes and higher derivative interactions. These ingredients exist in string theory. At higher orders in the derivative expansion of the spacetime effective action, the conformally Ricci-flat form of the internal space metric~\C{warpedmetric} is not preserved, but this form is a sufficiently good approximation for our discussion of radiation.  

Without some additional quantum ingredient, the semi-classical background~\C{warpedmetric} is part of a family of solutions obtained by rescaling the internal space $ds_{\I}^2 \rightarrow \lambda\, ds_{\I}^2$ for any $\lambda>0$ with an accompanying change in the warp factor. So there is a large volume limit for the internal space when $\lambda$ is large. In this limit, the warp factor approaches a constant, and the higher-dimensional spacetime approaches a product manifold. It is important to note, however, that the warp factor can still have regions of large variation in $\I$. 

The most tractable and heavily studied backgrounds $M$ preserve spacetime supersymmetry. The expectation is that spacetime supersymmetry is spontaneously broken below the compactification scale. For a  set of examples of this type, $\I$ is obtained from the geometry of a Calabi-Yau $4$-fold with some additional structure. Such spaces are complex K\"ahler Ricci-flat manifolds with potentially many shape and size parameters, which correspond to massless scalar fields in spacetime. The scalar fields that determine the complex structure of $\I$ typically get a mass from the fluxes that thread the space~\cite{Dasgupta:1999ss}.\footnote{See~\cite{Bena:2020xrh} for evidence that this might not be generically true for all the complex structure moduli when the number of such moduli is large.} This mass scale, $M_{\mathrm{flux}}$, can be significantly lighter than the Kaluza-Klein scale of the compactification, denoted $M_{KK}$. 

Let us get a rough feel for the numbers involved. If we assume an upper bound on the size of any compact dimension of roughly order microns, or equivalently eV, from gravitational bounds~\cite{Kapner:2006si} to approximately $10^{-18}\,  {\rm m}$ or a TeV from collider bounds~\cite{Zyla:2020zbs}, and six compact dimensions then the ten-dimensional Planck scale takes the range $M_p^{D=10} \sim 10 \, {\rm keV} - 10 \, {\rm TeV} $. Of course, the size of any compact dimensions might be much smaller than this upper bound. 
We expect scalars from the complex structure moduli to get masses of order
\begin{align}
M_{\mathrm{flux}} \sim {\left( M_{KK} \right)^3 \over M_s^2}, 
\end{align}
where $M_s$ is the string scale. For a string coupling of order one, the string scale and Planck scale are comparable: $M_s \sim M_{p}^{D=10}$. In this case,
\begin{align}
M_{\mathrm{flux}} \sim {\left( M_{KK} \right)^{3/2} \over M_p^{1/2}},
\end{align}
where $M_p$ is the observed four-dimensional Planck scale. The scalars then have a mass in the range of $10^{-14} - 10^4 \, {\rm eV}$ for a Kaluza-Klein scale ranging from $1 \, {\rm eV} - 1 \, {\rm TeV}$.\footnote{Masses at the very low end of this range will be constrained by bounds from superradiant instabilities from spinning black holes. This lower bound is in the range of $10^{-11} \, {\rm eV}$; see, for example~\cite{Stott:2018opm, Cardoso:2018tly}. For a recent discussion of superradiance in string theory, see~\cite{Mehta:2021pwf}.} This is a huge range of masses but it certainly includes masses light enough that we can simply ignore the mass and treat the scalar as massless for the purposes of detection by a gravitational wave detector. The last point to mention about the complex structure moduli is the number of such moduli. From known constructions of Calabi-Yau 4-fold geometries, there are examples with of $O(10^5)$ such moduli~\cite{Taylor:2017yqr, Halverson:2017ffz}.\footnote{The currently largest known value of the Hodge number, $h^{3,1}$, which determines the number of complex structure moduli for a Calabi-Yau $4$-fold is $303148$ found in~\cite{Candelas:1997eh, Lynker:1998pb}.   We would like thank Wati Taylor and Jim Halverson for discussions on moduli bounds.} 

There is one other notable feature of the flux compactifications described by~\C{warpedmetric}. Namely they are warped compactifications with a warp factor $e^{\varphi(y)}$ which can have a very large variation. Such compactifications can look very asymmetric because of the presence of strongly warped throats in the geometry~\cite{sethi-talk}. The primary reason for interest in such throats is to generate small scales from the Planck scale to solve the hierarchy problem in the spirit of the Randall-Sundrum model~\cite{Randall:1999ee}, although in the context of an actual compactification from string theory. 

In addition to generating hierarchies in the four-dimensional effective theory, this has potentially interesting consequences for exotic compact objects, specifically objects localized in higher dimensions. There is no complete understanding of how large the warp factor might become in flux vacua, largely because it is very difficult to find semi-classical compact flux solutions, which are necessarily supersymmetric backgrounds. However, it is reasonable to expect a variation in the warp factor at least large enough to account for the $O(10^{16})$ hierarchy between weak scale physics of $O(10^3) \, {\rm GeV}$ and Planck scale physics of $O(10^{19}) \, {\rm GeV}$. In principle, the variation of the warp factor could be much larger because the D3-brane tadpole found in F-theory on a Calabi-Yau $4$-fold~\cite{Sethi:1996es, Becker:1996gj}, which determines the maximum amount of background flux, can be as large as $O(10^4)$ in known examples. The background flux, together with gravitational curvature terms, source the harmonic equation satisfied by the warp factor. 

The upshot of this stringy top down look at compactified extra dimensions is that there can be many scalar fields with masses potentially below the Kaluza-Klein scale. We now turn to what kinds of compact objects might be sensitive to either these scalar fields, or directly to the existence of additional dimensions.

\subsection{Compact objects in higher dimensions}

\subsubsection*{\ul{\it Delocalized Compact Objects}}

In this work we want to study dynamical spacetimes which arise from the motion of compact objects. These objects might be stars or black holes in manifolds with compact extra dimensions. At a coarse level, there are two distinct categories of compact object we might study. The first are objects constructed strictly from the light degrees of freedom with masses below the Kaluza-Klein scale; for example, from the potentially light scalars discussed in section~\ref{cmpctst}. This class of compact object is essentially delocalized in the internal dimensions. We should be able to study the physics of these modes in four-dimensional effective field theory discussed in section~\ref{fourdeffectivefieldtheory}.

Surprisingly, even in this setting there are exotic compact objects that can support scalar hair, which is our basic signature of extra dimensions. The first are Bose stars reviewed in~\cite{Liebling:2012fv}: no particularly exotic ingredients are needed to construct Bose stars other than a complex scalar field. The scalar field is not static but the associated spacetime metric is static. It is interesting to note that the moduli scalar fields that arise in most string compactifications are naturally complex scalar fields because most such vacua give a low-energy supergravity theory.  Gravitational radiation from binary boson star systems has been studied in~\cite{Palenzuela:2007dm}.   

Closely related to Bose stars are gravitational atoms and molecules, which are clouds of scalar fields or massive vector fields surrounding a black hole, or a black hole binary~\cite{Baumann:2019eav, Ikeda:2020xvt}. Included in these configurations are Kerr black holes with scalar hair, which interpolate between Kerr black holes and rotating Bose stars~\cite{Herdeiro:2015waa}. This is already a rich phenomenology of exotic compact objects, which are sensitive to light scalar fields.

\subsubsection*{\ul{\it Circle compactification}}

The second category of compact object is at least partially localized in the internal directions. Our basic intuition follows from compactification on a circle of radius $R$. 
Black hole uniqueness theorems are considerably weaker above four dimensions, and it is useful to characterize the black objects we wish to study based on their localization properties. A black string solution is simply a $D=4$ black hole which knows nothing about the internal space. It is a delocalized solution admitting a space-like Killing vector generating rotations of the ${ S}^1$.

The other extreme is a black hole which is highly localized on the internal space, breaking the $U(1)$ isometry. 
Black holes with a size small compared to $R$ look locally like a $D=5$ Myers-Perry solution \cite{MYERS1986304}. Solutions with mass $M$ are dynamically stable only for a certain range of the ratio $M/R$ because of the Gregory-Laflamme instability~\cite{PhysRevLett.70.2837}. The entropy serves as a thermodynamic diagnostic for stability. For a fixed mass $M$, black strings have an entropy that scales like $S_{BS}\sim M^2$ while $D=5$ black holes have an entropy that scales like $S_{BH}\sim M^2 \sqrt{R/M}$~\cite{Gregory:2011kh}. For large $R$, the localized black hole configuration is the preferred solution.

Astrophysical black hole mergers detectable by LIGO have constituent masses of roughly $O(10)$ solar masses, which corresponds to a distance scale of $O(10^4) \, {\rm m}$. This is ten orders of magnitude larger than the best upper bound on the Kaluza-Klein scale.  $M$ is clearly much greater than the range of Kaluza-Klein scales discussed in section~\ref{cmpctst}, and therefore one should expect that the generic compact object will be delocalized.

For circle compactifications, the binary merger of black holes localized at a point was studied in~\cite{Durrer:2003rg, Du:2020rlx} using a point particle approximation. With no other ingredients, the massless degrees of freedom in four dimensions are a graviton, a Kaluza-Klein scalar and a graviphoton. The luminosity of gravitational waves released in the merger process is about $20\%$ less than the merger of four-dimensional black holes mainly because of scalar radiation produced in the merger. 

To see this consider $\B\times \Sp^1$ with coordinates $(t,x_{1},x_{2},x_{3}, y)$ and flat metric $ds^2=\eta_{\mu\nu}dx^{\mu}dx^{\nu}+dy^{2}$, where $y \sim y+2\pi R$. In linearized gravity, the stress-energy for a point particle of mass $m$ and world-line given by $X^{M}(\tau)$ with affine parameter $\tau$ is given by:
\begin{equation}
\label{ptparticlestress}
T^{MN}(X)=m\int d\tau \dot{X}^{M}\dot{X}^{N}\delta^{(5)}(X - X(\tau)).
\end{equation}
 The indices $(M,N,\ldots)$ run over all the spacetime dimensions while $(\mu, \nu, \ldots)$ run over four-dimensional quantities in accord with the conventions spelled out later in section~\ref{sec:conventions}. For a particle moving only in $\B$, $\dot{X}^{y}(\tau)=0$.

The massless scalar field in four dimensions is the zero mode of $\delta g_{yy} = h_{yy}$ where $g_{MN}$ is the full spacetime metric. By this we mean Fourier expand the fluctuation $h_{yy}$ in the $y$ direction and restrict to the zero mode. We will denote the zero mode by a barred quantity ${\ov h}_{yy}$. In linearized gravity, this is sourced by the zero mode of the stress tensor,
\begin{equation}
\label{phiyagi}
\Box_{\eta} \ov{h}_{yy} = -8\pi \bigg(\ov{T}_{yy}-\eta^{\mu \nu}\ov{T}_{\mu \nu}\bigg) ,
\end{equation}
where $\Box_{\eta}=\eta^{\mu\nu}\partial_{\mu}\partial_{\nu}$. For the stress-tensor given in~\C{ptparticlestress}, $\ov{T}_{yy}=0$ and the right hand side of~\C{phiyagi} is non-zero, leading to the mismatch with experiment. The situation gets worse with more compact dimensions. Taken at face value, this would seem to rule out this simple model of 
compact extra dimensions.

However, we do not expect astrophysical black holes to be localized in a model like this because of the Gregory-Laflamme instability: the black holes are much larger than any extra dimension. Much more likely is a completely delocalized black string wrapping the $y$ direction.  For a string with induced metric $\gamma_{ab} = \partial_a X^M\partial_b X^N g_{MN}$ and tension $\mu$, the stress-energy tensor is given by
\begin{align}\label{string_stress_energy}
    T^{M N} = \mu \int d \sigma \, d \tau \, \sqrt{-\gamma} \, \gamma^{ab} \partial_a X^M \partial_b X^N \, \delta^{(5)} ( X - X ( \sigma, \tau ) ) \, .
\end{align}
Choosing $g_{\mu \nu} = \eta_{\mu \nu}$ and fixing static gauge for the wrapped string $(\sigma \sim y, \tau \sim t)$ gives
\begin{align}
    T^{yy} = 2\pi \mu R \int d \tau \, \delta^{(4)} ( X - X ( \tau ) ) \, ,
\end{align}
with $2\pi\mu R = m$. 
This makes the right hand side of \C{phiyagi} vanish as we expect for a model that replicates a standard $D=4$ black hole. 

Using this observation we can actually construct a model for a $D=4$ particle, at the level of hydrodynamics, which interpolates between the black string and the completely localized black hole. Consider the stress tensor with affine parameter $\tau$ given by, 
\begin{equation}
\label{ptparticlestressmodified}
T^{\mu\nu}(X)=m\int d\tau  \dot{X}^{\mu}\dot{X}^{\nu} \delta^{(5)}(X - X (\tau) ) \, , \qquad  T^{yy}_\epsilon ( x ) = \epsilon m \int d \tau \, \delta^{(4)} ( X - X ( \tau ) ).
\end{equation}
This is conserved. It is a hybrid of a $D=5$ point particle with a uniform stress on the $y$ circle. For $\epsilon=0$, this is the $D=5$ point particle while for $\epsilon=1$, the right hand side of~\C{phiyagi} vanishes and the zero mode of $T^{\mu\nu}(X)$ coincides with the black string~\C{string_stress_energy}. For intermediate $\epsilon$, this will result in a $D=4$ particle with some scalar charge that will generate some scalar radiation. However, the amount is tunable. We would expect more complicated stress-energy distributions in the $y$ direction for configurations corresponding to arrays of $D=5$ black holes and non-uniform black strings. The upshot is that there are many potential stress tensors that could describe black objects in $\B\times \Sp^1$ with varying amounts of scalar charge from the $D=4$ perspective, whose dynamics can be made consistent with current observation.

The circle is a very special example of a compactification. For the more general warped backgrounds described in section~\ref{cmpctst}, there is an exciting possibility of novel phenomena. One might imagine localized black objects, analogous to the $D=5$ black hole just discussed, which are globally unstable because of a Gregory-Laflamme type argument, but which are nonetheless long lived because of the local behavior of the warp factor. It would be interesting to explore this possibility further.

\subsection{Signatures of compact dimensions}

In Section~\ref{fourdeffectivefieldtheory} we saw that memory can be used to detect new physics. More precisely, given a particular model of the stress-energy in a theory, gravitational observatories can make independent measurements of arm motion and of gravitational memory, and then compare these measurements; disagreement indicates a missing contribution to the stress-energy. Such a missing contribution could come from various sources, including additional light fields in the theory or a matter coupling to a Jordan frame metric which differs from the Einstein frame metric. However, for the purposes of the current work, we are most interested in the possibility that a discrepancy in these measurements could arise from the presence of compact extra dimensions.

In a theory with extra dimensions, we will show that the radiative degrees of freedom near $\scrip$ are encoded in a generalized news tensor written as $\mathcal{N}_{ab}$, where the indices $a, b$ now run over both the the asymptotic $2$-sphere $\Sp^{2}$ and the internal space $\I$. The components $\mathcal{N}_{A B}$ will encode the familiar Bondi news contribution $N_{AB}$ as well as an additional scalar breathing mode $N$ which give rise to gravitational radiation in the non-compact directions. However, we will see that a generic internal manifold will support additional radiative modes encoded in $\mathcal{N}_{A m}$ and $\mathcal{N}_{mn}$, which involve fluctuations in the directions of the internal manifold $\I$. Viewed from the perspective of a macroscopic observer in $\B$, the additional modes in $\mathcal{N}_{A m}$ and $\mathcal{N}_{mn}$ are precisely the radiative degrees of freedom for electromagnetic gauge-fields and light scalars, respectively. This implies that there is an electromagnetic memory effect and a scalar memory effect associated with these additional modes.

In theories with these extra modes arising from compact dimensions, the null stress energy appearing in equation (\ref{derivmemory-null}) receives additional contributions; one now has
\begin{align}
    \Ds^{A}\Ds^{B}\Delta_{AB} &= 2\Delta m(\theta) + 8\pi \int_{-\infty}^{\infty}du \, \bigg( \mathcal{T}^{(2)}_{uu}(u,\theta) + \frac{1}{32\pi}N_{AB}N^{AB} \bigg), \nonumber \\
    \mathcal{T}^{(2)}_{uu} ( u , \theta ) &\equiv T^{(2)}_{uu} ( u , \theta ) + \frac{1}{32 \pi} \left( \mathcal{N}_{Am} \mathcal{N}^{Am} + \mathcal{N}_{mn} \mathcal{N}^{mn} + N^{2}\right) \, .
\end{align}
Here $N$ is associated with a breathing mode of the internal space which is a scalar degree of freedom. Therefore, for a particular model for the null stress energy $T_{uu}^{(2)}$ that should contribute to memory, the presence of extra compact dimensions will generate a discrepancy between the predicted and measured memory effects. This discrepancy is captured in the four-dimensional effective stress tensor $\mathcal{T}^{(2)}_{uu}$, which includes the electromagnetic and scalar contributions from the higher-dimensional gravity modes. 

We can extract more data about these contributions from a different class of measurements. The ordinary electromagnetic and scalar memory effects generate a velocity kick for a suitable charged test particle. Even without any abelian charge or extra dimensions, gravity generates a similar velocity kick for a test particle. Likewise, in theories with extra dimensions, a particle with velocity in the internal directions will experience a velocity kick in $\B$ because of the passage of gravitational radiation in the internal space.

Measuring these velocity kicks requires a different experimental design than is typical for current gravitational observatories, which study geodesic deviation for pairs of point particles. Instead, if one can measure the trajectory of point particles -- even a single point particle -- undergoing geodesic motion, relative to a lab frame which is stationary in an appropriate sense, then one can in principle extract all of $\mathcal{N}_{Am}$ and a part of $\mathcal{N}_{mn}$ described in section~\ref{sec:memory_effect_compactified}. These additional sources of news are the primary signatures of extra dimensions we might hope to see with memory measurements alone.

\subsection{Conventions}\label{sec:conventions}

Unless otherwise specified, we work in units where $G=c=\hbar=1$, and follow the conventions of~\cite{Wald:106274}. Our metric signature is mostly positive and our sign convention for curvature is such that the scalar curvature of the round sphere metric is positive. The full $D$-dimensional spacetime manifold, denoted $M$, 
has the topology $M= \B\times \I$ where $\B$ is a four-dimensional Lorentzian manifold and $\I$ is a $(D-4)$-dimensional compact Riemannian manifold. Our index conventions are listed below:
\begin{itemize}
\item
Indices $(M,N,L,\dots)$ run over the full spacetime manifold $\M$ with metric $g_{MN}$ and covariant derivative $\nabla_{M}$. The Riemann tensor associated to the metric $g_{MN}$ is $R_{MNP}{}^{Q}$.

\item Indices $(\mu,\nu,\lambda,\dots)$ run over $\B$, and are raised and lowered with the asymptotic Minkowski metric $\eta_{\mu \nu}$. We denote the covariant derivative compatible with $\eta_{\mu \nu}$ by $\partial_{\mu}$.
\item Indices $(m,n,l,\dots)$ run over $\I$, and are raised and lowered with metric $\Imet_{mn}$. The covariant derivative compatible with $\Imet_{mn}$ is $\Dint_{m}$. The Riemann tensor of $\Imet_{mn}$ is $\IRiem_{mnp}{}^{q}$ which has vanishing Ricci: $\Imet^{mp}\IRiem_{mnp}{}^{q}=0$.\footnote{That $\Imet_{mn}$ is Ricci-flat follows from our fall-off ansatz given in \cref{ansatz} and the Einstein equations.} 

\item Indices $(A,B,C,\dots)$ run over $\Sp^{2}$, and are raised and lowered with the round metric $\Smet_{AB}$. The covariant derivative compatible with $\Smet_{AB}$ is $\Ds_{A}$.

\item Lastly indices $(a,b,c,\dots)$ run over  $\Sp^{2}\times \I$, and are raised and lowered with the product metric $\mathfrak{q}_{ab}$ given by $\mathfrak{q}=q \oplus \Imet$. 
\end{itemize}
Indices for tensors on $\M$ are raised and lowed with the asymptotic Ricci-flat product metric which we denote by a hat,  
\begin{equation}\label{hat_metric}
\G_{MN}dx^{M}dx^{N}=\eta_{\mu \nu}dx^{\mu}dx^{\nu}+\Imet_{mn}(y) dy^{m}dy^{n}, 
\end{equation}
where $x^{M}=\{x^{\mu},y^{m}\}$ are arbitrary coordinates on $\B$ and $\I$, respectively.  We also use these conventions to denote coordinates on submanifolds like $S^2$ or $S^2\times \I$, as well as components in a coordinate basis. We will use the same index notation for tensors which are intrinsic to a submanifold and the components of an ambient tensor along a submanifold; for example, the tensor $T^{MN}$ defined on the full spacetime $\M$ has angular components $T^{AB}(x,y)$ while the intrinsic tensor $t^{AB}(\theta)$ lives on $S^2$. We do not feel the potential confusion that might arise from doing this justifies introducing a new alphabet.

To simplify keeping track of powers of ${1\over r}$, we will expand tensors in a normalized basis, which in Bondi coordinates  is $\{du,dr,e^A=rd\theta^{A},dy^{m}\}$. This is a little different from the more common convention found in~\cite{Bieri_2014,Flanagan_2017,tahura2020bransdicke,Strominger:2017zoo}. As an explicit example consider the one-form on the sphere with coordinates $\theta^A$, 
\begin{align}
    V_\mu \, dx^\mu &= v_A (\theta) \, d \theta^A = \left( \frac{v_A(\theta)}{r} \right) \, \left( r \, d \theta^A \right) , 
    \label{angle_conventions}
\end{align}
for some $v_A(\theta)$. With this choice of basis, the $O({1\over r})$ term $V_A^{(1)} = v_{A}(\theta)$ is non-zero. 
When we perform asymptotic expansions near $\scri^+$, as in~\cref{etabondi}, we will use a superscript to indicate a term at a given order in $\frac{1}{r}$, keeping in mind the preceding convention for angular directions.
For example, a scalar field $\phi$ would be expanded as follows, 
\begin{align}
    \phi = \sum_{n=0}^{\infty} \frac{\phi^{(n)}}{r^n}. 
\end{align}

Lastly, given a tensor on $\I$ we can expand in eigenmodes of the appropriate Laplacian. It will be useful to denote the zero mode in such a harmonic expansion by a bar. For example, given a function $t(x^{\mu},y^{m})$ on $\M$ the zero mode is denoted by $\ov{t}(x)$. This zero mode solves $\Dint^{2}t=0$ where $\Dint^{2}\equiv \Imet^{mn}\Dint_{m}\Dint_{n}$ is the scalar Laplacian on $\I$. Similarly for a $1$-form $t_{M}(x,y)$ we denote the zero modes by $(\ov{t}_{\mu}(x,y),\ov{t}_{m}(x,y))$, while the zero modes of a symmetric $2$-tensor $t_{MN}(x,y)$ are denoted $(\ov{t}_{\mu\nu}(x,y), \ov{t}_{\mu m}(x,y), \ov{t}_{mn}(x,y))$. For Ricci-flat manifolds, this kind of harmonic decomposition simplifies considerably as we review in section~\ref{sec:linearized_review}.

\section{Review of Linearized Dimensional Reduction
}\label{sec:linearized_review}

The topics under discussion in this work are of potential interest to multiple communities, including string theorists, general relativists, quantum field theorists and gravitational wave astronomers. To make the work as self-contained as possible, we will review techniques that are more familiar to a specific community. 

The usual procedure of dimensional reduction is to start with a vacuum configuration which we take to be a $D$-dimensional product manifold, 
\begin{equation}
\M = \B \times \I ,
\end{equation}
where $\B$ is the non-compact Lorentzian spacetime, and $\I$ is the $(D-4)$-dimensional compact Riemannian internal space. We will also take $\I$ to be connected and closed (i.e. compact without boundary). $\M$ is equipped with the product metric 
\begin{equation}
\label{background}
 \G_{MN}dx^{M}dx^{N}=\eta_{\mu \nu}dx^{\mu}dx^{\nu}+\Imet_{mn}(y)dy^{m}dy^{n},
\end{equation}
where $\eta_{\mu \nu}$ is the Minkowski metric, $\Imet_{mn}(y)$ is a Ricci-flat metric on $\I$ and $x^{M}=\{x^{\mu},y^{m}\}$ are coordinates on $\B$ and $\I$, respectively. Our discussion does not involve fermions so we will not worry about issues like a spin structure.  

Let us consider pure gravity with the Einstein-Hilbert action on the total spacetime manifold $\M$:
\begin{align}
    S = \frac{1}{2 \kappa} \int_{\M} d^D x\, \sqrt{-g} \, R .
\end{align}
The supergravity theories that describe low-energy limits of string theory have additional fields, which we will ignore for the moment, to focus on the graviton. We will discuss dimensional reduction for {\it linearized} metric perturbations, which is the usual approach. This should be contrasted with our later discussion in subsection~\ref{subsec:reduction_nonlinear}\ near $\scrip$, which is for the full nonlinear theory.  

Consider a linearized perturbation of $\G_{MN}$ denoted $h_{MN}$. Let $\hat{\nabla}_{M}$ be the covariant derivative operator compatible with $\G_{MN}$. Imposing the gauge conditions\footnote{ \Cref{Lorgauge} is a special case of the Lorenz gauge. While Lorenz gauge is useful in studying radiation in {\em linearized} gravity with no null sources, we note that it is incompatible with the ${1\over r}$ fall-off of the metric in asymptotically null directions in a general radiating spacetime~\cite{Satishchandran:2019pyc}. The proof of \cite{Satishchandran:2019pyc} shows that harmonic gauge, which is the nonlinear generalization of Lorenz gauge, is incompatible with the fall-off conditions in $D$-dimensional non-compact spacetimes,  but the proof straightforwardly generalizes to cases with compact extra dimensions using the techniques and formulae in this paper.} 
\begin{equation}
 \label{Lorgauge}
 \hat{\nabla}^{M}h_{MN}=0 \quad \textrm{ and } \quad \G^{MN}h_{MN}=0
\end{equation}
yields the linearized Einstein equation in Lorenz gauge: 
\begin{equation}
\label{Lic}
 \Box_{\G}h_{MN}+2\hat{R}_{M}{}^{P}{}_{N}{}^{Q}h_{PQ}=0.
\end{equation}
Here $\Box_{\G}\equiv \G^{MN}\hat{\nabla}_{M}\hat{\nabla}_{N}$, $\,\hat{R}_{MPN}{}^{Q}$ is the Riemann tensor of the background metric $\G_{MN}$, and indices are raised and lowered with the background metric. The residual gauge freedom that preserves \C{Lorgauge} is given by 
\begin{equation}
\label{diff}
h_{MN}\to h_{MN}+\hat{\nabla}_{(M}\xi_{N)} \quad \textrm{ where } \quad \Box_{\G}\xi_{M}=0 \, , \quad \hat{\nabla}^{M}\xi_{M}=0.
\end{equation} 
Note that the exact (not asymptotic) symmetry group of \cref{background} is trivially the direct product of the Poincar\'e group $(\mathcal{P})$ and the isometry group ($\mathfrak{I}$) of $(\I,\Imet_{mn})$:
\begin{equation}
\label{Poinisom}
\mathcal{P} \times \mathfrak{I} .
\end{equation}
For background metric \cref{background}, the only non-vanishing components of the Riemann tensor are the internal components; therefore the Riemann tensor  is equivalent to $\IRiem_{mnp}{}^{q}$ on $(\Imet_{mn},\I)$.

Consider the projection of \cref{Lic} into $\B$ and rewrite $\Box_{\G}$ in terms of the derivative operator $\partial_{\mu}$ compatible with $\eta_{\mu\nu}$, and the covariant derivative operator $\Dint_{m}$ compatible with $\Imet_{mn}$. This yields  
\begin{equation}
\label{dab}
\Dint^{2}h_{\mu\nu} + \Box_{\eta}h_{\mu\nu}=0 ,
\end{equation}
where $\Dint^{2}\equiv \Imet^{mn}\Dint_{m}\Dint_{n}$ and $\,\Box_{\eta}\equiv \eta^{\mu\nu}\partial_{\mu}\partial_{\nu}$. Expanding $h_{\mu \nu}$ in terms of eigenfunctions of the Laplacian on $\I$, \cref{dab} yields an infinite tower of massive modes (one for each eigenvalue). The mass scale is set by the size of the compact extra dimensions. Since the goal of this paper is to study  radiation with compact extra dimensions we are interested in either massless fields, or fields with masses below the Kaluza-Klein scale; see the discussion in section~\ref{cmpctst}.


The massless modes $\ov{h}_{\mu \nu}$ are annihilated by the Laplacian and correspondingly satisfy a massless wave equation in $\B$:
\begin{equation}
\label{masspertgrav}
  \Dint^{2}\ov{h}_{\mu\nu}=0 \implies \Box_{\eta}\ov{h}_{\mu \nu}=0 .
\end{equation}
The zero-mode $\ov{h}_{\mu \nu}$ is harmonic on $\I$ and therefore independent of the internal coordinates $y$.
Projecting both indices of \cref{diff} into $\B$ shows that diffeomorphisms act on the zero mode $\ov{h}_{\mu \nu}$ by 
\begin{equation}
\ov{h}_{\mu\nu}(x^{\mu}) \rightarrow \ov{h}_{\mu\nu}(x^{\mu})+\partial_{(\mu}\ov{\xi}_{\nu)}(x^{\mu}) \quad \textrm{ where } \quad \Box_{\eta}\ov{\xi}_{\mu}=0 \textrm{, } \; \partial^{\mu}\ov{\xi}_{\mu}=0,
\end{equation}
and $\ov{\xi}_{\mu}$ is the  zero-mode of the projection of $\xi_{M}$ into $\B$. The massless spin-2 graviton arising from this reduction is $\ov{h}_{\mu\nu}$.



\subsection{Vector modes}
\label{subsec:linvec}
\par Analogously, we can study the vector perturbation $h_{\mu m}$ using the  linearized Einstein equation \C{Lic}. 
We again collect results here on the massless mode $\ov{h}_{\mu m}$ which satisfies 
\begin{equation}
\label{harmvec}
\Dint^{2}\ov{h}_{\mu m}=0  .
\end{equation}
Viewing $h_{\mu m}$ as a one-form on $\I$, we note that solutions to \cref{harmvec} are spanned by the space of one-forms $\ov{V}_{m}$ on $\I$ that satisfy 
\begin{equation}
\label{harmV}
\Dint^{2}\ov{V}_{m}=0.
\end{equation}
Equation \C{harmV} is a condition on $\ov{V}_m$ in terms of the coordinate Laplacian $\Dint^2$. For any compact manifold, the coordinate Laplacian on a one-form $V_{m}$ is related to the Hodge Laplacian $(\Delta^{(H)})$ on $V_m$ by the well known Weitzenb\"ock identity for one-forms:
\begin{equation}
\Dint^{2}V_{m} = -\Delta^{(H)}V_{m} + \Imet^{pn}\IRiem_{mp}V_{n}.
\end{equation}
Here $V_{m}$ is a one-form on $\I$ and $\IRiem_{mp}$ is the Ricci tensor of $(\Imet_{mn},\I)$. Therefore on any Ricci-flat manifold, the coordinate Laplacian can be replaced by (minus) the Hodge Laplacian when acting on one-forms. Solutions to \cref{harmV} are harmonic one-forms.  We now investigate the properties of solutions to \cref{harmV}. First recall the well-known Hodge decomposition of a one-form.
\begin{prop} 
\label{hodge}
Let $(\I,\Imet_{mn})$ be a compact Riemannian manifold. Any globally defined one-form $V_{m}$ can be uniquely decomposed as follows, 
\begin{equation}
\label{Vmdecomp}
V_{m}=\Dint_{m}S+v_{m},
\end{equation}
where $\Dint^{m}v_{m}=0$. We refer to $v_{m}$ and $S$ as the vector and scalar parts of $V_{m}$, respectively. 
\end{prop}

\par If $V_{m}$ is harmonic then $S$ must be a constant and consequently, $V_{m}$ is divergence free. Further a harmonic $V^m= \Imet^{mn}V_{n}$ is a Killing vector if $\I$ is Ricci-flat. To see this, let $\xi^{n}$ be a Killing vector on $\I$ i.e. $\xi_{m}=\Imet_{mn}\xi^{n}$ satisfies $\Dint_{(m}\xi_{n)}=0$. Applying $\Dint^{m}$ to Killing's equation and commuting the derivatives yields, 
\begin{align}
\label{Dke}
\Dint^{2}\xi_{m}+\Dint_{m}\Dint^{n}\xi_{n}-\IRiem_{m}{}^{n}\xi_{n}=0.
\end{align}
The second and third terms of \cref{Dke} both vanish since $\IRiem_{mn}=0$ and $\xi_{m}$ is divergence free by Killing's equation. Therefore if $\Imet^{mn}\xi_{n}$ is a Killing vector then $\xi_{m}$ is indeed harmonic.

To complete the correspondence we now show that if a one-form $\ov{V}_{m}$ is harmonic then $\Imet^{mn}\ov{V}_{n}$ is also a Killing vector \cite{bochner1946}. Contracting \cref{harmV} with $\ov{V}^{m}$ and integrating over $\I$ gives,  
\begin{equation}
\int_{\I}\Dint^{m}\ov{V}^{n}\Dint_{m}\ov{V}_{n}=0 \implies \Dint_{m}\ov{V}_{n}=0.
\end{equation}
Consequently, solutions to \cref{harmV} are covariantly constant and therefore Killing. The space of solutions to \cref{harmV} is therefore the space of Killing vectors on $\I$. The number of linearly independent harmonic one-forms on $\I$ is counted by the first Betti number, $b_{1}$, which is a topological invariant. 
The preceding observations can be summarized in the following lemma \cite{bochner1946}: 

\begin{lem}[Bochner]
\label{KVlem}
Let $(\I,\Imet_{mn})$ be a compact Ricci-flat Riemannian manifold. The space of harmonic one-forms is then in one-to-one correspondence with the space of Killing vectors, which are covariantly constant. The dimension of the space of Killing vectors is $b_{1}(\I)$.  
\end{lem}

In the case where $b_1 > 0$, the Ricci-flat space $\I$ of dimension $D-4$ can be written as a free quotient of $\mathbb{T}^k \times \widetilde{\mathcal{M}}_{\textrm{int}}^{D-4-k}$ where $\widetilde{\mathcal{M}}_{\textrm{int}}^{D-4-k}$ is also Ricci-flat~\cite{Fischer}. 
We can now give the general solution to \cref{harmvec}, 
\begin{equation}
\label{metric_vector_decomposition}
\ov{h}_{\mu m}(x^{\mu}, y^{m} )=\sum_{i=1}^{b_{1}}A_{\mu}^{(i)}(x^{\mu})\otimes \ov{V}_{m}^{(i)}(y^{m}) ,
\end{equation}
where $\{\ov{V}_{m}^{(i)}\}$ are the  $b_{1}$ linearly independent Killing vectors. The coefficients $A_{\mu}^{(i)}(x)$ define a set of $b_{1}$ graviphoton vector fields on $\B$. 
Furthermore, it follows from \cref{Lic,Lorgauge} that each vector field $A_{\mu}^{(i)}(x^{\mu})$ satisfies the wave equation and is divergence free on $\B$:
\begin{equation}
\Box_{\eta}A_{\mu}^{(i)}=0  \textrm{ and }\partial^{\mu}A_{\mu}^{(i)}=0.
\end{equation}
Projecting one index of \cref{diff} into $\B$ and one index into $\I$, and using \C{harmvec} implies that the gauge freedom of $\ov{h}_{\mu m}$ is 
\begin{equation}
\ov{h}_{\mu m}\longrightarrow \ov{h}_{\mu m} + \sum_{i=1}^{b_{1}}[\partial_{\mu }\lambda^{(i)}(x^{\mu})] \, \ov{V}_{m}^{(i)}( y^{m} ) ,
\end{equation}
where $\lambda(x^{\mu})$ is a smooth function on $\B$, which satisfies the wave equation. This is equivalent to an abelian gauge transformation on $A_{\mu}^{(i)}$,  
\begin{equation}
A_{\mu}^{(i)}(x^{\mu}) \to A_{\mu}^{(i)}(x^{\mu}) + \partial_{\mu}\lambda^{(i)}(x^{\mu}), \qquad \Box_{\eta}\lambda^{(i)}=0.
\end{equation}
The Lie algebra for these spin-1 massless gauge-fields is determined by the isometry group of $\I$. The isometry group is clearly abelian for Ricci-flat $\I$ since, by \Cref{KVlem}, any Killing vector is also covariantly constant and therefore the commutator of any two Killing vectors vanishes.  




\subsection{Scalar modes}
\label{subsec:linscalar}
\par We finally consider the perturbations $h_{mn}$ which satisfy
\begin{equation}
    \bm{D}^{2}h_{mn}+2\IRiem_{m}{}^{p}{}_{n}{}^{q}h_{pq} + \Box_{\eta} h_{mn}=0 .
\end{equation}
%
Therefore massless perturbations $\ov{h}_{mn}$  are spanned by the tensor fields on $\ov{T}_{mn}(y^{m})$ which satisfy 
\begin{equation}
\label{Tmn}
\Dint^{2}\ov{T}_{mn} + 2\IRiem_{m}{}^{p}{}_{n}{}^{q}\ov{T}_{pq}=0 \, .
\end{equation}
The operator acting on $\ov{T}_{mn}$ in \cref{Tmn} is the Lichnerowicz Laplacian. \Cref{Lorgauge} implies a further constraint on the allowed solutions to \cref{Tmn}. Expanding the divergence of $h_{MN}$ in terms of harmonic one-forms implies that
\begin{equation}
    \label{divT}
    \Dint^{m}\ov{T}_{mn}=0.
\end{equation}
The space of solutions to \cref{Tmn,divT} is the moduli space of infinitesimal deformations that preserve the vanishing of the Ricci tensor. This moduli space is known to be finite-dimensional~\cite{besse1987einstein}.


\par To further investigate the implications of \cref{Tmn,divT},  we first recall a well known result about the decomposition of symmetric tensors \cite{Hollands:2016oma}: 
\begin{prop}
\label{propsymm}
Let $(\I,\Imet_{mn})$ be a compact Riemannian Einstein space with dimension $D-4$, i.e., $\IRiem_{mn}=c\Imet_{mn}$, for some constant $c$, which includes the Ricci-flat case. Then any second rank, symmetric tensor field $T_{mn}$ can be uniquely decomposed as 
\begin{equation}
\label{Tmndecomp}
T_{mn}=t_{mn}+\Dint_{(m}W_{n)}+\bigg(\Dint_{m}\Dint_{n}-\frac{1}{D-4}\Imet_{mn}\Dint^{2}\bigg)S+\frac{1}{D-4}\Imet_{mn}U  , 
\end{equation}
where $\Dint^{m}t_{mn}=0=\Imet^{mn}t_{mn}$, $\Dint^{m}W_{m}=0$ and $U\equiv \Imet^{pq}T_{pq}$. We refer to $t_{mn},$ $W_{m}$ and $(S,U)$ as the tensor, vector and scalar parts of $T_{mn}$, respectively. 
\end{prop}
%
\par In keeping with our notation, we denote the tensor, vector and scalar parts of $\ov{T}_{mn}$ as $\ov{t}_{mn}$, $\ov{W}_{m}$, $\ov{S}$ and $\ov{U}$. This is in accord with our prior notation of denoting harmonic functions and harmonic one-forms with a bar since, as we shall see, the scalar and vector parts of $\ov{T}_{mn}$ are indeed harmonic. 
Taking the trace of \cref{Tmn} yields 
\begin{equation}
\label{Uconst}
\Dint^{2}\ov{U}=0  , 
\end{equation}
which implies that $\ov{U}$ is a constant. 
Taking the divergence of \cref{Tmndecomp} using \cref{Uconst,divT} then gives 
\begin{equation}
\label{divtmnzeroVS}
\frac{1}{2}\Dint^{2}\ov{W}_{n}=\frac{D-5}{D-4}\Dint_{n}\Dint^{2}\ov{S}.
\end{equation}
Taking another divergence of \cref{divtmnzeroVS} and using the fact that $W_{n}$ is divergence-free gives, 
\begin{equation}
\label{trfreesc}
(D-5)\Dint^{4}\ov{S}=0 .
\end{equation}
The case $D=5$ corresponds to a $1$-dimensional Ricci-flat compact space, namely $S^1$. In this case, $\ov{t}_{mn} = \ov{W}_n=\ov{S}=0$ and the only modulus is a rescaling of the metric. If  $D> 5$ then \cref{trfreesc} implies that $\ov{S}$ is a constant. \Cref{divtmnzeroVS} then requires that $W_{n}$ be harmonic and, by \Cref{KVlem}, it is therefore also Killing. Consequently, $\ov{T}_{mn}$ has no vector part. In addition, its scalar part is constant and determined by its trace. Any solution to \cref{Tmn,divT} can be uniquely decomposed in the form,  
\begin{equation}
\label{finaldecompTmn}
\ov{T}_{mn}=\ov{t}_{mn}+\frac{1}{D-4}\Imet_{mn}\ov{U} , 
\end{equation}
where $\ov{U}$ is a constant while $\ov{t}_{mn}$ is both trace-free and satisfies \cref{Tmn,divT}. The mode $\ov{U}$ is the overall breathing mode of the space. The $\ov{t}_{mn}$ are the volume-preserving moduli. 


\par Finally, we note the enormous simplification for the case of a torus where $\I=\mathbb{T}^{D-4}$. In this case, the Riemann tensor $\IRiem_{mnp}{}^{q}$ vanishes and the $\ov{T}_{mn}$ are constant. Including the overall volume modulus, there are ${1\over 2} (D-4)(D-3)$ metric moduli. 
We summarize these statements about the moduli space of Ricci-flat Riemannian manifolds in the following lemma: 
\begin{lem}
\label{lemTmn}
Let $(\I,\Imet_{mn})$ be a compact, Ricci-flat Riemannian manifold. The solutions $\ov{T}_{mn}$ to \cref{Tmn} can be uniquely decomposed as in \cref{finaldecompTmn} where $\ov{U}$ is a constant and $\ov{t}_{mn}$ satisfies $\Dint^{m}\ov{t}_{mn}=0=\Imet^{mn}\ov{t}_{mn}$. If $\I=\mathbb{T}^{D-4}$ then $\ov{t}_{mn}$ is constant. 
\end{lem}
Therefore, the space of massless linearized perturbations $\ov{h}_{mn}$ can be decomposed into a set of $\dm +1$ scalar fields
\begin{equation}
\label{scalarfields}
\ov{h}_{mn}=\frac{\Imet_{mn}}{D-4}\phi(x)+\sum_{i=1}^{\dm}\Phi^{(i)}(x)t_{mn}^{(i)}(y) , 
\end{equation}
where the scalar field $\phi(x)$ is associated to the volume mode or breathing mode $\ov{U}$, 
and $\dm$ is the dimension of the moduli space of volume preserving deformations. It is important to stress that these modes are guaranteed to be massless only in the linearized approximation with the exception of the volume mode $\phi$ which is exactly massless.

Finally, the linearized Einstein equations imply that the scalars $\phi$ and $\Phi^{(i)}$ satisfy the massless wave equation, 
\begin{equation}
\Box_{\eta}\phi =0 \quad \textrm{ and }\quad \Box_{\eta}\Phi^{(i)}=0.
\end{equation}
Diffeomorphisms of $\ov{h}_{mn}$ can only be generated by one-forms $\xi_{m}$ which change the perturbation by $\Dint_{(m}\xi_{n)}$. Using~\cref{hodge}, we decompose $\xi_{m}=\eta_{m} + \bm{D}_{m}\xi$ with $\bm{D}^n \eta_n=0$, which shows that $\eta_m$ can only affect $W_m$ of~\C{Tmndecomp}. Similarly, $\xi$ cannot affect the zero mode of $U$. 
Consequently the scalar fields $\phi$ and $\Phi^{(i)}$ in \cref{scalarfields} have no diffeomorphism freedom. 
\par 
The preceding discussion is a general analysis of the moduli space of linearized deformations of $\I$. However, the  precise enumeration of solutions to \cref{Tmn,divT} must be treated on a case-by-case basis for each choice of $\I$. In many cases of interest in string theory, $\I$ has special holonomy and one can say more about the count of solutions to \cref{Tmn,divT}. For example, if the internal manifold $\I$ is Calabi-Yau, one can use K\"ahler geometry to compute the dimension of the moduli space of metric deformations in terms of the Hodge numbers $h^{p, q}$ of $\I$; specifically $h^{1,1}$ and $h^{{{D-6}\over 2}, 1}$. 

There is a separate question of whether infinitesimal deformations can be promoted to finite deformations. For Calabi-Yau, $G_2$ and $Spin(7)$ spaces, all zero modes seen in a linear analysis survive to the full nonlinear theory~\cite{doi:10.1142/S0129167X04002296}. In this work, we only need the existence of a finite number of solutions for \cref{Tmn,divT}; we make no additional assumptions about $(\I,\Imet_{mn})$ besides Ricci-flatness. For general Ricci-flat $\I$, it is hard to determine whether the zero modes found at linear order remain massless in a fully nonlinear analysis. 

To either reach $\scrip$ or the actual physical location of the detector, a scalar mode must be either exactly massless or of sufficiently light mass and high-energy that we can approximate the mode as massless. For our analysis, we will need to use the condition that $R_{mn}(\Imet + h) =0$ to third order in $h$ where we only fluctuate the internal metric. This plays a role in Appendix~\ref{EinsEqn} for the asymptotic expansion of the solution in powers of ${1\over r}$ near $\scrip$. However, it is important to note that the asymptotic expansion is only applicable for metric fluctuations that are unobstructed and correspond to exactly massless fields. Let us denote the number of exactly massless volume-preserving scalar modes by $\dims$ in contrast with the number of massless modes $d_L$ in the linearized approximation.

\section{Compactified Isolated Systems} 
\label{sec:isolatedsystems}
\par We first need to define the class of Lorentzian spacetimes that we will study. Although we are motivated by string theory, we do not restrict to $10$ or $11$-dimensional spacetimes. Rather we consider 
$D$-dimensional spacetimes with $4$ non-compact spacetime dimensions and $D-4$ compact Riemannian extra dimensions, which represent `gravitational lumps' or localized metric configurations whose curvature grows weak in asymptotic null directions. Following standard terminology in the general relativity community, we refer to such spacetimes as compactified isolated systems, or simply as isolated systems. As discussed in section~\ref{cmpctst}, this class of metrics describes string compactifications on Ricci-flat spaces and approximates warped compactifications in the limit of large internal volume where the warping becomes small. 

First note that any metric $g_{MN}$ on $M=\B\times \I$ is of the form 
\begin{equation}
\label{genmet}
    ds^2 = g_{\mu \nu }(x,y)dx^{\mu} dx^{\nu} + 2 A_{\mu n}(x,y)dx^{\mu}dy^{n}+\varphi_{mn}(x,y)dy^{m}dy^{n},
\end{equation}
where $x^{\mu}$ and $y^{m}$ are arbitrary local coordinates on $\B$ and $\I$, respectively. We define the notion of an isolated system on a manifold $\M=\B\times \I$ by introducing a geometric gauge in coordinates adapted to outgoing null hypersurfaces. In these coordinates, we define a class of metrics which suitably tend to $\G_{MN}$ in asymptotically large null directions. These coordinates are defined in a manner analogous to the standard Bondi coordinates in four-dimensional asymptotically flat spacetimes. Since these coordinates are essential for the analysis of gravitational radiation, we briefly review their construction here. 

\par The Bondi coordinates are denoted $(u,r,\theta^{A},y^{m})$. In Bondi gauge $u$ is a function on spacetime such that surfaces of constant $u$ are outgoing null hypersurfaces. The coordinates $\theta^{A}$ are two arbitrary angular coordinates on $S^{2}$, and the $y^{m}$ are $D-4$ arbitrary coordinates on $\I$. In Bondi gauge, the normal co-vector $\nabla_{M}u$ is null $g^{MN}(\nabla_{M}u)(\nabla_{N}u)=0$ and we define the corresponding future directed null vector $K^{M}\equiv -g^{MN}\nabla_{N}u$. The $r$ coordinate is a `radial' coordinate which varies along the null rays. Note this is not a space-like coordinate but a null coordinate! In this gauge, the tangent to the null rays corresponds to the radial coordinate vector field. In summary, 
\begin{equation}
\label{K}
  K_{M}\equiv -\nabla_{M}u, \qquad K^{M}= \bigg(\frac{\partial}{\partial r}\bigg)^{M} \; \textrm{ and } \quad g_{MN}K^{M}K^{N}=0.  \qquad    \textrm{(Bondi gauge)}
\end{equation}
The angular coordinates $\theta^{A}$ and the internal coordinates $y^{M}$ are both chosen to be constant along these outgoing null rays so that $K^{M}\nabla_{M}\theta^{A}=-g^{MN}(\nabla_{M}u)(\nabla_{N}\theta)=0$ and $K^{M}\nabla_{M}y^{m}=-g^{MN}(\nabla_{M}u)(\nabla_{N}y^{m})=0$. These Bondi gauge conditions imply that the metric $g_{MN}$ satisfies:
\begin{equation}
\label{bondigaugeconditions}
  g_{rr}=0, \quad g_{rA}=0 \; \textrm{ and  }\quad A_{rm}=0 ,
\end{equation}
where $A_{\mu n}$ is defined in \cref{genmet}. The metric $g_{MN}$ in these coordinates is adapted to outgoing null hypersurfaces. Now we define an isolated system with compact extra dimensions which tends to the Ricci-flat metric \C{background}. In coordinates $(u,r,\theta^{A},y^{m})$ adapted to outgoing null directions, the asymptotic metric is given by 
\begin{align}
\label{etabondi}
    \G_{MN}dx^{M}dx^{N} &=\eta_{\mu \nu}dx^{\mu}dx^{\nu}+\Imet_{mn}dy^{m}dy^{n}, \nonumber \\
    &= -du^{2}-2dudr + r^{2}q_{AB}d\theta^{A}d\theta^{B} + \Imet_{mn}dy^{m}dy^{n}.
\end{align}
We define an isolated system as a metric $g_{MN}$ given by \cref{genmet} which, in coordinates $x^{\mu}=(u,r,\theta)$ and $y^{m}$, approaches the flat metric $\G_{MN}$ given by \cref{etabondi} in powers of ${1\over r}$ in the orthonormal frame described in section~\ref{sec:conventions}:
\begin{equation}
\label{ansatz}
    g_{\mu \nu}\sim \eta_{\mu \nu}+\sum_{n=1}^{\infty}r^{-n}h_{\mu \nu}^{(n)}, \quad A_{\mu n}\sim \sum_{n=1}^{\infty}r^{-n}A_{\mu n}^{(n)} \; \textrm{ and }\; \varphi_{mn}\sim \Imet_{mn} + \sum_{n=1}^{\infty}r^{-n}\varphi_{mn}^{(n)}.
\end{equation}
This is gauge-equivalent to the Bondi gauge choice\footnote{The original Bondi gauge conditions also impose that the ``radial'' coordinate correspond to an areal coordinate which imposes that $\partial_{r}(\textrm{det}(g_{AB}))$. Additionally, the fall-off $g_{ur}$ in Bondi gauge is such that $g_{ur}^{(1)}$ vanishes. We shall not impose these conditions in the general fall-off given by \cref{ansatz}}
\begin{equation}
  h^{(n)}_{rr}=0, \quad h^{(n)}_{rA}=0 \quad \textrm{ and  }\quad A^{(n)}_{rm}=0,  
\end{equation}
for all $n$. The symbol ``$\sim$'' in \cref{ansatz} denotes an asymptotic expansion. 
For convenience we have assumed an asymptotic expansion in ${1\over r}$ to all orders with the upper limit of the sums in \cref{ansatz} taken to be $\infty$. This is not strictly necessary for most of this analysis. The results obtained in    \cref{subsec:asympsymm,subsec:reduction_nonlinear,radera} require only that \cref{ansatz} be valid at order $n=1$. The results obtained in \cref{statera} require that \cref{ansatz} be valid up to order $n=3$. 

A full analysis of the validity of this ansatz would require examining global stability for a suitable class of initial data. Such an analysis was undertaken in \cite{Wyatt:2017tow, Andersson:2020fuz} where stability was proven in the case of supersymmetric compactifications. It would be interesting to study the asymptotic behavior of such solutions near null infinity and compare with the ansatz assumed here. 



\par As noted in \cref{sec:conventions}, our conventions for expanding the metric coefficients in powers of ${1\over r}$  differs from more common conventions. Usually the expansion coefficients refer to the powers of ${1\over r}$ which arise from the components of $g_{MN}$ in a coordinate basis. In our conventions spelled out in \cref{sec:conventions}, the metric expansion coefficients $g_{\mu \nu}^{(k)}$, $A_{\mu m}^{(k)}$ and $\varphi_{mn}^{(k)}$ all contribute to the {\em physical} fall-off rate of the metric $g_{MN}$ at order ${1\over r^{k}}$, as seen in any orthonormal frame. 
From the preceding discussion, Bondi gauge has a preferred geometric status in constructing the notion of an isolated system. We shall see, however, that Bondi gauge does not appear to be the preferred gauge when asymptotically solving the leading order Einstein equations with compact spatial directions, studied in \cref{radera,statera}.

We also need to specify the asymptotic fall-off of the stress-energy tensor. The inclusion of massive sources is straightforward since their stress-energy vanishes near $\scrip$. For massless sources, we demand that 
\begin{equation}
\label{stressfalloff}
    T_{MN} =\sum_{n=2}^{\infty}r^{-n}T_{MN}^{(n)} ,
\end{equation}
where the non-vanishing component of the leading order stress tensor are $T_{uu}^{(2)}$, $T_{um}^{(2)}$ and $T_{mn}^{(2)}$. This is consistent with the dominant energy condition. As we will see, the fall-off of $T_{\mu \nu}$ and $T_{\mu m}$ ensure finiteness of the energy flux and charge-current flux to $\scrip$. The fall-off of $T_{mn}$ agrees with the intuition from Kaluza-Klein reduction.



There is one further condition we will impose, which turns out to be easily satisfied by the most common forms of stress-energy. From our ansatz~\C{ansatz} and the analysis found in Appendix~\ref{EinsEqn}, we see that $ \int_{\I} \hat{g}^{mn}G_{mn}^{(2)}=0$. This turns out to be surprisingly nontrivial to demonstrate.  Einstein's equations then imply that the zero mode,  $\int_{\I} \hat{g}^{mn}T_{mn}^{(2)}$, vanishes. In fact, $G_{mn}^{(2)}$ is orthogonal to every exactly massless scalar fluctuation $t^{mn}$, not just the breathing mode of $\I$. Similarly, we will impose a stronger condition on the stress-energy tensor that $\int_{\I} t^{mn}T_{mn}^{(2)}$ vanishes for every exactly massless scalar fluctuation $t^{mn}$. This stronger version is also motivated from the analysis found in Appendix~\ref{EinsEqn}.

We can see whether this is a reasonable condition by examining a few typical sources of stress-energy. 
If one considers a $D$-dimensional scalar field $\phi$ with stress-tensor
\be 
T_{MN}= \nabla_{M}\phi \nabla_{N}\phi - \frac{1}{2}g_{MN}\nabla^{P}\phi \nabla_{P}\phi,
\ee 
and 
\be 
\phi = \phi^{(0)} + \frac{\phi^{(1)}(u,\theta, y)}{r} +\ldots ,
\ee 
then in this simple case, $\phi^{(1)}$ is harmonic on $\I$ and therefore constant in $y$. The leading non-vanishing stress-tensor component is then $T_{uu}^{(2)}=(\partial_{u}\phi^{(1)})^{2}$ and $T_{mn}^{(2)}=0$. If one generalizes this case by considering a $p$-form field strength $F$ with $D$-dimensional action $-\int_{M} {1\over 2 (p!)} \,  \, F_{M_1\ldots M_p} F^{M_1\ldots M_p}$, the stress-tensor takes the form:
\begin{equation}
    T_{MN} = {1\over 2 (p-1)!} \left( F_{M M_1 \ldots M_{p-1}} \tensor{F}{_N^{M_1}^{\ldots} ^{M_{p-1}}} - {1\over 2 p} g_{MN} F_{M_1 \ldots M_{p}} \tensor{F}{^{M_1}^{\ldots} ^{M_{p}}} \right). 
\end{equation}
In Kaluza-Klein reduction near $\scri$, $F=dA$ gives rise to massless spacetime fields associated to harmonic forms on $\I$ as
\begin{equation}\label{potentialexpansion}
A^{(1)}_{M_1\ldots M_{p-1}}(u,\theta,y) = 
\phi^{(1)}_{\mu_1\ldots \mu_q}(u, \theta) \, \omega_{m_{q+1}\ldots m_{p-1}}(y),     
\end{equation}
where $\omega\in H^{p-q-1}(\I, \mathbb{R})$ is a harmonic representative of the cohomology class. The field strength $F^{(1)}= d\phi^{(1)} \wedge \omega$, where at this order $d\phi^{(1)} = -\partial_u \phi^{(1)} \wedge K$ and the one-form $K$ is defined in~\C{K}. As noted in~\C{K}, $K$ is null with respect to the asymptotic metric so $ T^{(2)}_{mn}=0$ again as in the case of the scalar field. For these sources of stress-energy commonly found in string theory, we see a much stronger constraint on the asymptotic stress tensor than we assume; namely that 
\be 
T^{(2)}_{MN}= \frac{1}{2(p-1)!}\left(\partial_{u}\phi^{(1)}\right)^{2} K_{M}K_{N} \cdot |\omega|^2, 
\ee 
where $| \omega |^2 = \omega_{m_{q+1} \ldots m_{p-1}} \omega^{m_{q+1} \ldots m_{p-1}}$. Although in these cases of physical interest the stress tensor satisfies stronger conditions, in the body of this work we will only use the weaker assumptions of fall-off given by \cref{stressfalloff}.





Finally while we have defined isolated systems in the case where the spacetime is a product manifold, one can straightforwardly extend this definition to include a wider class of fibered metrics, including some gravitational instantons. For example, we could consider ${\mathbb R} \times {\rm TN}$ where TN refers the multi-Taub-NUT metric and ${\mathbb R}$ is time. This example is a particularly nice generalization of the circle compactification, which we will discuss in section~\ref{sec:circle}. The total space $\M$ is topologically ${\mathbb R^5}$, but the TN metric at spatial infinity is a Hopf fibration $S^1 \hookrightarrow S^3 \to S^2$. The Chern number of the fibration corresponds to the magnetic charge for the Kaluza-Klein gauge-field found from reducing the metric on the asymptotic $S^1$. The picture under Kaluza-Klein reduction on the asymptotic $S^1$ is a collection of particles located at the NUT singularities of the TN metric, which are magnetically-charged under the Kaluza-Klein gauge-field. 
While in this construction, TN appears only in the spatial metric and time is completely factorized, there have been studies of asymptotic symmetries and dual supertranslations where TN appears with the fibered $S^1$ identified with time~\cite{Kol:2019nkc}.


While we will primarily focus on the case of product manifolds, many of our results only require that the metric satisfy \cref{ansatz} locally in some neighborhood of null infinity. In particular, our results about the asymptotic dimensional reduction of the Weyl tensor, the local constraints on the radiative order metric and asymptotic symmetries, found in \cref{subsec:asympsymm,subsec:reduction_nonlinear,radera}, remain valid as long as the metric asymptotes to $\G_{MN}$ at $\scrip$. On the other hand, arguments that involve inversion of elliptic operators on the sphere or integrating Einstein's equations over retarded time, found in \Cref{statera,sec:memory_effect_compactified}, will need to be modified in the fibered case. In order to extend these results to the fibered case, it is more useful to work with manifestly gauge invariant quantities. In \Cref{app:linearized_gauge_invariant}, we provide an alternative, manifestly gauge invariant derivation of our results in {linearized gravity} using the Bianchi identity.

\section{Asymptotics near Null Infinity}\label{sec:asymptotic_behavior}

In this section we will analyze the asymptotic behavior of the spacetime for an isolated system near null infinity. We first collect some results regarding the asymptotic behavior of the Weyl tensor for { any} isolated system without imposing decay conditions. Unless stated otherwise, we consider a metric $g_{MN}$ which satisfies the asymptotic expansion \cref{ansatz} near null infinity and obeys Einstein's equations: 
\begin{equation}
\label{EE}
    R_{MN}-\frac{1}{2}g_{MN}R = 8\pi T_{MN}.
\end{equation}
 In \cref{subsec:reduction_nonlinear} we show that the Bianchi identity implies that the `electric' part of the Weyl tensor,~defined in \cref{elweyl},  at order ${1\over r}$ admits a dimensional reduction in a manner exactly analogous to the dimensional reduction given in \cref{sec:linearized_review}. 
 In \cref{statera,radera} we examine, in detail, the change in the metric caused by a `burst' of gravitational radiation. We characterize this `burst' by requiring that the metric be stationary at asymptotically early and late times. In \cref{radera}, we analyze Einstein's equations during the radiative epoch. In \cref{statera}, we investigate the implications of Einstein's equations during the stationary eras.
\subsection{Asymptotic reduction in nonlinear gravity}\label{subsec:reduction_nonlinear}
As shown in \Cref{sec:linearized_review}, {linearized} metric perturbations in  Lorenz gauge with background metric \C{background} reduce to a collection of gravitons, graviphotons and scalars. In the full nonlinear theory, we will show that the leading order electric Weyl tensor for any isolated system at null infinity admits a harmonic decomposition in a way analogous to linearized Kaluza-Klein analysis. 
This provides a gauge invariant description of radiation, Kaluza-Klein decomposed into spin-0, spin-1 and spin-2 components, in full nonlinear general relativity. 
\par We remind the reader that the Weyl tensor is related to the Riemann tensor, 
\begin{equation}
\label{WeylRiemSchout}
    C_{MNPQ}=R_{MNPQ}-2g_{[M[P}S_{Q]N]},
\end{equation}
where $S_{MN}$ is the Schouten tensor which, in terms of the Ricci tensor, is given by:
\begin{equation}
\label{Shout}
    S_{MN}=\frac{2}{D-2}R_{MN}-\frac{1}{(D-1)(D-2)}g_{MN}R.
\end{equation}
Since the Einstein tensor is divergence free, the Schouten tensor satisfies $\nabla^{M}S_{MN}=\nabla_{N}S$ where $S\equiv g^{MN}S_{MN}$. The uncontracted Bianchi identity is  
\begin{equation}
\label{bianchi}
\nabla_{[M}C_{NP]QR}=-2g_{[Q[N}\nabla_{M}S_{P]R]}.
\end{equation}
The nested notation appearing on the right hand side of~\C{bianchi} means antisymmetrize over $(N,M,P)$ and antisymmetrize over $(Q,R)$ separately. We will use this notation below. Contracting over $M$ and $Q$ and using the tracelessness of the Weyl tensor yields
\begin{equation}
    \label{divweyl}
    \nabla^{M}C_{MPQR}= (D-3)\nabla_{[Q}S_{R]P}.
\end{equation}
Applying $g^{MT}\nabla_{T}$ to \cref{bianchi}, commuting the derivatives and using \cref{divweyl,WeylRiemSchout} implies 
\begin{align}
\label{waveweyl}
\Box_{g}C_{NPQR}&=2(D-2)\nabla_{[N}\nabla_{[Q}S_{R]P]}-2g_{[Q[N}\Box_{g}S_{P]R]}+2g^{MT}g_{[Q[N}\nabla_{|T|}\nabla_{P]}S_{R]M}\nonumber \\
&-(D-2)g^{TM}S_{T[N}C_{P]MQR}+2g^{TM}S_{T[Q}C_{R][NP]M}-2g^{OM}g^{RT}S_{OR}g_{[Q[N}C_{P]|M|R]T}\nonumber \\
&+\frac{1}{2}g^{MT}S_{MT}C_{NPQR}+2g^{MT}S_{M[N}C_{P][QR]T}+2g^{MO}g^{TK}C_{M[NP]T}C_{OKQR} \nonumber \\
&+4g^{MO}g^{TK}C_{M[Q[N|T|}C_{P]|K|R]O},
\end{align}
where $\Box_{g}\equiv g^{MN}\nabla_{M}\nabla_{N}$. Therefore in any spacetime, the Weyl tensor satisfies the wave equation with source given by terms that are either products of the Weyl tensor, products of the Weyl tensor with the Schouten tensor or derivatives of the Schouten tensor. The asymptotic expansion of the metric given by \C{ansatz} implies the ${1\over r}$ expansion for the Weyl tensor: 
\begin{equation}
\label{weylexp}
    C_{NPQR}\sim \sum_{n=0}^{\infty}\frac{C^{(n)}_{NPQR}}{r^{n}}.
\end{equation}
After imposing Einstein's equations the only non-vanishing components of $C^{(0)}_{NPQR}$ is the Riemann tensor $\IRiem_{npqr}$ of the Ricci-flat asymptotic internal space $\I$ with metric $\Imet_{mn}$. Further, the Schouten tensor is defined in terms of the Ricci tensor in \cref{Shout} which, in turn, can be written in terms of the stress energy tensor by Einstein's equation \cref{EE}.

The asymptotic fall-off condition on the stress-tensor is given in \cref{stressfalloff}. 
This stress tensor fall-off directly implies an asymptotic expansion of the Schouten tensor, 
\begin{equation}
    \label{Shoutexp}
    S_{MN}\sim \sum_{n=2}^{\infty}\frac{S_{MN}^{(n)}}{r^{n}},
\end{equation}
where the sum starts at $O({1\over r^{2}})$ and $S_{MN}^{(2)}=\frac{2}{D-2}T_{MN}^{(2)}$.  We now show that \cref{waveweyl,divweyl} place strong constraints on the asymptotic behavior of the `electric part' of the Weyl tensor near null infinity. In particular, the leading order electric part of the Weyl tensor can be dimensionally-reduced in exactly the same manner as reviewed in \cref{sec:linearized_review}, but now in the full nonlinear theory. 
 The electric part of the Weyl tensor is defined as
\begin{equation}
\label{elweyl}
    E_{PR}\equiv C_{NPQR} \, n^N n^Q,
\end{equation}
where $n^{M}\equiv \left(\partial/\partial u \right)^{M}$. The properties of the Weyl tensor imply that the electric Weyl tensor is symmetric, tracefree and that its $u$-components vanish: 
\begin{equation}
\label{elweylprop}
  E_{MN}=E_{NM}, \quad g^{MN}E_{MN}=0 \quad \textrm{ and } \quad E_{uN}=0.
\end{equation} 
We note that $\underset{r\rightarrow \infty}{\lim} {E}_{MN}$ vanishes at fixed $u,\theta^{A}$ and $y^{m}$, and therefore the leading order electric Weyl tensor given by,
\begin{equation}
\label{elweyl1}
    \mathcal{E}_{MN}(u,\theta^{A},y^{m})\equiv \lim_{r\rightarrow \infty}rE_{MN}(r,u,\theta^{A},y^{m}),
\end{equation}
is gauge invariant. From the above relations, we now prove the following key lemma regarding the asymptotic dimensional reduction of $E_{MN}$. 
\begin{lem}[Asymptotic reduction of electric Weyl]
\label{nonlindimred}
Let $(M,g)$ be an isolated system whose metric $g_{MN}$ has an asymptotic expansion given by \cref{ansatz} and let $\mathcal{E}_{MN}$ be the leading order, electric Weyl tensor defined by \cref{elweyl,elweyl1}. $\mathcal{E}_{MN}$ satisfies the following properties:
\begin{enumerate}
    \item The components $\mathcal{E}_{uM}$ and $\mathcal{E}_{rM}$ vanish for any isolated system.
    \item The nonvanishing components satisfy 
    \begin{align}
        &\mathcal{E}_{AB}=\zm{\mathcal{E}}_{AB}(u,\theta), \quad \mathcal{E}_{Am}=\sum_{i=1}^{b_{1}}\mathcal{E}_{A}^{(i)}(u,\theta)\otimes \ov{V}_{m}^{(i)}(y^{m}), \\
        & \mathcal{E}_{mn}=-\frac{\Imet_{mn}}{D-4}q^{AB}\ov{\mathcal{E}}_{AB}(u,\theta)+\sum_{i=1}^{\dm}\mathcal{E}^{(i)}(u,\theta)\ov{t}_{mn}^{(i)}(y^{m}).\nonumber
    \end{align}
   The  $\ov{V}_{m}^{(i)}$ are a basis for the $b_{1}$ harmonic $1$-forms on $\I$, where $b_{1}$ is the first Betti number of $\I$. The $\ov{t}_{mn}^{(i)}$ are a basis of the $\dm$ symmetric, rank 2 tensors which satisfy the Lichnerowicz equation on $\I$ and $\bm{D}^{m}\ov{t}_{mn}^{(i)}=\Imet^{mn}\ov{t}_{mn}^{(i)}=0$, where $\dm + 1$ is the dimension of the moduli space. 
\end{enumerate}
\begin{proof}
That $\mathcal{E}_{uM}$ vanishes follows directly from the definition and properties of the electric Weyl tensor given in \cref{elweyl,elweylprop}. To prove that $\mathcal{E}_{rM}$ vanishes we note that contracting \cref{waveweyl} on the $N$ and $Q$ indices with $n^{N}$ and $n^{Q}$ gives the following equations for the electric Weyl tensor at order ${1\over r}$:
\begin{align}
\label{harmweyl}
    &\Dint^{2}\mathcal{E}_{\mu \nu}=0, \quad \Dint^{2}\mathcal{E}_{\mu n}=0 \quad \textrm{ and } \quad \bm{D}^{2}\mathcal{E}_{mn}+2\IRiem_{m}{}^{p}{}_{n}{}^{q}\mathcal{E}_{pq}=0.  
\end{align}
Since $\mathcal{E}_{MN}$ is gauge invariant we assume, without loss of generality, that the metric $g_{MN}$ is in a gauge such that the metric expansion coefficents $h_{rr}^{(1)}$, $h_{rA}^{(1)}$ and $h_{rm}^{(1)}$ all vanish. A straightforward calculation of the electric Weyl tensor using the metric in Bondi gauge implies that,
\begin{equation}
    \mathcal{E}_{rA}=0, \quad \mathcal{E}_{rr}=0 \quad \textrm{ and } \quad \mathcal{E}_{rm}=0. 
\end{equation}
Since $\mathcal{E}_{MN}$ is gauge invariant we conclude that $\mathcal{E}_{rM}$ vanishes for any isolated system. Applying $n^{P}$ and $n^{R}$ to the $P$ and $R$ components of \cref{divweyl} at order ${1\over r}$ and using the fact that $\mathcal{E}_{rM}$ vanishes gives
\begin{equation}
\label{harmweyldiv}
    \bm{D}^{n}\mathcal{E}_{A n}=0 \quad \textrm{ and }\quad \bm{D}^{m}\mathcal{E}_{mn}=0.
\end{equation}
\Cref{harmweyl,harmweyldiv} together with \Cref{lemTmn,KVlem} imply that $\mathcal{E}_{AB}$ and $\Imet^{mn}\mathcal{E}_{mn}$ are harmonic on $\I$, $\mathcal{E}_{Am}$ is spanned by harmonic $1$-forms $\ov{V}_{m}^{(i)}$ on $\I$, and the trace-free part of $\mathcal{E}_{mn}$ is spanned by $\ov{t}_{mn}^{(i)}$. Finally we note that 
\begin{equation}
    \Imet^{mn}\mathcal{E}_{mn}=-q^{AB}\mathcal{E}_{AB} ,
\end{equation}
which follows from the tracelessness of $\mathcal{E}_{MN}$ as well as the vanishing of $\mathcal{E}_{uM}$ and $\mathcal{E}_{rM}$.
\end{proof}
\end{lem}

\par \Cref{nonlindimred}  implies that the non-vanishing components of the leading order electric Weyl tensor, $\mathcal{E}_{MN}$, can be viewed as a tensor on $\Sp^{2}\times \I$. Let $\mathfrak{q}_{ab}$ be a $(D-2)$-dimensional product metric on $\Sp^{2}\times \I$ which, for arbitrary coordinates $x^{a}=\{\theta^{A},y^{m}\}$ on $\Sp^{2}\times \I$, is defined by\footnote{We faced an unfortunate choice in labeling combined coordinates for the sphere and the internal space. Either introduce a new letter or use $x^a$, which we hope the reader will not confuse with $x^\m$. We hope this choice is the lesser of two evils. All conventions are spelled out in section~\ref{sec:conventions}. }
\begin{equation}
    \mathfrak{q}_{ab} \, dx^{a} \, dx^{b}=q_{AB} \, d\theta^{A} \, d\theta^{B}+\Imet_{mn} \, dy^{m} \, dy^{n}.
\end{equation}
It is convenient to define a `news tensor' on $S^{2}\times \I$ which we denote  $\mathcal{N}_{ab}$, 
\begin{equation}\label{news}
    \mathcal{N}_{ab}\equiv \lim_{r\rightarrow \infty}r\bigg(\mathfrak{q}_{a}{}^{c}\mathfrak{q}_{b}{}^{d}-\frac{1}{D-2}\mathfrak{q}_{ab}\mathfrak{q}^{cd}\bigg)\partial_{u}\zm{g}_{cd},
\end{equation}
where $\zm{g}_{ab}$ is the zero mode of $g_{MN}$ along the $S^{2}\times \I$ directions. The components of $\mathcal{N}_{ab}$ satisfy 
\begin{equation}\label{news_conditions}
    \bm{D}^{2}\mathcal{N}_{AB}=0, \quad \bm{D}^{2}\mathcal{N}_{Am}=0 , \quad \bm{D}^{2}\mathcal{N}_{mn}+2\IRiem_{m}{}^{p}{}_{n}{}^{q}\mathcal{N}_{pq}=0, \quad \Imet^{mn}\mathcal{N}_{mn}=-q^{AB}\mathcal{N}_{AB},
\end{equation}
and the news therefore admits the decomposition, 
\begin{align}
    &\mathcal{N}_{AB}=N_{AB}(u,\theta)+\frac{1}{2}q_{AB}N(u,\theta), \quad  \mathcal{N}_{Am}=\sum_{i=1}^{b_{1}}N_{A}^{(i)}\otimes V_{m}^{(i)}(y^{m}), \label{NAB_and_NAm} \\
    &\mathcal{N}_{mn}=-\frac{\Imet_{mn}}{D-4}N(u,\theta)+\sum_{j=1}^{\dims}\mathcal{N}^{(j)}(u,\theta)~\ov{t}^{(j)}_{mn}(y^{m}), \label{Nmn}
\end{align}
where $N_{AB}$ is the trace-free projection of $\mathcal{N}_{AB}(u,\theta)$ and $N$ is the trace of $\mathcal{N}_{AB}$ on $S^{2}$ given by: 
\be 
N_{AB} = \bigg(q_{A}{}^{C}q_{B}{}^{D}-\frac{1}{2}q_{AB}q^{CD}\bigg)\mathcal{N}_{CD}(u,\theta) \quad \textrm{ and } \quad N=q^{AB}\mathcal{N}_{AB}(u,\theta).
\ee
\Cref{NAB_and_NAm,Nmn} give a decomposition of radiation in the full spacetime $\M$ into spin-2, spin-1 and spin-0 components. The four-dimensional Bondi news is related to the trace-free part $N_{AB}$, but note that $N_{AB}$ here is computed in $D$-dimensional Einstein frame. In section \ref{sec:frames}, we will discuss how the news and related observables are affected by the choice of frame.

The decomposition of the radiative modes given by \cref{Nmn} corresponds to the exactly massless modes arising from $\I$. The decomposition given by \Cref{nonlindimred} is a consequence of the leading order Bianchi identity and Einstein's equations. However, as we have spelled out in \cref{subsec:linscalar}, the space of truly massless modes is a subset of the modes enumerated in \Cref{nonlindimred}. The spin-2 mode, spin-1 modes and the scalar volume mode are truly massless. However, the number of truly massless volume-preserving scalars are $\dims\leq  \dm$. Therefore in \cref{Nmn}, we replaced $\dm$ with $\dims$. As we show in Appendix~\ref{EinsEqn}, if we had not done this truncation then our ansatz would not be consistent with Einstein's equations. 

Finally, a direct calculation of $\mathcal{E}_{MN}$ in terms of the metric implies that the non-vanishing components of $\mathcal{E}_{MN}$ can be compactly expressed in terms of $\mathcal{N}_{ab}$:
\begin{equation}
    \mathcal{E}_{ab}=-\frac{1}{2}\partial_{u}\mathcal{N}_{ab}.
\end{equation}
We refer to $\mathcal{N}_{ab}$ as the `news' tensor which is analogous to the Bondi news tensor in four dimensional asymptotically flat spacetimes. In such spacetimes, the null memory effect is determined by the squared Bondi news tensor integrated over retarded time, as discussed in section~\ref{fourdeffectivefieldtheory}. In \Cref{sec:memory_effect_compactified}, we prove that analogous statements hold for isolated systems with compact extra dimensions.

\subsection{Asymptotic analysis of the metric}
\label{radera}
We now analyze the leading order solution of Einstein's equations in the neighborhood of null infinity. We assume that the metric is initially in Bondi gauge which implies, in particular,
\begin{equation}
    h_{rr}^{(1)}=0, \quad h_{rA}^{(1)}=0 \quad \textrm{ and } \quad A_{rm}^{(1)}=0 ,
\end{equation}
where $A_{rm}$ is defined in~\C{genmet}. 
 Einstein's equation at leading order in ${1\over r}$ gives the following constraints: 
\begin{align}
    \textrm{$(uu;1)$ }&\quad \quad \quad \bm{D}^{2}h_{uu}^{(1)}+2\partial_{u}\bm{D}^{m}A_{mu}^{(1)}-\partial_{u}^{2}(q^{AB}h^{(1)}_{AB}+\Imet^{mn}\varphi_{mn}^{(1)})=0, \label{uu1}\\
    \textrm{$(ur;1)$ }&\quad \quad \quad \bm{D}^{2}h_{ur}^{(1)}=0, \label{ur1}\\
    \textrm{$(uA;1)$ }& \quad \quad \quad \bm{D}^{2}h_{uA}^{(1)}+\partial_{u}\bm{D}^{m}A_{Am}^{(1)}=0, \label{uA1}\\
    \textrm{$(AB;1)$ }&\quad \quad \quad \bm{D}^{2}h_{AB}^{(1)}=0, \label{AB1}\\
    \textrm{$(um;1)$ }&\quad \quad \quad \bm{D}^{2}A_{um}^{(1)}-\bm{D}_{m}\bm{D}^{n}A_{un}^{(1)}+\partial_{u}\bm{D}^{n}\varphi_{nm}^{(1)}+\partial_{u}\bm{D}_{m}(h_{ur}^{(1)} - q^{AB}h_{AB}^{(1)})  \label{um1}\\ 
    \quad &\quad \quad\quad  -\partial_{u}\bm{D}_{m}\Imet^{pq}\varphi_{pq}^{(1)}=0, \nonumber \\
    \textrm{$(Am;1)$ }&\quad \quad \quad\bm{D}^{2}A_{Am}^{(1)}-\bm{D}_{m}\bm{D}^{n}A_{An}^{(1)}=0, \label{Am1}\\
    \textrm{$(mn;1)$ }&\quad \quad \quad \bm{D}^{2}\varphi_{mn}^{(1)}+2\IRiem_{m}{}^{p}{}_{n}{}^{q}\varphi_{pq}^{(1)}-2\bm{D}_{(m}\bm{D}^{p}\varphi^{(1)}_{n)p}-2\bm{D}_{m}\bm{D}_{n}h_{ur}^{(1)} \label{mn1}\\ 
    \quad &\quad \quad\quad +\bm{D}_{m}\bm{D}_{n}(q^{AB}h_{AB}^{(1)}+\Imet^{pq}\varphi_{pq}^{(1)})=0.\nonumber 
\end{align}
The notation on the left hand side $(MN; k)$ refers to the $MN$ components of Einstein's equations at order ${1\over r^k}$. To solve these equations we want to find gauge choices, in a manner compatible with \cref{ansatz}, so that the following equations are true:
\begin{equation}
\label{radgauge}
    \bm{D}^{m}A_{um}^{(1)}=0, \quad \bm{D}^{m}A_{Am}^{(1)}=0 \, \textrm{ and }\, \varphi^{(1)}_{mn}=\Phi_{mn}+\bigg(\bm{D}_{m}\bm{D}_{n}-\frac{\Imet_{mn}}{D-4}\bm{D}^{2}\bigg)\Psi+\frac{\Imet_{mn}}{D-4}\phi,
\end{equation}
where $\bm{D}^{m}\Phi_{mn}=0=\Imet^{mn}\Phi_{mn}$, and $\phi(u,\theta)$ is constant on $\I$. We want to construct a diffeomorphism, specified by a vector-field, that preserves our asymptotic fall-off conditions and implements~\C{radgauge}. So we assume that the vector field has the form, 
\begin{equation}
    \xi_{M} \sim \frac{\xi_{M}^{(1)}(u,\theta,y)}{r}+ O\bigg(\frac{1}{r^{2}}\bigg),
\end{equation}
where we assume no $O(r^0)$ term in $\xi_M$. Under this diffeomorphism, the metric shifts by $g_{MN} \to g_{MN} + 2 \nabla_{(M} \xi_{N)}$. In order to achieve the gauge conditions of \cref{radgauge} the components of $\xi^{(1)}_{M}$ must satisfy 
\begin{align}\label{diffeomorphism_components}
    \bm{D}_m \xi_A^{(1)} = - A_{Am}^{(1)} \, , \quad  - \partial_u \xi_m^{(1)} + \bm{D}_m \xi_u^{(1)} = - A_{um}^{(1)} \, , \quad \bm{D}_{(m} \xi_{n)}^{(1)} = - \frac{1}{2}\varphi_{mn}^{(1)}.
\end{align}
To ensure that we preserve the Bondi gauge conditions at leading order, we set $\xi_{r}^{(1)}=0$. The first equation in (\ref{diffeomorphism_components}) implies that $\bm{D}^2 \xi_A^{(1)} = - \bm{D}^m A_{Am}^{(1)}$. The right side of this equation has no zero mode, and so we can solve for $\xi_A^{(1)}$. Next, using \Cref{propsymm}, we can decompose $\varphi^{(1)}_{mn}$ into tensor, vector and scalar parts:
\begin{align}
\label{varphimndecomp}
    \varphi^{(1)}_{mn} = \Phi_{mn} + \bm{D}_{(m} \zeta_{n)} + \left( \bm{D}_m \bm{D}_n - \frac{1}{D - 4} \Imet_{mn} \bm{D}^2 \right) \Psi + \frac{\Imet_{mn}}{D-4}\phi \, ,
\end{align}
where $\Imet^{mn}\Phi_{mn}=\bm{D}^{m}\Phi_{mn}=0$ and $\bm{D}^{m}\zeta_{m}=0$. Using \cref{hodge}, $\xi_m^{(1)} = \bm{D}_m \xi + \eta_{m}$ where $\bm{D}^{m}\eta_{m}=0$. Using these decompositions and taking the trace of the third equation in (\ref{diffeomorphism_components}) gives $ \bm{D}^2 \xi = - \frac{1}{2}\phi$. The zero-mode of $\phi$ is the obstruction to solving for $\xi$. Subtracting out the zero mode, we can solve $ \bm{D}^2 \xi = - \frac{1}{2}(\phi-\ov{\phi})$. With this choice of $\xi$, we can replace $\phi$ by $\bar{\phi}(u,\theta)$. Furthermore, we can choose $\eta_m = -\frac{1}{2}\zeta_m$, which eliminates the vector part of $\varphi^{(1)}_{mn}$. Finally, we consider the divergence of the second equation in (\ref{diffeomorphism_components}), $\bm{D}^2 \xi_u^{(1)} = - \bm{D}^m A_{um}^{(1)} + \partial_u \bm{D}^{2}\xi$. Since the right side of this equation has no zero mode, we can solve for $\xi_u^{(1)}$. This completes the specification of the diffeomorphism which implements \C{radgauge}.

The leading order Einstein equation (\cref{uu1,ur1,uA1,AB1,um1,Am1,mn1}) can now be directly solved. In this gauge, \cref{ur1,uA1,AB1} imply that $h_{ur}^{(1)},h_{uA}^{(1)}$ and $h_{AB}^{(1)}$ are constant on $\I$. Therefore, 
\begin{equation}
\label{zmodeuruAAB}
    h^{(1)}_{ur}=\ov{h}^{(1)}_{ur}(u,\theta), \quad h^{(1)}_{uA}=\ov{h}^{(1)}_{uA}(u,\theta), \quad h^{(1)}_{AB}=\ov{h}^{(1)}_{AB}(u,\theta). 
\end{equation}
\Cref{radgauge,zmodeuruAAB} imply that \cref{uu1}, which takes the form
\begin{equation}
\label{uu1gauge}
    \bm{D}^{2}h_{uu}^{(1)}=\partial_{u}^{2}(q^{AB}\ov{h}_{AB}^{(1)}+\phi),
\end{equation}
can be directly solved. Since the right hand side of \cref{uu1gauge} is in the kernel of the Laplacian $\bm{D}^{2}$, the left and right hand sides must both vanish implying 
\begin{equation}
\label{zmodeuu}
    h_{uu}^{(1)}=\ov{h}^{(1)}_{uu}(u,\theta) \quad \textrm{ and }\quad \partial_{u}^{2}(q^{AB}\ov{h}_{AB}^{(1)}+\phi)=0 .
\end{equation}
Applying $\Imet^{mn}$ to \cref{mn1} and using \cref{radgauge,zmodeuruAAB} yields 
\begin{equation}
\label{zmodeTF}
    (D-5)\bm{D}^{4}\Psi=0 ,
\end{equation}
which, by \Cref{propsymm}, implies that the trace-free scalar part of $\Phi_{mn}$ vanishes.\footnote{\Cref{zmodeTF} looks unconstrained for $D=5$ but that case is very special since the internal space is $S^1$ and the only term in~\C{varphimndecomp} is proportional to $\phi$.} Using our gauge conditions, harmonicity of the spacetime components $h_{\mu \nu}^{(1)}$ and that \cref{zmodeTF} implies $\bm{D}^{m}\varphi^{(1)}_{mn}=0$, the $(um;1)$ and $(Am;1)$ components of Einstein's equation imply that $A_{um}^{(1)}$ and $A_{Am}^{(1)}$ are harmonic with decomposition
\begin{equation}
    A_{um}^{(1)}=\sum_{i=1}^{b_{1}}A_{u}^{(1;i)}(u,\theta)\otimes \ov{V}_{m}^{(i)}(y^{m}) \, \textrm{ and }\, A_{Am}^{(1)}=\sum_{i=1}^{b_{1}}A_{A}^{(1;i)}(u,\theta)\otimes \ov{V}_{m}^{(i)}(y^{m}), 
\end{equation}
where $\ov{V}_{m}^{(i)}$ are a basis for harmonic one-forms on $\I$. Finally, \cref{radgauge,zmodeuruAAB,zmodeTF} imply that 
\begin{equation}
\label{zmodePhimn}
    \bm{D}^{2}\Phi_{mn}+2\IRiem_{m}{}^{p}{}_{n}{}^{q}\Phi_{pq}=0 \quad \textrm{ $\implies$ }\quad \Phi_{mn}=\sum_{i=1}^{\dims}\Phi^{(i)}(u,\theta)\ov{t}_{mn}^{(i)}(y^{m}) ,
\end{equation}
where $\ov{t}_{mn}^{(i)}$ are the $\dims$ trace-free, divergence-free, unobstructed  deformations of $\I$. Finally \cref{zmodeuu} implies that the sum $q^{AB}\ov{h}_{AB}+\phi$ can have, at most, linear-dependence on retarded time $u$. Einstein's equations at order ${1\over r^{2}}$, however, place a {\em stronger} constraint on the time-dependence of this quantity. In particular, a direct calculation of $q^{AB}$ applied to the zero mode of the trace-reversed Einstein equations implies that 
\begin{equation}\label{trace_constraint}
    \partial_{u}(q^{AB}\ov{h}_{AB}^{(1)}+\phi)=0 .
\end{equation}
We summarize our findings on the asymptotic behavior of the metric in the following lemma: 
\begin{lem}
\label{lemrad}
Let $(\M,g)$ be an isolated system in a gauge which satisfies our ansatz \cref{ansatz}. There exists a unique diffeomorphism which preserves our ansatz such that the leading order expansion coefficients of the metric have the following properties:
\begin{enumerate}
    \item The $\B$ metric components are harmonic on $\I$ and therefore satisfy 
    \begin{equation}
         h_{uu}^{(1)}=\ov{h}^{(1)}_{uu}(u,\theta), \quad h^{(1)}_{ur}=\ov{h}^{(1)}_{ur}(u,\theta), \quad  h^{(1)}_{uA}=\ov{h}^{(1)}_{uA}(u,\theta), \quad h^{(1)}_{AB}=\ov{h}^{(1)}_{AB}(u,\theta), 
    \end{equation}
     and the $h_{rr}^{(1)}$, $h_{rA}^{(1)}$ components vanish. 
    
    \item The components $A_{um}^{(1)}$ and $A_{Am}^{(1)}$ admit the decomposition 
    \begin{equation}
        A_{um}^{(1)}=\sum_{i=1}^{b_{1}}A_{u}^{(1;i)}(u,\theta)\otimes \ov{V}_{m}^{(i)}(y^{m}), \quad A_{Am}^{(1)}=\sum_{i=1}^{b_{1}}A_{A}^{(1;i)}(u,\theta)\otimes \ov{V}_{m}^{(i)}(y^{m}),
    \end{equation}
    and $A_{rm}^{(1)}$ vanishes. The $\ov{V}_{m}^{(i)}$ are a complete basis of $b_{1}$ linearly independent Killing vectors of $\I$ where $b_{1}$ is the first Betti number of $\I$. 
    
    \item The components $\varphi_{mn}^{(1)}$ satisfy 
    \begin{equation}
    \label{scexp}
        \varphi_{mn}^{(1)}=\frac{\Imet_{mn}}{D-4}\phi(u,\theta)+\sum_{i=1}^{\dims}\Phi^{(i)}(u,\theta)\ov{t}_{mn}^{(i)}(y^{m}),
    \end{equation}
    where $\phi\equiv \zm{\Imet^{mn}\varphi_{mn}^{(1)}}$ and the $\ov{t}_{mn}^{(i)}$ are a complete basis of $\dims$ symmetric, rank $2$ tensor fields which satisfy $\bm{D}^{m}\ov{t}_{mn}^{(i)}=0$, $\Imet^{mn}\ov{t}_{mn}^{(i)}=0$ and \cref{Tmn}. Furthermore, the metric satisfies $\partial_{u}(q^{AB}h_{AB}^{(1)}+\phi)=0$. 
\end{enumerate}
\end{lem}

Without loss of generality, we will assume this gauge in the remainder of this work. This gauge choice dramatically simplifies the analysis of the higher-dimensional Einstein equations by gauging away higher harmonics in the internal space. We note that any metric which admits an asymptotic expansion \C{ansatz}, and which satisfies the Einstein equations, can be put into this gauge. In this sense, our gauge choice is not an additional assumption but actually a consequence of the fall-off conditions and equations of motion.

In this gauge the news tensor, defined in~\C{news}, is very nicely related to the leading order metric by:
\begin{equation}\label{nicenews}
    \news_{ab}= \partial_{u}h_{ab}^{(1)}. 
\end{equation}
This expression for the news tensor identifies the gauge-invariant radiative degrees of freedom of the leading order metric, and  manifestly satisfies the relations spelled out in~\C{news_conditions}. 

\subsection{Asymptotic symmetries of compactified spacetimes}
\label{subsec:asympsymm}

In this section we investigate the asymptotic symmetries of spacetimes with compact extra dimensions. Before doing so, it will be convenient to further refine the gauge choice of \Cref{lemrad}. Note that the trace $q^{AB} h_{AB}^{(1)}$ is constrained by \cref{trace_constraint} so that $q^{AB} h_{AB}^{(1)}(u,\theta) = -\phi(u,\theta) + c ( \theta)$. We now show that there exists a residual gauge transformation, compatible with \Cref{lemrad}, which allows us to set $c = 0$. Performing a diffeomorphism parameterized by $\xi_M = c(\theta) K_M$, where $K_M$ is defined in \cref{K}, we see that the metric changes by
\begin{align}\label{residual_shift}
    h_{AB}^{(1)} \to h_{AB}^{(1)} + 2 c ( \theta ) q_{AB} \, , \quad  h_{uA}^{(1)} \to h_{uA}^{(1)} + \Ds_{A}c(\theta) \, ,
\end{align}
where $\Ds_A$ is the covariant derivative compatible with $q_{AB}$, defined in \cref{sec:conventions}. The shift in $h_{uA}^{(1)}$ does not affect the gauge fixed in \Cref{lemrad}, while the change in $h_{AB}$ is exactly of the form needed to eliminate $c ( \theta )$. Fixing this gauge, we may now assume that $c ( \theta ) = 0$ and therefore $q^{AB} h_{AB}^{(1)}$ has no further diffeomorphism freedom.

For an arbitrary dynamical spacetime the metric will not, generically, have any exact symmetries. However for given asymptotics, the spacetime will admit an asymptotic symmetry group. We define this group as the group of diffeomorphisms which preserve the gauge conditions in \Cref{lemrad} along with $q^{AB}h_{AB}^{(1)}=-\phi$. Since in this gauge, the metric decomposes into spin-2, spin-1 and spin-0 degrees of freedom there is a corresponding decomposition of the asymptotic symmetry group. The upshot of this is that we can consider the asymptotic symmetries of spin-2, spin-1 and spin-0 degrees of freedom separately.

To find the symmetry group of the spin-2 modes, we note that the $\B$ components of the leading order metric $h_{\mu \nu}^{(1)}$ are effectively in a Bondi-type gauge. The original Bondi gauge conditions on the leading order metric are $h_{rr}^{(1)} = h_{rA}^{(1)}=q^{AB}h_{AB}^{(1)}=0$. It then follows from Bondi's original analysis that the symmetry group that preserves these gauge conditions is the BMS group $\mathfrak{B}$ which we shall review shortly. We note that our gauge conditions also imply $h_{rr}^{(1)}=h_{rA}^{(1)}=0$. Additionally, we imposed $q^{AB}h_{AB}^{(1)}=-\phi$. Since $\phi$ has no residual gauge freedom this fixes $q^{AB}h_{AB}^{(1)}$. Therefore, the asymptotic symmetry group of the spin-2 degrees of freedom is the BMS group $\mathfrak{B}$. 

At this point as promised, we should recall some properties of the BMS group. The Lie algebra $(\mathfrak{bms})$ 
of $\mathfrak{B}$ contains an infinite-dimensional normal Lie subalgebra $\mathfrak{t}$, which contains the supertranslations. Explicitly, the elements of $\mathfrak{t}$ are
\begin{equation}
\label{supertrans}
\xi^{M}=-T(\theta)\bigg(\frac{\partial}{\partial u}\bigg)^{M}-\frac{1}{2}\Ds^{2}T(\theta)\bigg(\frac{\partial}{\partial r}\bigg)^{M}+\frac{1}{r}q^{AB}\Ds_{B}T(\theta)\bigg(\frac{\partial}{\partial \theta^{A}}\bigg)^{M}+\ldots  ,
\end{equation}
where the ``$\dots$'' denotes vector fields that vanish as $r\rightarrow \infty$ at fixed $u$, $\theta^{A}$ and $y^{m}$. The function $T(\theta)$ is smooth on the asymptotic $2$-sphere. If $T(\theta)$ is an $\ell=0$ spherical harmonic then \cref{supertrans} is an asymptotic time translation. If $T(\theta)$ is a linear combination of $\ell=1$ spherical harmonics then \cref{supertrans} is an asymptotic spatial translation. If $T(\theta)$ is orthogonal to the $\ell=0,1$ spherical harmonics then \C{supertrans} is called a {\em supertranslation} and, asymptotically, corresponds to the action of an infinitesimal, angle-dependent time translation. The quotient $\mathfrak{bms}/\mathfrak{t}=\mathfrak{so}(3,1)$ is the Lorentz Lie algebra, which correspond to conformal Killing vectors of $S^{2}$. At the level of group structure, the BMS group $(\mathfrak{B})$ is therefore the semi-direct product of the restricted Lorentz group $(\mathcal{L})$ and the infinite-dimensional supertranslation group $(\mathcal{T})$:
\begin{equation}
 \mathfrak{B}=\mathcal{L}\ltimes \mathcal{T}.
\end{equation}

We now turn to the spin-1 degrees of freedom. The diffeomorphisms that act on $A_{\mu m}^{(1)}$ and preserve our metric asymptotics~\C{ansatz} are generated by $\xi_{m}^{(0)}(\theta)$, which cannot depend on $u$. To preserve \Cref{lemrad}, $\xi_{m}^{(0)}$ must be harmonic on $\I$. Any such $\xi_{m}^{(0)}$ is a smooth function $S(\theta)$ multiplied by a Killing vector $\zm{V}^{m}(y)$ on $\I$,
\be\label{spin_one_vector_field}
\xi^{M}=S(\theta)\zm{V}^{m}(y)\bigg(\frac{\partial}{\partial y^{m}}\bigg)^{M} + \ldots,
\ee 
where the omitted terms again vanish as $r\rightarrow \infty$. There are $b_1$ Killing vectors on $\I$. In the limit as $r\rightarrow \infty$, the commutator of any two $\xi^M$ of the form (\ref{spin_one_vector_field}) vanishes so the asymptotic symmetry group generated by these vector fields is abelian. Let us denote this group of angle-dependent internal isometries by $\mathfrak{C}$. We note that elements of this group do not commute with Lorentz transformations in $\mathcal{L}$.

The remaining degrees of freedom are the spin-0 modes of~\C{varphimndecomp} given by the tensor modes $\Phi_{mn}$ describing the volume-preserving moduli, and the scalar mode $\phi$ which is the volume mode. There is no choice of asymptotic vector field which preserves our asymptotic conditions and the gauge conditions given in \Cref{lemrad} that can affect either $\Phi_{mn}$ or $\phi$. The only asymptotic diffeomorphism that can affect $\varphi^{(1)}_{mn}$ is of the form ${\xi_{m}^{(1)}\over r} + \ldots,$ but all of this gauge freedom has already been used to implement the gauge of \Cref{lemrad}. Thus there is no remaining diffeomorphism freedom for these modes.

Therefore,  the enlarged asymptotic symmetry group $(\mathfrak{G})$ is the semi-direct product of $\mathfrak{B}$ with the abelian group $\mathfrak{C}$:
\begin{equation}
 \mathfrak{G}=\mathfrak{B}\ltimes \mathfrak{C} .
\end{equation}
We note that this asymptotic symmetry group is identical to the asymptotic symmetry group of asymptotically flat Einstein-Maxwell-scalar theory where $\mathfrak{C}$ is replaced with the asymptotic symmetries of the electromagnetic field \cite{Bonga:2019bim}. Therefore, $\mathfrak{C}$ has the natural interpretation as the asymptotic symmetry group of the graviphotons. 

\begin{figure}[H]
\centering
\includegraphics[width=.9\textwidth]{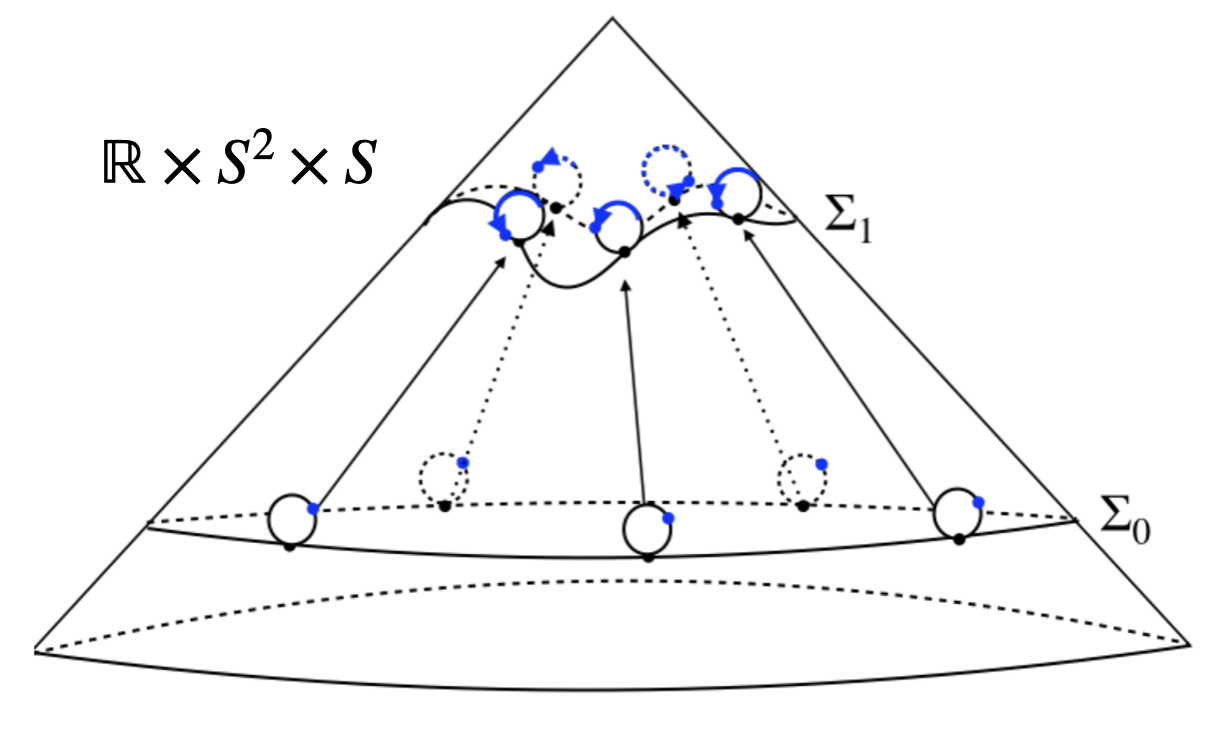}
\caption{The figure illustrates the action of a supertranslation and an angle-dependent internal isometry on the asymptotic sphere. We chose $\I = S^1$ for simplicity. Null infinity is an incoming null surface with 
topology $\mathbb{R}\times\Sp^{2}\times S^1$ whose cross sections are asymptotically large spheres. A point in $\mathbb{R}\times S^{2}$ (highlighted in black) and a point on $\I=S^1$, where the $S^1$ is represented by a circle, specifies a point on null infinity. At leading order in ${1\over r}$ supertranslations only act on $\B$ while angle-dependent internal isometries act only on $\I$. Given a constant $u$ cut of null infinity, labeled $\Sigma_{0}$, a supertranslation acts by $u\rightarrow u+T(\theta)$ and an angle-dependent internal isometry acts by $y \rightarrow y + S(\theta)$. 
The composition of these group actions takes the cut $\Sigma_{0}$ into the cut $\Sigma_{1}$.}
\label{superintisom}
\end{figure}

Finally we will give the action of elements of $\mathfrak{G}$ on $\scrip$, which has the topology of $\mathbb{R} \times S^2 \times \I$. An element of this asymptotic symmetry group moves a point $(u,\theta,y)$ to $(\tilde{u},\tilde{\theta}, \tilde{y})$ as
\begin{align}
\tilde{u}&=\omega (\theta)[u+T(\theta)] \, , \\
\tilde{\theta}^{A}&=\sigma(\theta) \, , \\
\tilde{y}^{m}&=\rho(y,\theta) \, ,
\end{align}
where $\sigma:S^{2}\rightarrow S^{2}$ acts by a conformal isometry of the $2$-sphere given by $\sigma^{\ast}q_{AB} =\omega^2 q_{AB}$. Similarly, at each fixed angle, the map $\rho(\cdot,\theta):\I\rightarrow \I$ acts as an isometry of the internal space: $\rho^\ast \Imet_{mn} = \Imet_{mn}$. An illustration of the combined action of a supertranslation with an angle-dependent internal isometry is given in figure \ref{superintisom}. Finally we note that, in terms of the leading order metric $h_{MN}^{(1)}$, the infinitesimal action  of the composition of a supertranslation and an angle-dependent internal isometry is 
\begin{align}
&h_{AB}^{(1)}(u,\theta,y)\rightarrow h_{AB}^{(1)}(u,\theta,y) +T(\theta)N_{AB}(u,\theta)+ \bigg(\Ds_{A}\Ds_{B} - \frac{1}{2}q_{AB}\Ds^{2}\bigg)T(\theta) ,  \label{supertranshAB}\\ &A_{Am}^{(1)}(u,\theta,y)\rightarrow A_{Am}^{(1)}(u,\theta,y)+ \Ds_{A}S(\theta)\otimes \ov{V}_{m}(y).\label{superintAAm}
\end{align}
So the composition of a supertranslation and an angle-dependent isometry only affects the zero-modes of the leading order metric.

\section{Bursts of Radiation}
\par Building on our discussion of the radiative degrees of freedom and the corresponding asymptotic symmetries in section~\ref{sec:asymptotic_behavior}, we now examine the response of the asymptotic spacetime metric to a burst of radiation. We study the metric near $\scrip$ by analyzing Einstein's equation in a ${1\over r}$ expansion. We consider spacetimes which are stationary at early times, undergo a period where there is a significant amount of gravitational radiation for a finite range of retarded time, and then approach stationarity at asymptotically late times. It was pointed out in \cite{Satishchandran:2019pyc}, at early or late times, that the metric corresponding to a collection of inertially moving {\em massive} bodies is stationary at order ${1\over r}$, but will generically be non-stationary at higher orders in ${1\over r}$. In particular, it was shown quite generally, that the behavior of the $\ell$-th multipole moment for the metric of a static compact object at some time $t=u+r$ behaves as 
\begin{equation}
    h_{MN}\sim \frac{(u+r)^{\ell }}{r^{\ell+1}}\sim \frac{1}{r} + \frac{\ell u }{r^{2}} +\dots 
\end{equation}
near $\scrip$ where $g_{MN}=\eta_{MN}+h_{MN}$ and the behavior in the internal space has been suppressed. Therefore a generic, boosted compact object will be stationary at leading order in $1/r$ but will generically be non-stationary at subleading orders in $1/r$. This non-stationarity for $\ell=1$ can be removed by boosting to the center of mass frame where the matter is at rest. However, $h_{MN}$ is generically non-stationary at sub-leading orders in $1/r$ if one has incoming or outgoing compact objects at early or late times.

However, for simplicity, we will investigate null memory effects caused entirely by the flux and scattering of incoming and outgoing gravitational radiation, and no ordinary memory. To impose this condition we assume the {\em stronger} stationarity conditions of \cite{Satishchandran:2019pyc}. Specifically we assume there exists a gauge in which the metric satisfies the following stationarity conditions at asymptotically early and late times:
\begin{equation}
\label{statmetric}
\partial_{u}h_{MN}^{(n)}\rightarrow 0 \quad \textrm{ as $u\rightarrow \pm\infty$} \quad  \textrm{ for all $n\geq 1$} .
\end{equation}
We will further require that the stress energy vanish in a neighborhood of null infinity at early and late times at the following orders: 
\begin{equation}
\label{statstress}
T_{MN}^{(n)}\rightarrow 0 \quad \textrm{ as $u\rightarrow \pm\infty$ \quad for all $n\leq 3$.} 
\end{equation}
This is not terribly restrictive: the condition includes all stress-energy with compact support and most isolated systems studied in the literature. 

This section is laid out as follows: in \cref{statera} we examine the constraints from Einstein's equation on the metric in the stationary eras. In \cref{subsec:changeinmet}, we use our results from \cref{radera,statera} to integrate Einstein's equations to obtain gauge invariant information about the change in the metric between the stationary eras caused by the passage of gravitational radiation to $\mathscr{I}^{+}$. As we shall see, certain components of the change in the metric correspond precisely to the composition of a supertranslation with an angle-dependent isometry.

\subsection{Stationary eras }
\label{statera}
We first investigate the behavior of the metric in a stationary era. Our stationarity conditions turn out to imply constraints on the angular behavior of the metric at leading order in ${1\over r}$. It is useful to note that \Cref{hodge} applies to any closed Riemannian manifold and \Cref{propsymm} applies to any compact Riemannian Einstein space, and therefore they both apply to the $2$-sphere equipped with the round metric $q_{AB}$. 

\begin{rem}
\label{remsphere}
\Cref{hodge,propsymm} apply to any compact Riemannian manifold. For example with the round metric $q_{AB}$  on the $2$-sphere then $(\Sp^{2},q_{AB})$ is a compact Riemannian Einstein space with $c=1$. Therefore, \Cref{hodge,propsymm} apply to both a one form $V_{A}$ and a second rank, symmetric tensor field $T_{AB}$ on $\Sp^{2}$. Therefore, $V_{A}$ and $T_{AB}$ can be both be decomposed uniquely as in \cref{Vmdecomp,Tmndecomp} where the covariant derivative is now the derivative operator $\Ds_{A}$ compatible with metric $q_{AB}$. There is no `tensor part' since there are no divergence-free, trace-free tensors on $\Sp^{2}$.  Furthermore, any divergence free vector $v_{A}$ on $\Sp^{2}$ can be written as the `curl' of a scalar function $P$, i.e.,  $v_{A}=\epsilon_{A}{}^{B}\Ds_{B}P$. This is sometimes called the `magnetic parity' or `parity odd' part of the vector. Finally, any rotationally invariant operator (such as $\Ds^{2}\equiv q^{AB}\Ds_{A}\Ds_{B}$) acting on a one-form or a symmetric tensor preserves this decomposition. 
\end{rem}
\par Given \Cref{remsphere}, we now determine the metric constraints from Einstein's equations in a stationary era. We adopt the gauge described in \Cref{lemrad}. The analysis of Einstein's equations in a stationary era is greatly simplified by further fixing the gauge of the metric at $O({1\over r^{2}})$. In \Cref{EinsEqn}, we prove that one can put the metric in a gauge compatible with the stationarity conditions \C{statmetric} and \C{statstress} and the gauge of \Cref{lemrad} so that Einstein's equations imply that
\begin{align}
\label{statgauge1}
    h^{(2)}_{\mu \nu}=\ov{h}^{(2)}_{\mu \nu}(\theta), \quad A^{(2)}_{\mu m}=\sum_{i=1}^{b_{1}}A_{\mu}^{(2;i)}(\theta)\otimes \ov{V}_{m}^{(i)}(y^{m}), 
\end{align}
and
\begin{equation}
\label{statgauge2}
    \varphi_{mn}^{(2)}=\Phi^{(2)}_{mn}(\theta,y)+\bigg(\bm{D}_{m}\bm{D}_{n} - \frac{\Imet_{mn}}{D-4}\bm{D}^{2}\bigg)\Psi^{(2)}(\theta,y^{m})+\frac{\Imet_{mn}}{D-4}\ov{\phi}^{(2)}(\theta).
\end{equation}
Aside from special cases like $\I=\mathbb{T}^{k}$, neither $\Psi^{(2)}$ nor $\Phi_{mn}^{(2)}$ are zero modes on $\I$. 

We now analyze Einstein's equations in a stationary era in the gauge of \Cref{lemrad} with the constraints~\C{statgauge1} and~\C{statgauge2} imposed. 
The zero mode of Einstein's equations at order ${1\over r^{3}}$, after a lengthy calculation described in \Cref{EinsEqn}, yields 
\begin{align}
     \textrm{$(uu;3)$ }&\quad \quad \quad \Ds^{2}h_{uu}^{(1)}=0, \label{uu3}\\
     \textrm{$(ur;3)$ }&\quad \quad \quad \Ds^{2}h_{ur}^{(1)}=0, \label{ur3}\\
      \textrm{$(uA;3)$ }&\quad \quad \quad [\Ds^{2}-1]h_{uA}^{(1)}-\Ds_{A}\Ds^{B}h_{uB}^{(1)}-\Ds_{A}(h_{uu}^{(1)}-h_{ur}^{(1)})=0, \label{uA3}\\
      \textrm{$(rr;3)$ }& \quad\quad \quad  \phi-2h_{ur}^{(1)}=0, \label{rr3}\\
      \textrm{$(rA;3)$ }&\quad \quad \quad \Ds_{A}h_{ur}^{(1)}-\Ds_{A}\phi^{(1)}=0,\label{rA3}\\
      \textrm{$(AB;3)$ }&\quad \quad \quad [\Ds^{2}-2]h_{AB}^{(1)}-2\Ds_{(A}\Ds^{C}h_{B)C}^{(1)}+2\Ds^{C}h_{Cu}^{(1)}q_{AB}+\Ds_{A}\Ds_{B}q^{CD}h_{CD}^{(1)} \nonumber \\
      \quad &\quad \quad\quad +q_{AB}q^{CD}h_{CD}^{(1)}+[\Ds_{A}\Ds_{B}-q_{AB}](\phi-2h_{ur}^{(1)})=0, \label{AB3} \\
      \textrm{$(um;3)$ }&\quad \quad \quad \Ds^{2}A_{u}^{(1;i)}=0, \label{um3} \\
       \textrm{$(Am;3)$ }&\quad \quad \quad [\Ds^{2}-1]A_{A}^{(1;i)}+\Ds_{A}A_{u}^{(1;i)}=0,\label{Am3} \\
      \textrm{$(mn;3)$ }&\quad \quad \quad \Ds^{2}\phi=0 \textrm{ and }\Ds^{2}\Phi_{mn}^{(i)}=0, \label{mn3}
\end{align}
where the coefficients $A_{u}^{(1;i)},A_{A}^{(1;i)}$ and $\Phi_{mn}^{(i)}$ are defined in \Cref{lemrad}. In \cref{mn3}, the $\Phi^{(i)}$ are the $\dims$ exactly massless modes as discussed in \cref{subsec:linscalar}. Additionally, the $(rm;3)$ components of Einstein's equations vanish. \Cref{uu3,ur3,rr3,um1} imply that $h_{uu}^{(1)},h_{ur}^{(1)},\phi,\Phi_{mn}^{(i)}$ and $A_{u}^{(1;i)}$ are spherically symmetric and 
\begin{equation}
\label{phihurstat}
    \phi=2h_{ur}^{(1)}. 
\end{equation}
Consequently, the left hand side of \cref{rA3} vanishes. Using \Cref{propsymm} and \Cref{remsphere}, one can write
\begin{equation}
    A_{A}^{(1;i)}(\theta) = \Ds_{A}S^{(i)}(\theta)+\epsilon_{A}{}^{B}\Ds_{B}R^{(i)}(\theta),
\end{equation}
\begin{equation}
    h_{uA}^{(1)}(\theta) = \Ds_{A}P(\theta)+\epsilon_{A}{}^{B}\Ds_{B}F(\theta),
\end{equation}
and 
\begin{equation}
\label{hABdecomp}
    h_{AB}^{(1)}(\theta)=\epsilon_{(A}{}^{C}\Ds_{B)}\Ds_{C}W(\theta)+\bigg(\Ds_{A}\Ds_{B}-\frac{q_{AB}}{2}\Ds^{2}\bigg)T(\theta)+\frac{q_{AB}}{2}U(\theta).
\end{equation}
Applying $\epsilon^{CA}\Ds_{C}$ to \cref{uA3,rA3,Am3} yields 
\begin{equation}
  \Ds^{2}R^{(i)}(\theta)=0, \quad  \Ds^{2}F(\theta)=0 \quad \textrm{ and } \quad (\Ds^{2}+2)\Ds^{2}W(\theta)=0,
\end{equation}
and therefore the magnetic parity parts of $A_{A}^{(1;i)}$, $h_{uA}^{(1)}$ and $h_{AB}^{(1)}$ vanish.\footnote{The operator $(\Ds^{2}+2)\Ds^{2}$ annihilates the $\ell=0,1$ spherical harmonics. Let $\tilde{W}$ be the projection of $W$ into the subspace spanned by $\ell =0,1$ spherical harmonics. That $\tilde{W}$ is annihilated by the operator in \cref{hABdecomp} (i.e. $\epsilon_{(A}{}^{C}\Ds_{B)}\Ds_{C}\tilde{W}=0$) follows from the fact that any function that is a linear combination of $\ell=0,1$ spherical harmonics satisfies $\Ds_{A}\Ds_{B}\tilde{W}=-q_{AB}\tilde{W}$.} Applying $q^{AB}$ to \cref{AB3} yields a relation between $U(\theta),T(\theta)$ and $P(\theta)$:
\begin{equation}
    \Ds^{2}U(\theta)-\Ds^{2}(\Ds^{2}+2)T(\theta)+4\Ds^{2}P(\theta)=0.
\end{equation}

\noindent
We summarize the above results in the following lemma: 
\begin{lem}
\label{lemstat}
Let $(M,g)$ be an isolated system that satisfies both our ansatz~\C{ansatz} in a gauge compatible with \Cref{lemrad} and our stationarity conditions. There exists a unique diffeomorphism which preserves these gauge and stationarity conditions such that the leading order expansion coefficients satisfy the following relations:
\begin{enumerate}
    \item The $\B$ metric components satisfy:
    \begin{align}
       &  h^{(1)}_{uu}=c_{1}, \quad h^{(1)}_{ur}=c_{2}, \quad h^{(1)}_{uA}=\Ds_{A}P(\theta), \\  & h^{(1)}_{AB}=\bigg(\Ds_{A}\Ds_{B}-\frac{q_{AB}}{2}\Ds^{2}\bigg)T(\theta)+\frac{q_{AB}}{2}U(\theta), 
    \end{align}
    and $h_{rr}^{(1)}=0=h_{rA}^{(1)}$. Here $c_{1}$ and $c_{2}$ are constants, the functions $P(\theta), T(\theta)$ and $U(\theta)$ are smooth functions on $\Sp^{2}$ and are related by, 
    \begin{equation}
            \Ds^{2}U(\theta)-\Ds^{2}(\Ds^{2}+2)T(\theta)+4\Ds^{2}P(\theta)=0.
    \end{equation}
    \item The $A_{\mu m}^{(1)}$ components satisfy: 
    \begin{equation}
        A_{um}^{(1)}=\sum_{i=1}^{b_{1}}Q^{(i)}~\ov{V}_{m}^{(i)}(y^{m}), \quad A_{Am}^{(1)}=\sum_{i=1}^{b_{1}}\Ds_{A}S^{(i)}(\theta)\otimes \ov{V}_{m}^{(i)}(y^{m})
    \end{equation}
    and $A_{rm}^{(1)}=0$. The $Q^{(i)}$ are constants and the functions $S^{(i)}(\theta)$ are smooth functions on $\Sp^{2}$. 
    \item The internal space components satisfy:
    \begin{equation}
        \varphi_{mn}^{(1)}=\frac{\Imet_{mn}}{D-4}\phi +\sum_{i=1}^{\dims}\Phi^{(i)}\, t_{mn}^{(i)}(y^{m}),
    \end{equation}
    where $2c_{2}=\phi$ and the coefficients $\Phi^{(i)}$ are constants. 
\end{enumerate}
\end{lem}
\noindent
This discussion captures the leading order behavior of the metric near $\scrip$ for  stationary objects in the bulk; for example, stars or black holes with possible scalar hair. 

\subsection{Change in the metric coefficients after the burst of radiation} \label{subsec:changeinmet}

Now that we have determined the radiative degrees of freedom in \Cref{lemrad}, and the metric component constraints 
from the requirement of stationarity at asymptotically early and late times in \Cref{lemstat},  we now integrate the leading order Einstein equations to prove the following theorem: 
\begin{thm}
\label{changemetthm}
Let $(\M,g)$ be an isolated system which satisfies our ansatz and stationarity conditions. Let $g_{MN}$ be in the gauge described by \Cref{lemrad,lemstat} and satisfy Einstein's equation with stress energy $T_{MN}$ satisfying \cref{statstress} and the dominant energy condition. 
\begin{enumerate}
    \item The change in the metric coefficient $h_{AB}^{(1)}$ is
    \begin{equation}
        \Delta h^{(1)}_{AB}(\theta)=\bigg(\Ds_{A}\Ds_{B}-\frac{1}{2}q_{AB}\Ds^{2}\bigg)T(\theta)-\frac{1}{2}q_{AB}\Delta \phi  ,
    \end{equation}
    where $\Delta \phi = \Delta \left( \Imet^{mn}\varphi_{mn}^{(1)} \right)$ is a constant; specifically, it cannot be a function of $\theta$. The function $T(\theta)$ is a smooth function on $\Sp^{2}$ determining an asymptotic supertranslation (eq.~\ref{supertranshAB}) which satisfies 
    \begin{equation}
        \Ds^{2}(\Ds^{2}+2)T(\theta)=4\Delta h_{uu}^{(1)}-2\Delta \phi -16\pi \mathcal{F}(\theta) ,
    \end{equation}
    where $\Delta h_{uu}^{(1)}$ is a constant, $\mathcal{F}(\theta)\leq 0$ is 
    \begin{equation}
    \label{nullfluxuu}
        \hspace{-15pt}\mathcal{F}(\theta) = -\frac{1}{ \textrm{{\em Vol}}(\I)}\int_{\mathbb{R}\times \I}du \, d\mu_{\I} \, \bigg(T_{uu}^{(2)}(u,\theta,y)+\frac{1}{32\pi}\mathcal{N}^{ab}\mathcal{N}_{ab}(u,\theta,y)\bigg)
    \end{equation}
    and $d\mu_{\I}$ is the volume measure of $(\Imet_{mn},\I)$. 
    
    \item The change in the metric coefficient $A_{Am}^{(1)}$ is 
    \begin{equation}
    \label{DeltaAAm}
        \Delta A_{Am}^{(1)}(\theta,y^{m})=\sum_{i=1}^{b_{1}}\Ds_{A}S^{(i)}(\theta)\otimes \ov{V}_{m}^{(i)}(y^{m}) 
    \end{equation}
    where $\ov{V}_{m}^{(i)}$ are a basis of $b_{1}$ harmonic one-forms on $\I$ and the coefficients $S^{(i)}$ are a set of smooth functions of $\Sp^{2}$ which are parameters of an asymptotic internal isometry and satisfy 
    \begin{equation}
        \Ds^{2}S^{(i)}(\theta)=\Delta Q^{(i)} + 16\pi \mathcal{J}^{(i)}(\theta) ,
    \end{equation}
    where the $Q^{(i)}$ are constants and
    \begin{align}
    \label{nullfluxum}
        &\mathcal{J}^{(i)}(\theta) = \frac{1}{\textrm{{\em Vol}}(\I)}\int_{\mathbb{R}\times \I}du d\mu_{\I}\, T^{(2)}_{um}(u,\theta,y^{m})\ov{V}^{(i)m}(y^{m}).
    \end{align}
    
    \item The change in the metric coefficient $\varphi_{mn}^{(1)}$ is 
    \begin{equation}
    \label{Deltaphimn}
        \Delta \varphi_{mn}^{(1)}(y^{m})=\frac{\Imet_{mn}}{d-4}\Delta \phi + \sum_{i=1}^{\dims}\Delta \Phi^{(i)}~\ov{t}_{mn}^{(i)}(y^{m}) ,
    \end{equation}
    where $\Delta \phi$ and $\Delta \Phi^{(i)}$ are constants, and the $\ov{t}_{mn}^{(i)}$ are a basis of $\dims$ symmetric, divergence free two tensors on $\I$ which satisfy the Lichnerowicz equation. 
\end{enumerate}
\begin{proof}
We assume that the metric $g_{MN}$ is in a gauge compatible with \Cref{lemrad,lemstat}. The `zero mode' of the $(\mu \nu ; 2)$  components Einstein's equation at order $1\over r^{2}$ (see \Cref{asymptotics}), yields 
\begin{align}
    \textrm{$(uu;2)$ }&\quad \quad \quad \partial_{u}\Ds^{A}h_{Au}^{(1)}+\partial_{u}h_{ur}^{(1)}-\partial_{u}h_{uu}^{(1)}=8\pi \ov{T}_{uu}^{(2)}+\frac{1}{4}\zm{\mathcal{N}^{ab}\mathcal{N}_{ab}}-\frac{1}{2}\partial_{u}\bigg(h^{(1)AB}N_{AB}\label{uu2}\\
    &\quad \quad \quad+2A^{(1)Am}N_{Am}+\zm{\varphi^{(1)mn}N_{mn}}-\partial_{u}\zm{h}_{rr}^{(2)}-\partial_{u}q^{AB}\zm{h}^{(2)}_{AB}-\partial_{u}\zm{\Imet^{mn}\varphi_{mn}^{(2)}}\bigg) \nonumber \\
    \textrm{$(ur;2)$ }&\quad \quad \quad \partial_{u}\phi-2\partial_{u}h_{ur}^{(1)}=\partial_{u}^{2}\zm{h}_{rr}^{(2)} \label{ur2} \\
   \textrm{$(uA;2)$ } &\quad \quad \quad \partial_{u}\Ds^{B}h_{BA}^{(1)}-2\partial_{u}h_{uA}^{(1)}+\partial_{u}\Ds_{A}h_{ur}^{(1)}=\partial_{u}^{2}\ov{h}_{rA}^{(2)} \label{uA2}
\end{align}
and the $(rr;2)$, $(rA;2)$ and $(AB;2)$ components of Einstein's equation vanishes. Integrating \cref{ur2} together with our stationarity conditions \cref{statmetric} implies that 
\begin{equation}
\label{Dphihur}
    \Delta \phi=2\Delta h_{ur}^{(1)}
\end{equation}
which agrees with \cref{phihurstat}. \Cref{lemstat} implies that $\Delta \phi$ is spherically symmetric. Furthermore we note that, by \Cref{lemrad} 
\begin{equation}
    \partial_{u}(q^{AB}h_{AB}^{(1)}+\phi)=0 \implies \Delta \phi=-\Delta U
\end{equation}
where $U=q^{AB}h^{(1)}_{AB}$ in the stationary eras. Combining \cref{uu2,uA2} yields 
\begin{equation}
    \partial_{u}\Ds^{A}\Ds^{B}h_{AB}^{(1)}=2\partial_{u}h_{uu}^{(1)}-(\Ds^{2}+2)\partial_{u}h_{ur}^{(1)}+16\pi \zm{T}_{uu}^{(2)}+\frac{1}{2}\zm{\mathcal{N}^{ab}\mathcal{N}_{ab}}-\partial_{u}C_{1} ,
\end{equation}
where $C_{1}$ denotes a collection of terms which vanish in the stationary eras. Integrating with respect to retarded time, using \cref{statmetric} and using the decomposition of $h_{AB}^{(1)}$ in the stationary eras given by \Cref{lemstat} yields
\begin{equation}
    \mathscr{D}^{2}(\mathscr{D}^{2}+2)\Delta T(\theta)=4\Delta h_{uu}^{(1)}-2\Delta \phi^{(1)}-16\pi \mathcal{F}(\theta)
\end{equation}
where $\mathcal{F}$ is the total flux of stress energy and news squared to null infinity given by \cref{nullfluxuu}. That $\mathcal{F}\leq 0$ follows from the positivity of $T_{uu}^{(2)}$ due to the dominant energy condition and the positivity of $\mathcal{N}^{ab}\mathcal{N}_{ab}$.  

\par The zero mode of the $(\mu m;2)$ components of Einstein's equation at order $1\over r^{2}$ can be extracted by taking the zero mode of the $(\mu m;2)$ components contracted with the orthonormal basis vectors $\Imet^{mn}\ov{V}_{n}^{(i)}$ on $\I$. The $(rm;2)$ and $(Am;2)$ components of Einsteins equation vanish and the zero mode of the $(um;2)$ components yield 
\begin{align}
    \textrm{$(um;2)$ }\quad \quad \partial_{u}\Ds^{A}A_{A}^{(1;i)}-\partial_{u}A_{u}^{(1;i)}=\int_{\I}&\big(16\pi T_{um}^{(2)}\ov{V}^{(i)m}+\partial_{u}C_{2}) \label{um2}
\end{align}
where $A_{A}^{(i)}(u,\theta)$ and $A_{u}^{(i)}(u,\theta)$ are defined in \Cref{lemrad} and $C_{2}$ vanishes in the stationary eras. Integrating \cref{um2} and using \cref{statmetric,statstress} and using the decomposition of $A_{Am}^{(1)}$, $A_{um}^{(1)}$ in the stationary era given by \Cref{lemstat} as well as the decomposition of $\varphi_{mn}^{(1)}$ and $N_{mn}$ given by \Cref{lemrad} and \cref{Nmn} respectively yields the desired relation
\begin{equation}
    \mathscr{D}^{2}\Delta S^{(i)}(\theta)=\Delta Q^{(i)} + 16\pi \mathcal{J}^{(i)}(\theta)
\end{equation}
where the $\mathcal{J}^{(i)}(\theta)$ are defined by \cref{nullfluxum}. Finally, the $(mn;2)$ components of Einstein's equation place no further constrains on the change in $\varphi_{mn}^{(1)}$ and therefore, \Cref{lemrad,lemstat} imply that $\Delta \varphi_{mn}^{(1)}$ is given by \cref{Deltaphimn}. 

\par That $T(\theta)$ and the $S^{(i)}(\theta)$ generate an asymptotic supertranslation and an asymptotic angle-dependent internal isometry between the stationary eras follows from \Cref{supertranshAB,superintAAm} and that $\mathcal{N}_{ab}=0$ in the stationary eras. 
\end{proof}
\end{thm}
\par We finally consider the spherical harmonic dependence of the change in the metric coefficients $\Delta h_{AB}^{(1)},\Delta h_{Am}^{(1)}$ and $\Delta \varphi_{mn}^{(1)}$. We first note that, by \Cref{lemstat}, $\Delta \varphi_{mn}^{(1)}$ is clearly spanned only by $\ell=0$ spherical harmonics. By \Cref{propsymm}, if $T(\theta)$ is spanned by $\ell=0,1$ spherical harmonics then $\Ds_{A}\Ds_{B}T(\theta)=-q_{AB}T(\theta)$. Therefore, it follows that the tracefree part of $\Delta h_{AB}^{(1)}$ on $S^{2}$ is orthogonal to the $\ell=0,1$ spherical harmonics. Furthermore, by the form of \cref{DeltaAAm}, we have that $\Delta A_{Am}^{(1)}$ is orthogonal to the $\ell=0$ spherical harmonics. 

\section{The Memory Effect in Compactified Spacetimes}\label{sec:memory_effect_compactified}
\subsection{Unification of memory effects}\label{sec:unification}

We now explore the geometric interpretation of \Cref{changemetthm} in terms of the memory effect, which is an observable quantity. Physically, the memory effect is the permanent relative displacement of a system of test particles, initially at rest, caused by the passage of a burst of gravitational radiation. The relative displacement of test particles is governed by the geodesic deviation equation 
\begin{equation}
\label{geodev}
(v^{M}\nabla_{M})^{2}\xi^{N}=-R_{MPQ}{}^{N}v^{M}v^{Q}\xi^{P} ,
\end{equation}
where $v^{M}$ is the tangent vector of the worldline of the particle, $\xi^{M}$ is the deviation vector and $R_{MPQ}{}^{N}$ is the Riemann tensor. We are interested in the displacement of test particles located near future null infinity and shall determine the leading order memory effects in a $1\over r$ expansion in a neighborhood of null infinity. 

\par We consider a spacetime where the metric near future null infinity is stationary at leading order in $1\over r$, at asymptotically early and late retarded times. In this subsection, we will simplify and integrate \cref{geodev} to derive an explicit formula for the memory effect. This discussion is a modification of a similar analysis found in \cite{Satishchandran:2017pek}. There are subtle differences when one considers compact internal manifolds, which makes the argument worth revisiting.

\par Consider an array of initially stationary test particles in a neighborhood of null infinity, which we model as a congruence of time-like geodesics whose tangents $v^{A}$ initially point in the $(\partial /\partial u)^{M}$ direction. In a neighborhood of null infinity, the spacetime metric deviates from the Ricci-flat direct product metric (\ref{background}) at order $1\over r$. Consequently, the geodesic equation implies that $v^{M}$ differs from the corresponding integral curve of $(\partial /\partial u)^{M}$ only at order $1\over r$ and therefore $u$ will differ from an affine parameterization beginning at this order. 

For an arbitrary internal manifold, the curvature is generically non-vanishing at infinity. Nevertheless, these considerations imply that the quantity $R_{MPQ}{}^{N}v^{M}v^{Q}$ in \cref{geodev} does vanish at infinity and is only non-vanishing at order $1\over r$. Therefore, the deviation of $v^{M}$ from $(\partial/\partial u)^{M}$ in \cref{geodev} can only affect $\xi^{N}$ at order $1\over r^{2}$ and faster fall-off. Finally, by \cref{WeylRiemSchout}, the Riemann tensor differs from the Weyl tensor at $O\big(\frac{1}{r^{2}}\big)$ since the stress energy falls off like $\frac{1}{r^{2}}$. Since we are only considering the memory effect at leading order in $1\over r$, we can replace $v^{M}$  with $(\partial /\partial u)^{M}$ and $R_{PML}{}^{N}v^{P}v^{L}$ with the electric Weyl tensor $\mathcal{E}_{M}{}^{N}$ (as defined in \cref{elweyl1}) in \cref{geodev} which yields
\begin{equation}
\label{geodevweyl}
\frac{\partial^{2}}{\partial u^{2}}\xi^{M}=-\mathcal{E}^{M}{}_{N}\xi^{N} 
\end{equation}
Indices on the right hand side of \cref{geodevweyl} are raised and lowered with the asymptotic metric $\hat{g}_{MN}$.
\Cref{geodevweyl} implies that $\xi^{M}$ differs from the integral curve of its initial value $\xi^{M}_{0}$ at order $1\over r$ and we may replace $\xi^{M}$ by its initial value in the right hand side of \cref{geodevweyl}. Thus, at leading order in $1\over r$, we have 
\begin{equation}
\label{geodevweyl0}
\frac{\partial^{2}}{\partial u^{2}}\xi^{(1)M}=-\mathcal{E}^{M}{}_{N}\xi^{N}_{(0)} ,
\end{equation}
where $\xi_{M}^{(1)}$ is the deviation vector at $O\big(\frac{1}{r}\big)$. Integrating \cref{geodevweyl0} twice, we obtain 
\begin{equation}
\xi^{(1)M}\bigg\vert^{u=\infty}_{u=-\infty}=\Delta^{M}{}_{N}\xi^{N}_{(0)} ,
\end{equation}
where 
\begin{equation}
\label{memory}
\Delta_{MN}\equiv -\int_{-\infty}^{\infty}du^{\prime}\int_{-\infty}^{u^{\prime}}du^{\prime\prime}\mathcal{E}_{MN}.
\end{equation}
We refer to $\Delta_{MN}$ as the {\em memory tensor}. This characterizes the memory effect as a linear map on the initial displacement to the change in the relative separation. Further, as noted in \Cref{nonlindimred}, the only non-vanishing components of ${\mathcal E}_{MN}$ are ${\mathcal E}_{ab} = -{1\over 2} \partial_u \mathcal{N}_{ab}$ where $a,b$ are along $S^2 \times \I$. This gives a simpler manifestly gauge-invariant relation for the memory, 
\begin{equation}
\label{memoryandnews}
\Delta_{ab}(\theta,y)\equiv \frac{1}{2}\int_{-\infty}^{\infty}du \, \mathcal{N}_{ab}(u,\theta,y).
\end{equation}
From~\C{memoryandnews}, it follows that
\begin{equation}
\Delta_{ab}=\Delta_{ba}, \quad  \mathfrak{q}^{ab}\Delta_{ab}=q^{AB}\Delta_{AB }+\hat{g}^{mn}\Delta_{mn}=0, 
\end{equation}
and clearly $\Delta_{ab}$ is time-independent. Additionally from \cref{news_conditions}, we see that
\begin{equation}
\Dint^{2}\Delta_{AB}=0, \quad \Dint^{2}\Delta_{Am}=0, \quad \Dint^{2}\Delta_{mn}+2\IRiem_{m}{}^{p}{}_{n}{}^{q}\Delta_{pq}=0.
\end{equation}
Using arguments identical to those in the proof of \Cref{nonlindimred}, we see that $\Delta_{AB}$ is independent of internal coordinates $y^{m}$ and $\Delta_{Am}$ and $\Delta_{mn}$ can be uniquely decomposed in a basis of harmonic $1$-forms $\ov{V}_{m}^{(i)}$ and Lichnerowicz zero modes $\ov{t}_{mn}^{(i)}$, respectively,  
\begin{equation}
\label{memdecomp}
\Delta_{Am}=\sum_{i=1}^{b_{1}}\Delta^{(i)}_{A}(\theta)\otimes \ov{V}_{m}^{(i)}(y)\; \textrm{ and }\;  \Delta_{mn}=\sum_{i=1}^{\dims}\Delta^{(i)}(\theta)\ov{t}_{mn}^{(i)}(y)+\frac{1}{D-4}\Imet_{mn}\Imet^{pq}\Delta_{pq}(\theta).  
\end{equation}
The $\Delta_{A}^{(i)}$ are a collection of $b_{1}$ $1$-forms on $\Sp^{2}$, and the $\Delta^{(i)}$ are smooth functions on $\Sp^{2}$.  

We now provide a geometric interpretation of \Cref{changemetthm}. In the gauge given in~\Cref{lemrad}, the news tensor can be expressed in terms of the leading order metric~\cref{nicenews}. This provides a direct relation  between the change in the $h_{AB}^{(1)},h_{Am}^{(1)}$ and $h_{mn}^{(1)}$ before and after the radiation epochs:
\begin{equation}\label{memory_to_change_in_metric}
\Delta_{AB}=\frac{1}{2}\Delta h_{AB}^{(1)}, \quad \Delta_{Am}=\frac{1}{2}\Delta A_{Am}^{(1)} \quad \textrm{ and } \quad\Delta_{mn}=\frac{1}{2}\Delta \varphi_{mn}^{(1)} \, .
\end{equation}
Using the results of \Cref{changemetthm} we can now relate the memory to the change in the metric due to a burst of radiation. We first note that certain metric components appearing in \Cref{changemetthm} can be directly related to definitions of the Bondi mass aspect and electric charge aspect in $\B$. 
\be 
m_{B}\equiv -\frac{1}{2}\mathcal{E}_{rr}^{(3)}=\frac{1}{2}h_{uu}^{(1)} \text{ and }Q^{(i)}\equiv F_{ur}^{(2;i)}=A_{u}^{(1;i)} \text{ (in a stationary era)},
\ee 
where $F=dA$ using the exterior derivative on $\B$ and $A_{\mu m}$ is defined in \cref{genmet}. 
Using the results of \Cref{changemetthm} and \cref{memdecomp} we see that
\begin{align}
    & \Ds^{A}\Ds^{B}\Delta_{AB} = 2\Delta m_{B}-\frac{1}{2}\Delta \phi- 8\pi \mathcal{F}(\theta),\quad q^{AB}\Delta_{AB}=-\frac{1}{2}\Delta \phi, \label{massflux} \\
    &\Ds^{A}\Delta_{A}^{(i)}(\theta)=\frac{1}{2}\Delta Q^{(i)} + 8\pi \mathcal{J}^{(i)}(\theta), \label{chargeflux} \\
    &\Delta^{(i)}=\frac{1}{2}\Delta \Phi^{(i)} \; \text{ and }\; \hat{g}^{mn}\Delta_{mn} = \frac{1}{2}\Delta \phi. \label{scalarflux} 
\end{align}
In analogy with the decomposition of the news in \cref{NAB_and_NAm,Nmn} we can decompose the flux $\mathcal{F}(\theta)$ into gravitational, electromagnetic and scalar contributions to the flux: 
\be 
\label{totflux}
\mathcal{F}(\theta) = \mathcal{F}_{\text{GR}}(\theta) + \mathcal{F}_{\text{EM}}(\theta) + \mathcal{F}_{S}(\theta) ,
\ee 
where 
\begin{align}
    \mathcal{F}_{\text{GR}}(\theta) =& -\int_{\mathbb{R}} du  \,  \bigg(\ov{T^{(2)}_{uu}}+ \frac{1}{32\pi }N^{AB}N_{AB}\bigg) , \label{GRflux}\\ 
    \mathcal{F}_{\text{EM}}(\theta)=&-\sum_{i=1}^{b_{1}}\int_{\mathbb{R}} du \, \mathcal{N}^{(i)A}\mathcal{N}^{(i)}_{A} , \label{EMflux}\\
    \mathcal{F}_{\text{S}}(\theta) =&-2\int_{\mathbb{R}} du ~N^{2} - \sum_{j=1}^{\dims}\int_{\mathbb{R}}du ~(\mathcal{N}^{(j)})^{2} \label{Sflux}.
\end{align}
From the point of view of reduction, \cref{GRflux} corresponds to the flux of four-dimensional gravitational radiation energy as well as null stress energy. \Cref{EMflux} corresponds to the flux of electromagnetic energy and \cref{Sflux} is the flux of scalar energy where the first term is the contribution from the volume mode and second term is the contribution from the volume-preserving moduli. 

We can give a physical interpretation to these relations, which express memory in terms of fluxes. First consider \cref{massflux}. The spherically symmetric part of the left hand side vanishes. The  right hand side defines a change in the spherically symmetric part of the mass aspect. It is reasonable to view 
\be 
m = m_{B}-{1\over 4} \phi  \qquad \text{ (in a stationary era) } \, ,
\ee 
as the mass since the change in this quantity is determined by the energy flux to $\scrip$ in analogy with the four-dimensional result~\C{derivmemory-null}. Similarly,  $Q^{(i)}$ is the electric charge for each asymptotic gauge-field $A^{(i;1)}_{\mu}$ since $\Delta Q^{(i)}$ is determined by the charge flux to $\scrip$. Via~\C{scalarflux}, scalar memory is defined by the change in the scalar charge, given by the coefficient of the ${1\over r}$ term in the expansion of the field near $\scrip$, between early and late times. In this case, there is no integrated flux term.   

The memory effect $\Delta_{AB}$ corresponds to the permanent relative angular displacement of a pair of freely falling test masses. $\Delta_{Am}$ corresponds to the displacement in the internal space directions (i.e. along Killing directions) for a pair of test masses that are initially angularly displaced. If the test masses had some initial displacement in the internal space then, due to a change in scalar charge, the relative displacement in the internal space will change by an amount $\Delta_{mn}$. Physically, the internal space is small and therefore relative displacements of test masses into the internal space are undetectable. Nevertheless, the four-dimensional scalar and electromagnetic memory effects are usually described in terms of velocity kicks~\cite{Bieri:2013hqa,Tolish:2014bka}. We should be able to recover this way of observing memory from the higher-dimensional gravitational picture.

\par To see how this emerges, consider the geodesic motion of a test particle with velocity $v^{M}$ 
\be 
\label{geodeqn}
v^{M}\nabla_{M}v^{N}= 0 ,
\ee 
which follows from varying the point-particle action
\begin{align}\label{point_particle_action}
    S = - m \int \sqrt{ - \G_{M N} ( x ) \, dx^M \, dx^N } \, .
\end{align}
This equation of motion \cref{geodeqn} describes the motion of a point particle following a timelike geodesic. We consider the case where the tangent $v^{M}$, initially $v^{M}_{(0)}$, is of the form
\begin{gather}\label{initialv}
    v^{M}_{(0)}\equiv c_1 \bigg(\frac{\partial}{\partial u}\bigg)^{M}+ c_2 \ov{V}^{m}(y)\bigg(\frac{\partial}{\partial y^{m}}\bigg)^{M} ,  \\[8pt]
    c_1^2 = \frac{1 + \sqrt{ 1 + 4 q^2}}{2} \, , \qquad c_2^2 = \frac{- 1 + \sqrt{ 1 + 4 q^2}}{2} \, ,
\end{gather}
where $\ov{V}^{m}(y)$ is a unit normalized Killing vector, which is automatically geodesic on $\I$: 
\be 
\ov{V}^{m}\bm{D}_{m}\ov{V}^{n}=0 \text{ and }\Imet^{mn}\ov{V}_{m}\ov{V}_{n}=1 .
\ee 
This characterizes an initially stationary test particle with charge $q$ determined by the velocity in the internal direction at some early time $u=u_0$. The vector field $\ov{V}^{m}$ must be Killing to ensure the test particle is constructed from zero modes of the internal space. Since our discussion is purely classical, we will not worry about quantization conditions on the internal momentum, which force such momenta to be of order the Kaluza-Klein scale.  

We are interested in the velocity kick of this test particle relative to a preferred class of asymptotic, stationary observers, which will define our lab frame. To define a time-like vector field $v_{M}^{\text{lab}}$, we Lie-transport the tangent vector $v_M^{(0)}$, so that $v_M^{\text{lab}}$ in our coordinates agrees with the trivial extension $v_M^{(0)}$  for all $u>u_{0}$. We note that this is an accelerated reference frame, which implies that it differs from geodesic evolution of $v_M^{(0)}$ at order $\frac{1}{r}$:
\be 
\label{VMexp}
v_{M} = v_{M}^{\text{lab}}+\frac{v_{M}^{(1)}(u,\theta,y)}{r}+O\bigg(\frac{1}{r^{2}}\bigg) .
\ee 
Expanding \cref{geodeqn} in powers of $\frac{1}{r}$ and integrating the geodesic equation a straightforward computation yields in the gauge described by \Cref{lemrad} that the non-vanishing components of the velocity kick are $\Delta v^{A(1)}$ and $\Delta v^{r(1)}$. 
\be \label{kickone}
\Delta v^{(1)}_{A}(u,\theta)=c_{1}^{2}\int_{-\infty}^{u}du^{\prime} \partial_{u^{\prime}}h_{uA}^{(1)}+\frac{q}{2} \int_{-\infty}^{u}du^{\prime}\mathcal{N}_{Am}\ov{V}^{m} .
\ee 
The first term on the right hand side of~\C{kickone} is not proportional to the charge! Rather it is finite as $q \rightarrow 0$ and corresponds to a purely gravitational velocity kick. This effect actually has nothing to do with the compact internal space and is present in just $\B$. It would be very interesting to explore the potential observability of this effect. The second term is the electromagnetic kick we expect. Note that $\mathcal{N}_{Am}\ov{V}^{m}$ is independent of $y$ because of~\cref{NAB_and_NAm}.
Similarly, the radial velocity kick
\be 
\Delta v^{r(1)}(u,\theta) = \frac{c_{2}^{2}}{2}\int_{-\infty}^{u}du^{\prime}\mathcal{N}_{mn}\ov{V}^{m}\ov{V}^{n},
\ee 
is sensitive to radiation from the specific scalar zero modes associated to the torus component in the decomposition theorem of~\cite{Fischer}.

The total velocity kicks in the angular and radial directions, respectively, are given by
\begin{align}
   \Delta v_A (\theta) &\equiv \lim_{u \to \infty} \Delta v_A^{(1)}(u,\theta)  \, , \\
   \Delta v^r(\theta) &\equiv  \lim_{u \to \infty} \Delta v^{r(1)}(u,\theta)  \, .
\end{align}
Using \cref{uA2} we find that the integrand of the first term in \cref{kickone} can be expressed in terms of an integral of the news:
\be 
\label{huaident}
\partial_{u}h_{uA}^{(1)}=\frac{1}{2}\Ds^{B}N_{BA}+\frac{1}{4}\Ds_{A}N+\frac{1}{2}\Ds_{A}h_{ur}^{(1)}-\frac{1}{2}\partial_{u}^{2}\ov{h}_{rA}^{(2)} .
\ee 
Integrating \cref{huaident} and using \cref{Dphihur} implies that 
\be 
\Delta h_{uA}^{(1)}(\theta) = \frac{1}{2}\int_{\mathbb{R}} du \Ds^{B}N_{BA} .
\ee 
Using \cref{memoryandnews} yields the total velocity kick in terms of the memory 
\begin{align}
\Delta v_{A}(\theta) =& c_{1}^{2}\Ds^{B}\Delta_{BA}+q\Delta_{Am}\ov{V}^{m} ,\\ 
\Delta v^{r}(\theta) =& c_{2}^{2}\Delta_{mn}\ov{V}^{m}\ov{V}^{n}   .
\end{align}

This leaves the question of how to detect radiation for moduli associated to the simply-connected component of $\I$. It appears that directly detecting such radiation requires a more sophisticated detector, but we can make one comment on this issue. In principle, a detector can measure $\mathcal{N}_{AB}, \mathcal{N}_{Am}$ and the torus contribution to $\mathcal{N}_{mn}$ by the motion of the arms of a LIGO-like detector and the motion of a charged test particle. Squaring these contributions gives us all of \cref{totflux} except any unknown null stress-energy, including contributions from additional moduli. We can use the measured fluxes to compute what should be the dominant contribution to the right hand side of \cref{massflux}. Assuming the size of the ordinary memory effect compared with the radiation contribution is still small, and there is a sizeable discrepancy between the observed gravitational memory and the flux computation, we can place upper bounds on the possible contribution of any additional moduli.

\subsection{The circle case}\label{sec:circle}

The original beauty of Kaluza-Klein theory was a unification of electromagnetism, gravity and scalar field theory in a single $5$-dimensional theory of gravity compactified on a circle. Let us revisit this beautiful and simple example to unify the separately studied notions of memory for gravity~\cite{Zeldovich:1974gvh,Christodoulou:1991cr}, electromagnetism~\cite{Bieri:2013hqa,Pasterski:2015zua,Bieri:2011zb} and scalar theories~\cite{Tolish:2014bka} in the framework of $5$-dimensional gravity using the discussion of section \ref{sec:unification}. 

Let us take a spacetime metric with an exact $U(1)$ isometry,
\begin{equation}
    \G_{MN} dx^M dx^N = g_{\mu\nu}dx^\mu dx^\nu + e^{2\varphi(x)} (dy+ A_\mu(x) dx^\mu)^2,  
\end{equation}
where $y \sim y + 2\pi L$ and $\varphi \rightarrow 0$ at infinity. Reducing the $D=5$ Einstein-Hilbert action with zero cosmological constant on $y$ gives the $4$-dimensional action,
\begin{equation}\label{circle_reduced_EH_action}
    S = {1 \over 16\pi G} \int d^4x \, e^{\varphi(x)} \sqrt{g} \left( R - \frac{1}{4} e^{2 \varphi} F_{\mu \nu} F^{\mu \nu} + \partial_\mu \varphi \partial^\mu \varphi \right),  
\end{equation}
where $F=dA$. This is a special case of $\I$ that we studied earlier in the frame we have assumed in our discussion so far, which is not Einstein frame! The $1\over r$ terms in the expansion of $A_\mu$ and $e^{2\varphi(x)}$ can be identified with $A_{\mu y}^{(1)}$ and $\varphi_{yy}^{(1)}$ defined in \cref{genmet} and discussed in the preceding sections.

Specializing \cref{point_particle_action} to the case of a $\B \times S^1$ gives the geodesic equation,
\begin{align}\label{geoqen}
    \frac{d^2 x^M}{d \tau^2} + \tensor{\Gamma}{^M_N_P} \frac{d x^N}{d \tau} \frac{dx^P}{d \tau} = 0 \, ,
\end{align}
with the Christoffel symbols given to leading order in $\frac{1}{r}$ by
\begin{align}
   \Gamma^{C}_{uu}=q^{CD} \partial_{u}h_{uD}^{(1)} \quad ,\quad \Gamma^{C}_{uy }=\frac{1}{2} q^{CD}\partial_{u}A^{(1)}_{D} \quad ,\quad \Gamma^{r}_{yy} =\frac{1}{2}\partial_{u}\phi^{(1)} , 
\end{align}
where $\phi^{(1)}=2\varphi^{(1)}$. Assuming an initial $v^M_{(0)}$ of the form~\cref{initialv} gives the following leading order equations of motion, 
\be 
\partial_{u}v^{M(1)}=-c_{1}^{2}\Gamma_{uu}^{M}-2q\Gamma^{M}_{uy}-c_{2}^{2}K^{M}\Gamma^{r}_{yy},
\ee 
where $K^{M}\equiv \left( \frac{\partial}{\partial r} \right)^M$. In this case, the time-dependent behavior of the angular and radial velocity kicks for a particle with charge $q$, which might vanish, is determined using
\be
\partial_{u}v^{C;(1)}=-c_{1}^{2}q^{CD}\partial_{u}h_{uD}^{(1)}-qF_{uA}^{(1)} \, , \quad \partial_{u}v^{r;(1)} = -\frac{c_{2}^{2}}{2}\partial_{u}\phi^{(1)}. 
\ee 
Using the analysis of section~\ref{sec:unification}, the total velocity kick from the far past ($u\rightarrow -\infty)$ to the far future ($u\rightarrow +\infty)$ is given by
\be 
\Delta v_{A} = c_1^2 \Ds^{B}\Delta_{BA} + q \Delta_{Ay}\, , \quad \Delta v^{r} = c_2^2 \Delta_{yy},
\ee
where $\Delta_{BA},  \Delta_{By}$ and $\Delta_{yy}$ are found in \cref{memory_to_change_in_metric}.

One final comment: in the context of subleading soft photon theorems, there are proposals to permit gauge transformations in abelian gauge theory that grow linearly with $r$ near $\scrip$~\cite{Campiglia:2016hvg,Laddha:2017vfh}. This is an interesting possibility, although the asymptotic behavior of the gauge parameter no longer defines a $U(1)$ group element. In the Kaluza-Klein context, allowing such gauge transformations becomes a statement about higher-dimensional gravity, which would generalize the class of diffeomorphisms normally permitted, assuming such a generalization is sensible. It would be interesting to explore this embedding further.

\subsection{Color memory} \label{colormemory}

While most of the analysis in this paper assumes a Ricci-flat $\I$, we cannot resist sketching how color memory studied in~\cite{Pate:2017vwa,Strominger:2017zoo} should also emerge from Kaluza-Klein reduction. The starting point is a higher-dimensional gravity theory which admits a space with non-abelian isometries. We will assume a $D-4$ sphere for simplicity. Let us take an action, 
\begin{equation}
    S = {1\over 2 \kappa_D^2} \int d^Dx  \sqrt{-g} \left(R - 2\Lambda - |F_{D-4}|^2 \right), 
\end{equation}
where $F_{D-4}$ is a $D-4$-form field strength. Compactifying this theory on $S^{D-4}$ with radius $L$ gives an effective four-dimensional potential for the radius $L$ of the form:
\begin{equation}
    V_{\rm eff} = {2\Lambda \over L^{D-4}} - {(D-4)(D-5) \over L^{D-2}} + {N^2 \over L^{3(D-4)}}. 
\end{equation}
Here we assume the sphere metric is $L^2 ds^2_{S^{D-4}}$, where $ds^2_{S^{D-4}}$ is the metric for a sphere of unit volume. The parameter $N$ is proportional to the amount of quantized $F_{D-4}$ flux through the sphere. Since this is a classical gravity theory, we can chose $\Lambda$ conveniently to ensure the resulting spacetime is flat Minkowski. Under this condition, the potential has a minimum with $L$ growing with $N$. This is all we need. We have engineered Minkwoski spacetime from a compactification with non-abelian isometries. In this case, the identity component of the isometry group is $SO(D-3)$. 

Let us return to the geodesic equation~\cref{geoqen} for a test particle with velocity along the sphere. The novelty in this case, by comparison with the Ricci-flat case, is that the internal velocity vector can rotate as higher-dimensional gravitational radiation passes by. In the Ricci-flat case, the Christoffel symbols along Killing directions vanish. For spaces with non-abelian isometry groups, like the sphere, this is no longer true. From a four-dimensional perspective, the color charge would therefore appear to change because of a burst of radiation, in agreement with~\cite{Pate:2017vwa}. 
\subsection{Frames}\label{sec:frames}

The final issue we need to address is the choice of frames. As illustrated in the circle example of section~\ref{sec:circle}, the natural four-dimensional frame that corresponds to studying radiation in terms of the $D$-dimensional metric is not Einstein frame. Let us parametrize the volume mode or breathing mode of the internal metric in analogy with the circle case,
\begin{equation}
    ds^2_{\I} = e^{2\varphi (x)} \G_{mn} dy^m dy^n, 
\end{equation}
where $\varphi \rightarrow 0$ at infinity. To connect with our earlier discussion, note that $\phi=2 (D-4)\varphi^{(1)}$ where $\phi$ is defined in \Cref{lemrad}. Reducing to four dimensions gives an effective action of the form, 
\begin{equation}
    S = {1\over 16\pi G} \int d^4x \, e^{(D-4) \varphi}\sqrt{-g} R + \ldots, 
\end{equation}
where the omitted terms involve scalar and vector fields whose kinetic terms typically depend on $\varphi$. 
Our analysis in terms of $\G$ gives formulae for memory in this frame. To convert to Einstein frame with a canonical Einstein-Hilbert action, we need to perform one conformal transformation and use the relations described in section~\ref{fourdeffectivefieldtheory}. The Einstein frame metric is defined by
\begin{align}
    g_{\mu \nu}^{(E)} &=  e^{(D-4) \varphi} \, g_{\mu \nu},  \nonumber \\
    &= \left( 1 + (D-4) \frac{\varphi^{(1)}}{r} + \cdots \right) g_{\mu \nu} = \eta_{\mu \nu}+\frac{h_{\mu \nu}^{(1)}}{r} + (D-4) \frac{\varphi^{(1)}}{r} \eta_{\mu \nu} + \ldots,  \\
    &=\eta_{\mu \nu}+\frac{h_{\mu \nu}^{(1)}}{r} + \frac{1}{2}\frac{\phi}{r} \eta_{\mu \nu} + \ldots .
\end{align}
Therefore the leading order metric in Einstein frame is
\be 
h_{\mu \nu}^{(1;E)} = h_{\mu \nu}^{(1)}+\frac{1}{2}\phi\eta_{\mu \nu}
\ee 
and so the Einstein news tensor is 
\be 
\mathcal{N}_{AB}^{(E)} = \mathcal{N}_{AB}-\frac{1}{2}Nq_{AB} = N_{AB}.
\ee 
Thus the Einstein news tensor is equivalent to the trace-free Bondi news tensor -- an Einstein frame observer is insensitive to the overall breathing mode as we expect \cite{tahura2020bransdicke}. The components of electromagnetic and scalar radiative degrees of freedom are unchanged: 
\be 
\mathcal{N}^{(E)}_{Am}=\mathcal{N}_{Am} \; \text{ and } \;
\mathcal{N}^{(E)}_{mn}=\mathcal{N}_{mn}  .
\ee 
The memory effects as viewed by such an Einstein frame observer are then given by
\be 
\Delta^{(E)}_{AB} = \Delta_{AB} - \frac{1}{2} q_{AB} \left( q^{CD}\Delta_{CD}\right), \quad  \Delta_{Am}^{(E)} = \Delta_{Am} \; \text{ and } \; \Delta_{mn}^{(E)}= \Delta_{mn} .
\ee 

\section*{Acknowledgements}

We would like to thank Mark Stern and Robert Wald for helpful discussions. C.~F. and S.~S. are supported in part by NSF Grant No. PHY1720480 and PHY2014195. C.~F. is also supported by U.S. Department of Energy grant DE-SC0009999 and by funds from the University of California. G.~S. is supported in part by NSF grants PHY 15-05124 and PHY18-04216 to the University of Chicago.

\newpage
\appendix

\section{Asymptotic Expansion of Einstein's Equations}
\label{EinsEqn}

In this Appendix, we collect some technical results regarding the asymptotic Einstein equations and the decay of certain components of the Ricci tensor that will be used ubiquitously in this paper. To simplify our analysis we assume that the metric is in the gauge described by \Cref{lemrad}.

\subsection{Constraints on the asymptotic expansion} \label{asymptotics}

\par  It is more convenient for our analysis to examine the trace-reversed Einstein equations given by, 
\begin{equation}\label{treversed}
  R_{MN}=8\pi\mathcal{T}_{MN} ,
\end{equation}
where $\mathcal{T}_{MN}$ is the trace-reversed stress tensor:
\begin{equation}
  \mathcal{T}_{MN}=T_{MN}-\frac{1}{D-2}g_{MN} \left(g^{PQ}T_{PQ}\right).
\end{equation}
It is useful to split the Ricci tensor into a linear and nonlinear part using the metric split $\G_{MN} + h_{MN}$ for some chosen $\G$. We define the nonlinear part of the Ricci tensor as 
\begin{equation}
\label{nonlinRicci}
  \mathcal{R}_{MN}\equiv R_{MN}-\widetilde R_{MN}  , 
\end{equation}
where $R_{MN}$ is the Ricci tensor and  $\widetilde R_{MN}$ is the linearized Ricci tensor defined below: 
\begin{equation}
  \widetilde R_{MN}\equiv -{1\over 2} \left( \Box_{\G} h_{MN}+2\hat{R}_{M}{}^{P}{}_{N}{}^{Q}h_{PQ}-2\hat{\nabla}_{(M}\hat{\nabla}^{P}h_{N)P}+\hat{\nabla}_{M}\hat{\nabla}_{N}h \right).
\end{equation}
On the right hand side, all differential operators along with Riemann are defined with respect to $\G$. In the Appendices, we will denote the linearized version of objects with a tilde, just as $\widetilde{R}_{MN}$ is the linear part of $R_{MN}$.

In our analysis we defined $\G$ in~\C{etabondi} while $h_{MN}$ is given by the collection of functions $(h_{\mu\nu}, A_{\mu n}, \varphi_{mn})$ appearing in~\C{ansatz}. We will expand~\C{treversed} to find a series of recursion relations of the form: (linearized Ricci) = (stress-energy) - (non-linear Ricci). We find the following relations:   
\begin{align}
 \label{uun} & [\Ds^{2}+(n-1)(n-2)]h_{uu}^{(n-1)}+2(n-1)\partial_{u}h_{uu}^{(n)}+\Dint^{2}h_{uu}^{(n+1)}+\partial_{u}^{2}(h^{(n+1)} +\phi^{(n+1)}) \cr &-2\partial_{u}\psi_{u}^{(n+1)}=-16\pi \mathcal{T}_{uu}^{(n+1)}+2\mathcal{R}_{uu}^{(n+1)},
\end{align}
\begin{align}
 \label{urn} &[\Ds^{2}+n(n-3)]h_{ur}^{(n-1)}+2h_{uu}^{(n-1)}-2\Ds^{A}h_{Au}^{(n-1)}+2(n-1)\partial_{u}h_{ur}^{(n)}+\Dint^{2}h_{ur}^{(n+1)}+n\psi_{u}^{(n)}\nonumber \\
  & -\partial_{u}\psi_{r}^{(n+1)}-n\partial_{u}(h^{(n)}+\phi^{(n)})=-16\pi \mathcal{T}_{ru}^{(n+1)}+2\mathcal{R}_{ur}^{(n+1)},
\end{align}
\begin{align}
 \label{uAn} &[\Ds^{2}+(n-1)(n-2)-1]h_{uA}^{(n-1)}-2\Ds_{A}(h_{uu}^{(n-1)}-h_{ur}^{(n-1)})+2(n-1)\partial_{u}h_{uA}^{(n)}+\Dint^{2}h_{uA}^{(n+1)}\nonumber \\
  &-\Ds_{A}\psi_{u}^{(n)}-\partial_{u}\psi_{A}^{(n+1)}+\Ds_{A}\partial_{u}(h^{(n)}+\phi^{(n)})=-16\pi \mathcal{T}_{uA}^{(n+1)}+2\mathcal{R}_{uA}^{(n+1)}, 
\end{align}
\begin{align}
 \label{rrn} &[\Ds^{2}+(n-1)(n-2)-4]h_{rr}^{(n-1)}+4h_{ur}^{(n-1)}+2q^{AB}h_{AB}^{(n-1)}-4\Ds^{A}h_{Ar}^{(n-1)}+2(n-1)\partial_{u}h_{rr}^{(n)}  \nonumber \\
  &+\Dint^{2}h_{rr}^{(n+1)}+2n\psi_{r}^{(n)}+n(n-1) (h^{(n-1)} +\phi^{(n-1)})=-16\pi \mathcal{T}_{rr}^{(n+1)}+2\mathcal{R}_{rr}^{(n+1)}, 
\end{align}
\begin{align}
 \label{rAn} &[\Ds^{2}+(n-1)(n-2)-5]h_{rA}^{(n-1)}+4h_{uA}^{(n-1)}-2\Ds_{A}(h_{ur}^{(n-1)}-h_{rr}^{(n-1)})-2\Ds^{B}h_{BA}^{(n-1)}+\Dint^{2}h_{rA}^{(n+1)}\nonumber \\
  &+2(n-1)\partial_{u}h_{rA}^{(n)}-\Ds_{A}\psi_{r}^{(n)}+n\psi_{A}^{(n)}-(n-1)\Ds_{A}(h^{(n-1)}+\phi^{(n-1)})\cr 
  &=-16\pi \mathcal{T}_{rA}^{(n+1)}+2\mathcal{R}_{rA}^{(n+1)},
\end{align}
\begin{align}
 \label{ABn} &[\Ds^{2}+(n-1)(n-2)-2]h_{AB}^{(n-1)}-4\Ds_{(A}h_{B)u}^{(n-1)}+4\Ds_{(A}h_{B)r}^{(n-1)}+2(n-1)\partial_{u}h_{AB}^{(n)}+\Dint^{2}h_{AB}^{(n+1)}\nonumber \\
  &-2\Ds_{(A}\psi_{B)}^{(n)}-2(\psi_{r}^{(n)}-\psi_{u}^{(n)})q_{AB} +(\Ds_{A}\Ds_{B}-(n-1)q_{AB})(h^{(n-1)}+\phi^{(n-1)})
  \nonumber \\
  &-q_{AB}\partial_{u}(h^{(n)}+\phi^{(n)})+2(h_{rr}^{(n-1)}-2h_{ur}^{(n-1)}+h_{uu}^{(n-1)})q_{AB}=-16\pi \mathcal{T}_{AB}^{(n+1)}+2\mathcal{R}_{AB}^{(n+1)}, 
\end{align}
\begin{align}
 \label{um} &[\Ds^{2}+(n-1)(n-2)]A_{um}^{(n-1)}+2(n-1)\partial_{u}A_{um}^{(n)}+\Dint^{2}A_{um}^{(n+1)}-\Dint_{m}\psi_{u}^{(n+1)}-\partial_{u}\psi_{m}^{(n+1)}\nonumber \\
  &+\Dint_{m}\partial_{u}(h^{(n+1)}+\phi^{(n+1)})=-16\pi \mathcal{T}_{um}^{(n+1)}+2\mathcal{R}_{um}^{(n+1)},
\end{align}
\begin{align}
 \label{rm} &[\Ds^{2}+n(n-3)]A_{rm}^{(n-1)}+2A_{um}^{(n-1)}-2\Ds^{A}A_{Am}^{(n-1)}+2(n-1)\partial_{u}A_{rm}^{(n)}+\Dint^{2}A_{rm}^{(n+1)}+n\psi_{m}^{(n)}\nonumber \\
  &-\Dint_{m}\psi_{r}^{(n+1)}-n\Dint_{m}(h^{(n)}+\phi^{(n)})=-16\pi \mathcal{T}_{rm}^{(n+1)}+2\mathcal{R}_{rm}^{(n+1)},
\end{align}
\begin{align}
 \label{Am} &[\Ds^{2}+(n-1)(n-2)-1]A_{Am}^{(n-1)}-2\Ds_{A}(A_{um}^{(n-1)}-A_{rm}^{(n-1)})+2(n-1)\partial_{u}A_{Am}^{(n)}+\Dint^{2}A_{Am}^{(n+1)}\nonumber \\
  &-\Ds_{A}\psi_{m}^{(n)}-\Dint_{m}\psi_{A}^{(n+1)}+\Dint_{m}\Ds_{A}(h^{(n)}+\phi^{(n)})=-16\pi\mathcal{T}_{Am}^{(n+1)} +2\mathcal{R}_{Am}^{(n+1)},
\end{align}
\begin{align}
 \label{mn} &[\Ds^{2}+(n-1)(n-2)]\varphi_{mn}^{(n-1)}+2(n-1)\partial_{u}\varphi_{mn}^{(n)}+\Dint^{2}\varphi_{mn}^{(n+1)}+2\IRiem_{m}{}^{p}{}_{n}{}^{q}\varphi_{pq}^{(n+1)}-2\Dint_{(m}\psi_{n)}^{(n+1)}\nonumber \\
  &+\Dint_{m}\Dint_{n}(h^{(n+1)}+\phi^{(n+1)})=-16\pi\mathcal{T}_{mn}^{(n+1)} +2\mathcal{R}_{mn}^{(n+1)}.
\end{align}
Here we have defined 
\begin{equation}
  \psi_{M}\equiv \partial^{N}h_{NM}, \qquad h^{(n)} \equiv \eta^{\mu\nu} h_{\mu\nu}^{(n)}, \qquad \phi^{(n)} \equiv \Imet^{mn} \varphi_{mn}^{(n)},   
\end{equation}
so that 
\begin{equation}
\label{psiu}
  \psi_{u}^{(n)}=\Ds^{A}h_{Au}^{(n-1)}+(3-n)(h_{ur}^{(n-1)}-h_{uu}^{(n-1)})-\partial_{u}h_{ur}^{(n)}+\Dint^{m}A_{um}^{(n)}  , 
\end{equation}
\begin{equation}
   \label{psir}
  \psi_{r}^{(n)}=\Ds^{A}h_{Ar}^{(n-1)}+(3-n)(h_{rr}^{(n-1)}-h_{ur}^{(n-1)})-q^{AB}h_{AB}^{(n-1)}-\partial_{u}h_{rr}^{(n)}+\Dint^{m}A_{rm}^{(n)}  , 
\end{equation}
\begin{equation}
  \psi_{A}^{(n)}=\Ds^{B}h_{BA}^{(n-1)}+(4-n)(h_{rA}^{(n-1)}-h_{uA}^{(n-1)})-\partial_{u}h_{rA}^{(n)}+\Dint^{m}A_{Am}^{(n)}  , 
\end{equation}
\begin{equation}
  \psi_{m}^{(n)}=\Ds^{A}A_{Am}^{(n-1)}+(3-n)(A_{rm}^{(n-1)}-A_{um}^{(n-1)})-\partial_{u}A_{rm}^{(n)}+\Dint^{n}\varphi_{nm}^{(n)} \, .
\end{equation}
In the body of this work, we will need the expansion of Einstein's equations to order ${1\over r^2}$, and to order ${1\over r^3}$ for the special case of a stationary era. 

\par A direct calculation of $\mathcal{R}_{MN}^{(2)}$ in the gauge of \Cref{lemrad} shows that the non-vanishing components of $\mathcal{R}_{MN}^{(2)}$ can be written entirely in terms of the news~\cref{nicenews}. Explicitly the non-vanishing components of $\mathcal{R}_{MN}^{(2)}$ are given by, 
\begin{align}
   \mathcal{R}_{uu}^{(2)}=&-\frac{1}{4}\mathcal{N}^{ab}\mathcal{N}_{ab}+\frac{1}{2}\partial_{u}\big(h^{(1)}_{ab}\mathcal{N}^{ab}\big), \label{Ruu2}  \\
   \mathcal{R}^{(2)}_{um}=&\frac{1}{4}(\bm{D}_{m}\Phi_{pq})\mathcal{N}^{pq}-\frac{1}{2}\Phi^{pn}\bm{D}_{p}\mathcal{N}_{mn}-\frac{1}{2(D-4)}\phi \bm{D}^{n}\mathcal{N}_{nm}+\frac{1}{2}\bm{D}_{m}(\Phi^{np}\mathcal{N}_{np}), \label{Rum2} \\ 
   \mathcal{R}_{mn}^{(2)}=&-\frac{1}{4}(\bm{D}_{(m}\Phi^{pq})(\bm{D}_{n)}\Phi_{pq})+(\bm{D}^{p}\Phi^{q}{}_{m})(\bm{D}_{[p}\Phi_{q]n})+\frac{1}{2}\Phi^{pq}\bm{D}_{p}\bm{D}_{q} \Phi_{mn} \label{Rmn} \\ \nonumber 
   &+\frac{1}{2(D-4)}\phi \bm{D}^{2} \Phi_{mn}+\frac{1}{4}\bm{D}_{m}\bm{D}_{n}(\Phi^{pq}\Phi_{pq}),
\end{align}
where the product in \cref{Ruu2} is explicitly given by
\be 
h_{ab}^{(1)}\mathcal{N}^{ab} = h^{(1)}_{AB}\mathcal{N}^{AB}+A_{Am}^{(1)}\mathcal{N}^{Am}+\Phi_{mn}\mathcal{N}^{mn}+\frac{1}{D-4}\, \phi\, \hat{g}^{mn}\mathcal{N}_{mn},
\ee 
and the scalars $\Phi_{mn}(u,\theta,y)$ and $\phi(u,\theta)$ are defined in \Cref{lemrad}. The remaining components of $\mathcal{R}_{MN}^{(2)}$ vanish. In \Cref{subsec:changeinmet}, the zero modes of the nonlinear parts of the Ricci tensor appear as ``flux'' terms for the change in  metric. More precisley, we find that the zero modes of $ \mathcal{R}_{uu}^{(2)}$ and  $\mathcal{R}_{um}^{(2)}$ determine the change in the metric due to a burst of radiation. The zero mode of \cref{Ruu2} is manifestly non-vanishing unless $\mathcal{N}_{ab}=0$. To determine the zero mode of $\mathcal{R}_{um}^{(2)}$ we contract with a Killing vector $\ov{V}^{m}$ of $(\I,\Imet_{mn})$ and integrate over $\I$: 
\begin{align}
    \int_{\I} \mathcal{R}_{um}\ov{V}^{m}=&  \frac{1}{4} \int_{\I}\bigg[\mathcal{N}^{pq}(\ov{V}^{m}\bm{D}_{m}\Phi_{pq}) \cr & -2\bm{D}_{p}\bigg(\Phi^{pn}\mathcal{N}_{mn}\ov{V}^{m}+\frac{\phi \mathcal{N}^{p}{}_{m}\ov{V}^{m}}{D-4}-\ov{V}^{p}\Phi^{mn}\mathcal{N}_{mn}\bigg)\bigg], \\
    =&\frac{1}{4}\int_{\I}\mathcal{N}^{mn}\pounds_{\bar{V}}\Phi_{mn},
\end{align}
where in the first line we used the fact that $\Phi_{pq}$ is divergence free, $\phi$ is constant on $\I$ and that $\bar{V}^{m}$ is covariantly constant to write the last three terms in \cref{Rum2} as a total derivative. In the second line we used the fact that $\bar{V}^{m}$ is covariantly constant to write the directional derivative in terms of the Lie derivative. However the decomposition theorem of~\cite{Fischer} states that $\I$ is a free quotient of a Riemannian product of a torus and a connected Ricci-flat space with vanishing $b_1$. For such a product, $\pounds_{\ov{V}}\Phi_{mn}=0$ since $\ov{V}$ is one of the torus isometries. 

At this stage, we want to check whether our ansatz~\C{ansatz} of an expansion in powers of ${1\over r}$ makes sense as an asymptotic expansion. This might seem fairly reasonable because in both  pure gravity and Maxwell-Einstein, there exists a large class of solutions which are smooth at $\scrip$ in a particular gauge~\cite{Chrusciel:2002vb}.\footnote{Note that starting with smooth initial data on a Cauchy surface and evolving that data does not generically lead to a solution with an analytic expansion in ${1\over r}$ near $\scrip$. Rather $\log(r)$ terms can be generated at subleading orders in ${1\over r}$ even in pure gravity~\cite{SEDP_1989-1990____A15_0}. However, there exists a class of initial data in pure gravity that guarantee $C^k$ differentiability at $\scrip$ for any $k$~\cite{Chrusciel:2002vb}. } 
However, this is not the case for a scalar field in four dimensions with null sources~\cite{Satishchandran:2017pek}. A scalar field $\phi$ in Minkowski spacetime satisfying
\begin{equation}\label{scalarsource}
    \Box_{\eta}\phi = J, 
\end{equation}
where $J$ is a source, does not admit a ${1\over r}$ expansion near $\scrip$ when $J \sim {1\over r^2}$, which is a configuration with finite flux through $\scrip$. Rather one must include $\frac{\log(r)}{r^{n}}$ terms in the expansion. This is without dynamical gravity. 

In our case, there is a general obstruction to integrating in from $\scrip$. Namely, if a specific scalar fluctuation of $\I$ is obstructed, or equivalently gets a mass at some order beyond the linearized approximation, then our ansatz is simply not valid for that mode. The mode could never propagate to $\scrip$, which we implicitly assume in our ansatz. We can see this obstruction emerge in the ${1\over r}$ expansion. Consider the $mn$ component of the vacuum Einstein's equations at order ${1\over r^{2}}$, i.e., \cref{mn} for $n=1$ and $\mathcal{T}_{mn}^{(2)}=0$: 
\be 
\label{mn2varphi}
\bm{D}^{2}\varphi_{mn}^{(2)} + 2\IRiem_{m}{}^{p}{}_{n}{}^{q}\varphi_{pq}^{(2)} - 2 \bm{D}_{(m}\psi^{(3)}_{n)} + \bm{D}_{m}\bm{D}_{n}(h^{(3)}+\phi^{(3)}) = 2\mathcal{R}_{mn}^{(2)}.
\ee 
After contracting both sides with a tensor field $t^{mn}(y)$  which is annihilated by Lichnerowicz, it is straightforward to check that the right hand side vanishes. We therefore get the following nonlinear obstruction to our ansatz, 
\be 
\label{Rmn2obst}
\int_{\I}t^{mn}\mathcal{R}_{mn}^{(2)} = 0. 
\ee 
It is straightforward to check that the volume mode, as expected, is unobstructed. Letting $t_{mn}=\Imet_{mn}(y)$ in \cref{Rmn2obst} and using \cref{Rmn} gives 
\be 
\Imet ^{mn}\mathcal{R}_{mn}^{(2)} = \bigg(\frac{1}{4}\bm{D}^{m}\Phi^{pq}\bm{D}_{m}\Phi_{pq} - \frac{1}{2}\bm{D}^{p}\Phi^{qm}\bm{D}_{q}\Phi_{pm} + \frac{1}{4}\bm{D}^{2}\Phi^{2}\bigg),
\ee 
where $\Phi^{2}=\Phi_{mn}\Phi^{mn}$. Integrating over $\I$, 
\begin{align}
\int_{\I}\Imet^{mn}\IRiem_{mn}^{(2)} =& \frac{1}{2}\int_{\I}\bigg(\frac{1}{2}\bm{D}^{m}\Phi^{pq}\bm{D}_{m}\Phi_{pq} - \bm{D}^{p}\Phi^{qm}\bm{D}_{q}\Phi_{pm}\bigg), \\
=&\frac{1}{2}\int_{\I}\bigg( -\frac{1}{2}\Phi^{pq}\bm{D}^{2}\Phi_{pq} + \Phi^{qm}\bm{D}^{p}\bm{D}_{q}\Phi_{pm}\bigg), \\
=&\frac{1}{2}\int_{\I}\bigg(\IRiem^{mpnq}\Phi_{mn}\Phi_{pq}-\IRiem^{mpnq}\Phi_{mn}\Phi_{pq} + \Phi^{qm}\bm{D}_{q}\bm{D}^{p}\Phi_{pm}\bigg), \\
=&~0,
\end{align}
where we have used
$$ \bm{D}^{2}\Phi_{mn}+2\IRiem_{m}{}^{p}{}_{n}{}^{q}\Phi_{pq}=0,$$
and that $\Phi_{mn}$ is divergence-free. As we spelled out in section \ref{subsec:linscalar}, the space of {\it exactly} massless modes $\dims\leq \dm$ is smaller than the kernel of Lichnerowicz. The exactly massless volume-preserving moduli satisfy \cref{Rmn2obst}. Thus, as in \Cref{lemrad}, we truncate the linearized massless moduli to exactly massless moduli and obtain a solution consistent with our ansatz and Einstein's equations at order ${1\over r^{2}}$.  As we will see in section \ref{statRicci}, this truncation also ensures that our ansatz is consistent with Einstein's equations at order ${1\over r^{3}}$. We fully expect that restricting to exactly massless modes is necessary to obtain a solution to Einstein's equations to all orders in ${1\over r}$, however we have not attempted to show this here. Note that this discussion motivates our imposing a similar condition on $T_{mn}^{(2)}$; namely, that $T_{mn}^{(2)}$ be orthogonal to the $\dims+1$ exactly massless scalar modes. 

\subsection{Going to the stationary era gauge}

We now want to show that a metric in the gauge of \Cref{lemrad} can be further restricted at order ${1\over r^2}$ in a stationary era. Specifically,
\begin{align}
    h^{(2)}_{\mu \nu}=\ov{h}^{(2)}_{\mu \nu}(\theta), \qquad &A^{(2)}_{\mu m}=\sum_{i=1}^{b_{1}}A_{\mu}^{(2;i)}(\theta)\otimes \ov{V}_{m}^{(i)}(y^{m}) 
\end{align}
and
\begin{equation}
\label{varphi2}
    \varphi_{mn}^{(2)}=\Phi^{(2)}_{mn}(\theta^{A},y^{m})+\bigg(\bm{D}_{m}\bm{D}_{n} - \frac{\Imet_{mn}}{d-4}\bm{D}^{2}\bigg)\Psi^{(2)}(\theta^{A},y^{m})+\frac{\Imet_{mn}}{d-4}\ov{\phi}^{(2)}(\theta).
\end{equation}
Note that $\varphi_{mn}^{(2)}$ is missing a vector term shown in~\Cref{propsymm}, and $\ov{\phi}^{(2)}$ is constant on $\I$. To achieve this gauge we first make a gauge transformation that is compatible with our ansatz~\C{ansatz}, stationarity conditions and \Cref{lemrad}. We choose a gauge vector field of the form, 
\be
\xi_{M}\sim \frac{\xi_{M}^{(2)}(\theta,y)}{r^{2}} + O\bigg(\frac{1}{r^{3}}\bigg),
\ee 
where $\xi_{M}$ is a $u$-independent gauge transformation. By an analysis similar to the proof of \Cref{lemrad} we see that $\bm{D}^{m}A^{(2)}_{\mu m}=0$ is divergence free and  $\varphi^{(2)}_{mn}$ admits the decomposition given in \cref{varphi2}. In a stationary era, $\mathcal{R}^{(2)}_{\mu \nu }=\mathcal{R}^{(2)}_{\mu n}$ and $T_{MN}^{(2)}=0$. Therefore:
\begin{align}
    \textrm{$(\mu \nu;2)$ }&\quad \quad \quad \bm{D}^{2}h_{\mu \nu}^{(2)}=0, \label{munu2stat}\\
    \textrm{$(\mu m;2)$ }&\quad \quad \quad \bm{D}^{2}A_{\mu m}^{(2)}= 0,   \label{mum2stat}\\ 
    \textrm{$(mn;2)$ }&\quad \quad \quad \bm{D}^{2}\varphi_{mn}^{(2)} + 2\IRiem_{m}{}^{p}{}_{n}{}^{q}\varphi_{pq}^{(2)}+\bm{D}_{m}\bm{D}_{n}(-2h_{ur}^{(2)}+h_{rr}^{(2)}+q^{AB}h_{AB}^{(2)}) \label{mn2stat} \\ 
  \quad &\quad \quad\quad -2\bm{D}_{(m}\bm{D}^{p}\varphi^{(2)}_{n)p}=2\mathcal{R}_{mn}^{(2)}.\nonumber 
\end{align}
We conclude that 
\be 
h_{\mu \nu}^{(2)}=\ov{h}^{(2)}_{\mu \nu}(u,\theta) \quad \textrm{ and } \quad A^{(2)}_{\mu m} = \sum_{i=1}^{b_{1}}A_{\mu}^{(2)}(u,\theta) \otimes \ov{V}_{m}(y). 
\ee 
Using these relations we now study $(mn;2)$. Taking the trace of  $(mn;2)$ gives\footnote{Just to remind the reader, $\Phi_{mn}$ without a superscript denotes the leading order term as in~\C{radgauge}.} 
\be 
-2 \bm{D}^{m}\bm{D}^{n}\varphi^{(2)}_{mn} =  \frac{1}{2}\bm{D}^{m}\Phi^{pq}\bm{D}_{m}\Phi_{pq}- \bm{D}^{m}\Phi^{pq}\bm{D}_{p}\Phi_{mq} + \frac{1}{2}\bm{D}^{2}(\Phi^{pq}\Phi_{pq}),
\ee 
which yields the following equation for $\Psi^{(2)}$:
\be 
\bigg(\frac{D-5}{D-4}\bigg)\bm{D}^{4}\Psi^{(2)}=-\frac{1}{4}\bm{D}^{m}\Phi^{pq}\bm{D}_{m}\Phi_{pq} + \frac{1}{2}\bm{D}^{m}\Phi^{pq}\bm{D}_{p}\Phi_{mq}-\frac{1}{4}\bm{D}^{2}(\Phi^{pq}\Phi_{pq}).
\ee 
We note that the above analysis implies that the right hand side has no zero modes and therefore, we can solve for $\Psi^{(2)}$ in terms of $\Phi_{mn}$.  After solving for $\Psi^{(2)}$ we can then solve for $\Phi_{mn}^{(2)}$: 
\begin{align} \label{phiequation}
    &L[\Phi^{(2)}_{mn}]=-L[\mathfrak{D}_{mn}\Psi^{(2)}]+ 2 \bigg(\frac{D-5}{D-4}\bigg) \bm{D}_{m}\bm{D}_{n}\bm{D}^{2}\Psi^{(2)}-\frac{1}{4}(\bm{D}_{(m}\Phi^{pq})(\bm{D}_{n)}\Phi_{pq}) \\ \nonumber 
    &+(\bm{D}^{p}\Phi^{q}{}_{m})(\bm{D}_{[p}\Phi_{q]n}) + \frac{1}{2}\Phi^{pq}\bm{D}_{p}\bm{D}_{q}\Phi_{mn} + \frac{1}{4}\bm{D}_{m}\bm{D}_{n}(\Phi^{pq}\Phi_{pq})+\frac{\phi\bm{D}^{2}\Phi_{mn} }{2(D-4)}.
\end{align}
 Here $L[\cdot ]$ is the Lichnerowicz operator and $\mathfrak{D}_{mn} \equiv  \big(\bm{D}_{m}\bm{D}_{n}-\frac{\Imet_{mn}}{D-4}\bm{D}^{2}\big)$. As in our discussion of section~\ref{asymptotics}, we again truncate to exactly massless scalar fluctuations for which the right hand side of~\cref{phiequation} has no Lichnerowicz zero modes. This guarantees solvability of \cref{phiequation}. 
On a generic Ricci flat manifold, $\psi^{(2)}$ will not be harmonic and $\Phi_{mn}^{(2)}$ does not satisfy the Lichnerowicz equation. In the special case of $\I = \mathbb{T}^{k}$, we see that $\bm{D}_{m}\Phi_{pq}=0$ and  
\be 
\bm{D}^{2}\Psi^{(2)}=0 \quad \textrm{ $\implies$ }\quad \bm{D}^{2}\Phi^{(2)}_{mn} + 2 \IRiem_{m}{}^{p}{}_{n}{}^{q}\Phi^{(2)}_{pq} =0  \,\,  \text{ for }  \I = \mathbb{T}^{k}.
\ee

\subsection{Ricci in a stationary era}
\label{statRicci}
The last result we want to record is the behavior of the nonlinear part of the Ricci tensor at order ${1\over r^3}$. By a lengthy but straightforward calculation, the following components of the nonlinear part of the Ricci tensor vanish in a stationary era and in our gauge at order ${1\over r^3}$ : 
\begin{equation}
\mathcal{R}_{\mu \nu}^{(3)}=0 \quad \textrm{ and } \quad \mathcal{R}_{um}^{(3)}=0  \textrm{ in a stationary era}, 
\end{equation} 
and the nonvanishing components are 
\begin{align}
&\mathcal{R}_{rm}^{(3)}=-\Dint_{m}(\Phi^{pq}\Phi_{pq})+\frac{1}{2}\Dint_{p}(\Phi^{pq}\Phi_{qm}) \quad \textrm{ in a stationary era}  , \label{rthreeone} \\ 
&\mathcal{R}_{Am}^{(3)}=\frac{1}{4}\Dint_{m}\Ds_{A}(\Phi^{pq}\Phi_{pq})-\frac{1}{2}\Dint_{p}(\Phi^{pq}\Ds_{A}\Phi_{mq}) \quad \textrm{ in a stationary era} . \label{eq:rthree}
\end{align}
Finally, the $\mathcal{R}_{mn}^{(3)}$ component is given by 
\begin{align}
    \mathcal{R}_{mn}^{(3)}=&-\frac{1}{2}\Dint_{(m}\Phi^{pq}\Dint_{n)}\varphi^{(2)}_{pq}+(\Dint^{p}\Phi^{q}{}_{(m})(\Dint_{|p|}\varphi^{(2)}_{n)q})+\frac{1}{2}\Dint_{m}\Dint_{n}(\Phi_{pq}\varphi^{(2)pq})\\ \nonumber 
    &-\Dint_{p}(\Phi^{pq}\Dint_{(m}\varphi^{(2)}_{n)q})+\frac{1}{2}\Dint_{p}(\Phi^{pq}\Dint_{q}\varphi_{mn}^{(2)})+\frac{1}{2}\Dint_{s}\big[\Phi^{s}{}_{P}\Phi^{Pq}\Xi_{mnq}\big]\\ \nonumber 
    &-\frac{1}{2}\Dint_{m}\big[\Phi^{s}{}_{P}\Phi^{Pq}\Xi_{nsq}\big]-\frac{1}{2}\Imet^{kq}\Phi^{ls}\Xi_{msq}\Xi_{kls}+\frac{1}{2}\Imet^{kq}\Phi^{ls}\Xi_{lnq}\Xi_{kms} \\ \nonumber 
    &+ \textrm{non-zero modes},
\end{align}
where $\Xi_{mrq}\equiv 2\Dint_{(m}\Phi_{r)q}-\Dint_{q}\Phi_{mr}$ and `non-zero modes' refers to modes orthogonal to the Lichnerowicz zero modes. Again this obstruction to solving Einstein's equations is generically non-trivial for a Ricci-flat space, but $\int_{\I} t^{mn}\mathcal{R}_{mn}^{(3)} =0$ if $t_{mn}$ is an exactly massless fluctuation, and hence the obstruction vanishes.  
Note that for the special case of $\I = \mathbb{T}^k$, $\mathcal{R}_{mn}^{(3)}=0$.

\section{A Gauge Invariant Derivation of Memory in Linearized Gravity with Compact Extra Dimensions}\label{app:linearized_gauge_invariant}

In this section we will derive the memory effect in linearized gravity for isolated systems with compact extra dimensions using the Bianchi identity. In particular we shall assume, in any neighborhood of null infinity, there exists a gauge in which the metric admits an asymptotic expansion of the form (\ref{ansatz}). 
We now derive the memory effect in a manifestly gauge invariant way using the Bianchi identity for the asymptotic Weyl tensor. Since we shall be working with gauge invariant quantities, we shall only need that the expansion (\ref{ansatz}) is valid in any local neighborhood of null infinity. 

\par We denote the linearized Weyl tensor by $ \widetilde{C}_{MNPQ}$. The linearized Bianchi identity is 
\begin{equation}
\label{linbianchi}
\partial_{[M}\widetilde{C}_{NP]QR}=0.
\end{equation}
The linearized electric Weyl tensor is defined as
\begin{equation}
\widetilde{E}_{PR}\equiv \widetilde{C}_{NPQR}n^{N}n^{Q} ,
\end{equation}
where $n^{N}\equiv (\partial /\partial u)^{N}$. \Cref{nonlindimred} applies to the leading order linearized electric Weyl tensor, which has non-vanishing components $\widetilde{\mathcal E}_{AB}$ and $\widetilde{\mathcal E}_{A m}$ that are harmonic on $\I$. The component $\widetilde{ \mathcal{E}}_{mn}$ satisfies the Lichnerowicz equation on $\I$. Finally, we again have that $q^{AB}\widetilde{\mathcal{E}}_{AB}=\Imet^{mn}\widetilde{\mathcal{E}}_{mn}$. 

We now compute the memory effect from the Bianchi identity. We recall that
\begin{equation}
\label{memlinweyl}
\widetilde\Delta_{MN}=\int_{-\infty}^{\infty}du^{\prime}\int_{-\infty}^{u^{\prime}}du^{\prime\prime} \, \widetilde{\mathcal E}_{MN} .
\end{equation}
We start with the scalar memory effect. Since $\widetilde\Delta_{mn}$ satisfies the Lichnerowicz equation we can expand $\widetilde\Delta_{mn}$ as 
\begin{equation}
\widetilde\Delta_{mn}=\sum_{i=1}^{d_L}\widetilde{\Delta}^{(i)}T_{mn}^{(i)}+\frac{1}{D-4}\Imet_{mn}\Imet^{pq}\widetilde\Delta_{pq} ,
\end{equation}
in terms of $d_L$ trace-free, divergence-free symmetric tensors $T_{mn}^{(i)}$ which satisfy the Lichnerowicz equation. Note that $\Imet$ is defined in~\C{hat_metric}. We note that $\Delta \Phi^{(i)}$ and $\Delta \phi$ in \Cref{changemetthm} are actually gauge invariant quantities and therefore, the derivation of scalar memory is exactly analogous to the derivation in the nonlinear theory: 
\begin{equation}
\widetilde\Delta^{(i)}=\frac{1}{2}\Delta \Phi^{(i)} \, \textrm{ and }\, \Imet^{mn}\widetilde\Delta_{mn}=\frac{1}{2}\Delta \phi .
\end{equation}
For the scalar case, working with gauge invariant variables does not buy us much.

To derive the electromagnetic memory effect, we note that an explicit computation using the linearized metric yields 
\begin{equation}
\label{linweylF}
\ov{ \widetilde C_{\mu \nu \rho m}}=\sum_{i=1}^{b_{1}}\partial_{\rho}F_{\nu \mu}^{(i)}(x^{\mu})\otimes V_{m}^{(i)}(y^{m}) ,
\end{equation}
where the bar on the left hand side denotes a projection to zero modes as described in section~\ref{sec:conventions}. Viewing the left hand side as a $1$-form in the internal space, this means projecting to harmonic $1$-forms on $\I$ in agreement with the expression on the right hand side. $F^{(i)}_{\mu \nu}$ is the field strength for the graviphoton associated to $V_{m}^{(i)}$. 
This field strength is now {\em gauge invariant} and $\partial_{\mu}$ is the derivative operator compatible with the flat metric $\eta_{\mu \nu}$. 

Since the Weyl tensor is trace-free and satisfies the first Bianchi identity, it follows that $F_{\mu \nu}^{(i)}$ satisfies 
\begin{equation}
\partial^{\mu}F_{\mu \nu}^{(i)}=0 \textrm{ and } \partial_{[\mu}F_{\nu \sigma]}^{(i)}=0
\end{equation}
for all $i$.  We then expand $F_{\mu \nu}^{(i)}$ in powers of $1\over r$ near null infinity as given by \cref{ansatz}.  Using \Cref{lemrad}, the the only non-vanishing component of $F_{\mu \nu}^{(i)}$ at order $1\over r$ is $F_{u A}^{(i;1)}$ which, by \cref{linweylF}, is directly related to $\widetilde{\mathcal E}_{Am}$ in the following way,
\begin{equation}
\label{linelAmF}
 \widetilde{\mathcal{E}}_{Am}=-\sum_{i=1}^{b_{1}}\partial_{u}F_{uA}^{(i;1)}(u,\theta)\otimes \ov{V}_{m}^{(i)}(y^{m}).
\end{equation}  The divergence equation for $F_{\mu \nu}^{(i)}$ at order $1\over r^{2}$ constrains the angular divergence of $F_{uA}^{(i;1)}$, 
\begin{equation}
\Ds^{A}F_{uA}^{(i;1)}=\partial_{u}F_{ur}^{(i;2)} .
\end{equation}
Similarly, applying $\epsilon^{AB}$ the Bianchi identity for $F_{\mu \nu}^{(i)}$ at order $1\over r^{2}$ yields 
\begin{equation}
 \epsilon^{AB}\Ds_{A}F_{uB}^{(i;1)}=\partial_{u}\epsilon^{AB}F_{AB}^{(2;i)} .
\end{equation}
Therefore, using \cref{linelAmF,memlinweyl} we find that
\begin{equation}
 \epsilon^{AB}\Ds_{A}\widetilde\Delta_{B}^{(i)}=\Delta\left(\epsilon^{AB}F_{AB}^{(2;i)}\right) \quad \textrm{ and }\quad \Ds^{A}\widetilde\Delta_{A}^{(i)}=\Delta\left(F_{ur}^{(2;i)}\right) .
\end{equation}
On the right hand side, $\Delta$ means the change in the quantity from $u=-\infty$ to $u=+\infty$. 

\par Finally we turn to the gravitational memory effect arising from asymptotic dimensional reduction. Using the fact that the Weyl tensor is divergence-free and satisfies the homogeneous wave equation one can show that the zero mode of $\widetilde{C}_{\mu\nu\rho\sigma }$ satisfies 
\begin{equation}
\label{BianchilinWeyl}
\partial_{[\mu}\ov{\widetilde{ C}_{\nu \rho ]\sigma \kappa}}=0.
\end{equation}
We first focus on the relevant equations for $\ov{\widetilde{E}_{\mu \nu}}$. By analogous manipulations that led to \cref{divweyl,waveweyl} we find that 
\begin{equation}
\label{divlinwavelin}
 \partial^{\mu}\ov{\widetilde{ E}_{\mu \nu}}=0 \quad \textrm{ and }\quad \Box_{\eta} \ov{\widetilde{ E}_{\mu \nu}}=0.
\end{equation}
Therefore, the $\B$ components of the linearized electric Weyl tensor satisfy the {\em same} equations as the components of the linearized electric Weyl tensor in flat spacetime. One major difference is that, when one has compact extra dimensions, $\eta^{\mu \nu}\widetilde{E}_{\mu \nu}$ is non-vanishing. In flat spacetime this quantity does vanish but, in the presence of compact extra dimensions, the tracelessness of the Weyl tensor implies that $\eta^{\mu \nu}\widetilde{E}_{\mu \nu}$ vanishes if only if $\Imet^{mn}\widetilde{E}_{mn}$ vanishes. This is a crucial difference that leads to contributions from the breathing mode of $\I$ to the observed gravitational memory in this frame. We will discuss the choice of frame in section~\ref{sec:memory_effect_compactified}. 
Because of this subtlety we shall explicitly derive the memory effects implied by the system of equations given in \cref{divlinwavelin}.

\par We now expand $\widetilde{E}_{\mu\nu}$ in powers of $1\over r$. The explicit recursion relations relating Weyl tensor components order by order in $1\over r$ can be found in \cite{Satishchandran:2017pek}. By \Cref{lemrad} the only non-vanishing component of $\wcE_{\mu \nu}$ is $\wcE_{AB}$. Since the trace $q^{AB}\wcE_{AB}$ is equivalent to $-\Imet^{mn}\wcE_{mn}$ we shall focus on the trace-free part of $\wcE_{AB}$ on the $2$-sphere. Applying $q^{CA}\Ds_{A}$ to the angle component of the divergence equation in \cref{divlinwavelin} at order $1\over r^{2}$ yields

\begin{equation}
\label{divTFAB}
\Ds^{A}\Ds^{B}\textrm{TF}\big[\wcE_{AB}\big]=-\frac{1}{2}\Ds^{2}q^{AB}\wcE_{AB}+\partial_{u}\Ds^{A}\ov{\widetilde{E}^{(2)}_{Ar}}, 
\end{equation}
where $\textrm{TF}\big[\cdot \big]$ takes a symmetric $2$-tensor on $\Sp^{2}$ and projects out the trace: $T_{AB} \rightarrow T_{AB} - {1\over 2} q_{AB} \left( q^{CD} T_{CD}\right)$. 

The $r$-component of the divergence equation in \cref{divlinwavelin} at order $1\over r^{3}$ gives, 
\begin{equation}
\label{divAr}
 \Ds^{A}\ov{\widetilde{E}_{Ar}^{(2)}}=\ov{q^{AB}\widetilde{E}_{AB}^{(2)}}+\partial_{u}\ov{\widetilde{E}_{rr}^{(3)}}.
\end{equation}
Finally applying $q^{AB}$ to the angle-angle components of the wave equation in \cref{divlinwavelin} at order $1\over r^{3}$ gives 
\begin{equation}
\label{qABEAB3}
[ \Ds^{2}-2]q^{AB}\ov{\widetilde{\mathcal E}_{AB}}+2\partial_{u}q^{AB}\ov{\widetilde{ E}_{AB}^{(2)}}=0 .
\end{equation}
\Cref{divTFAB,divAr,qABEAB3} imply that 
\begin{equation}
\label{deltaEAB}
 \Ds^{A}\Ds^{B}\textrm{TF}\big[\ov{\widetilde{\mathcal E}_{AB}}\big]=[\Ds^{2}-1]\ov{\Imet^{mn}\widetilde{ \mathcal E}}_{mn}+\partial_{u}^{2}\ov{\widetilde{ E}_{rr}^{(3)}} ,
\end{equation}
where we used that fact that $q^{AB}\widetilde{E}_{AB}=-\Imet^{mn}\widetilde{ E}_{mn}$. 

\par \Cref{deltaEAB} constrains the scalar part of $\textrm{TF}[\widetilde{\mathcal E}_{AB}]$ on the $2$-sphere. We now consider the vector part. The vector part of the angle-angle components of memory are determined by the magnetic Weyl tensor on $\B$ given by, 
\begin{equation}
 \widetilde B_{\mu \nu}\equiv \frac{1}{2}\epsilon^{\rho \sigma }{}_{\mu}\widetilde C_{\rho \sigma \nu u} ,
\end{equation}
where $\epsilon_{\mu \nu \rho}$ is the spatial volume form on $\B$ which is related to the volume element on $\B$ by $\epsilon_{\mu \nu \rho }=\epsilon_{u \mu \nu \rho}$; indices are raised with the background flat metric $\eta_{\mu \nu}$. The magnetic Weyl tensor is symmetric, has vanishing $u$-components and, by the first Bianchi identity, is traceless: 
\begin{equation}
 \widetilde B_{u\nu}=0,\quad \widetilde B_{\mu \nu}=\widetilde B_{\nu \mu} \quad \textrm{ and }\quad \eta^{\mu \nu}\widetilde B_{\mu \nu}=0.
\end{equation}
Furthermore, the linearized Bianchi identity and the fact that all components of the linearized tensor satisfies the wave equation implies that 
\begin{equation}
\label{divBboxB}
 \partial^{\mu}\widetilde B_{\mu \nu}=0 \quad \textrm{ and } \quad \Box \widetilde B_{\mu \nu}=0.
\end{equation}
Therefore, the linearized magnetic Weyl tensor satisfies the same relations as the linearized magnetic Weyl tensor in flat spacetime. In contrast to the $\B$ components of the linearized electric Weyl tensor, the magnetic Weyl tensor is traceless. The system of equations given by \cref{divBboxB} are therefore { identical} to their analogous equations in flat spacetime. The derivation of the vector part of memory for perturbations in flat spacetime has been treated previously in~\cite{Bieri_2014}. Since these computations are identical to the derivation of the vector part of $\widetilde\Delta_{AB}$, we will not repeat this analysis here. \Cref{divBboxB} implies the following fall-off for the magnetic Weyl tensor components: 
\begin{equation}
\label{magfalloff}
 \widetilde B_{AB}\sim O \left( {1\over r} \right), \quad \widetilde B_{r\mu}\sim O \left( {1\over r^{2}} \right) , \quad \widetilde B_{rr}\sim O \left( {1\over r^{3}} \right) .
\end{equation}
The final result from analyzing \cref{divBboxB} together with \cref{magfalloff} is
\begin{equation}
\label{magweylABrr}
 \Ds^{A}\Ds^{B}\widetilde B_{AB}^{(1)}=\partial_{u}^{2} \widetilde B_{rr}^{(3)} ,
\end{equation}
where $\widetilde B_{AB}^{(1)}=-\left({1\over 2}\right)\epsilon_{A}{}^{C}\widetilde{\mathcal{E}}_{CB}$ and, explicitly, $\widetilde B_{rr}^{(3)}=\left({1\over 2}\right)\epsilon^{AB}\widetilde C_{ABru}^{(3)}$.

After integrating \cref{magweylABrr,deltaEAB} and using the fact that $q^{AB}\widetilde\Delta_{AB}=-\Imet^{mn}\widetilde\Delta_{mn}$ we find that 
\begin{equation}
\label{TFdeltaAB}
\Ds^{A}\Ds^{B}\textrm{TF}[\widetilde\Delta_{AB}]= \frac{1}{2}[\Ds^{2}-1]\Delta \phi-\Delta \left(\ov{ \widetilde{E}_{rr}^{(3)}}\right) , 
\end{equation}
\begin{equation}
\label{curltracedeltaAB}
 \epsilon^{CA}\Ds_{C}\Ds^{B}\widetilde\Delta_{AB}=-\Delta\left( \widetilde B_{rr}^{(3)} \right) \quad \textrm{ and } \quad q^{AB} \widetilde{\Delta}_{AB}=-\frac{1}{2}\Delta \phi .
\end{equation}
\Cref{TFdeltaAB,curltracedeltaAB} are consistent with the linearized form of \cref{massflux} since, by \Cref{lemstat}, $\Delta \left( \widetilde B_{rr}^{(3)} \right)$ vanishes and $\Delta \phi$ is spherically symmetric under the strong stationarity conditions we imposed. 

\newpage
\bibliographystyle{utphys}
\bibliography{master}

\end{document}